\documentclass[preprints,review,accept,moreauthors,pdftex]{Definitions/mdpi} 

\firstpage{1} 
\pubvolume{XXXXXXXXXX}
\issuenum{0}
\articlenumber{0}
\pubyear{2021}
\copyrightyear{2021}
\history{Received: date; Accepted: date; Published: date}

\usepackage{amssymb}
\usepackage{amsmath}
\usepackage{longtable}
\usepackage{ulem}

\newcommand{\nucleus}[3]{^{#2} _{#3} \text{#1}}

\renewcommand{\paragraph}[1]{\medskip\noindent\textbf{#1}}

\Title{The Gamma-Ray Window to Intergalactic Magnetism}

\Author{Rafael~{Alves Batista}$^1$\orcidA{} and Andrey~{Saveliev}$^{2,3}$\orcidB{}}

\AuthorNames{Rafael {Alves Batista} and Andrey {Saveliev}}

\address{%
$^{1}$ \quad Radboud University Nijmegen, Department of Astrophysics/IMAPP, 6500 GL Nijmegen, The Netherlands \\
$^{2}$ \quad Immanuel Kant Baltic Federal University,
Institute of Physics, Mathematics and Information Technology,
236016 Kaliningrad, Russia \\
$^{3}$ \quad Lomonosov Moscow State University,
Faculty of Computational Mathematics and Cybernetics,
119991 Moscow, Russia \\
}

\corres{Correspondence: r.batista@astro.ru.nl, andrey.saveliev@desy.de}

\abstract{
One of the most promising ways to probe intergalactic magnetic fields (IGMFs) is through gamma rays produced in electromagnetic cascades initiated by high-energy gamma rays or cosmic rays in the intergalactic space. Because the charged component of the cascade is sensitive to magnetic fields, gamma-ray observations of distant objects such as blazars can be used to constrain IGMF properties. Ground-based and space-borne gamma-ray telescopes deliver spectral, temporal, and angular information of high-energy gamma-ray sources, which carries imprints of the intervening magnetic fields. This provides insights into the nature of the processes that led to the creation of the first magnetic fields and into the phenomena that impacted their evolution. Here we provide a detailed description of how gamma-ray observations can be used to probe cosmic magnetism. We review the current status of this topic and discuss the prospects for measuring IGMFs with the next generation of gamma-ray observatories.
}

\keyword{intergalactic magnetic fields; high-energy gamma rays; electromagnetic cascades}

\begin{document}

\tableofcontents

\section{Introduction}

The advent of imaging air Cherenkov telescopes (IACTs) enabled the study of very-high-energy (VHE; $E \gtrsim 1 \; \text{TeV}$) processes involving gamma rays with unprecedented precision. With small angular resolutions ($\theta_\text{psf} \sim 0.1^\circ$), IACTs such as the High-Energy Stereoscopic System (H.E.S.S.)~\cite{hinton2004a}, the Major Atmospheric Gamma Imaging Cherenkov (MAGIC)~\cite{magic2016c, magic2016a}, and the Very Energetic Radiation Imaging Telescope Array System (VERITAS)~\cite{weekes2002a, krennrich2004a} provide a unique view of the gamma-ray universe above TeV energies. 
These observations are supplemented, at higher energies, by measurements with water-Cherenkov detectors such as the High Altitude Water Cherenkov Experiment (HAWC)~\cite{Abeysekara:2013tza}, the Astrophysical Radiation with Ground-based Observatory at YangBaJing (ARGO-YBJ)~\cite{disciascio2015a}, the Large High Altitude Air Shower Observatory (LHAASO)~\cite{lhasso2019a}, and the Tibet Air Shower Experiment (Tibet-AS$\gamma$)~\cite{tibet2007a}. The launch of the Fermi Large Area Telescope~\cite{fermi2009a} (Fermi-LAT) in 2008 was undoubtedly one of the most important landmarks in gamma-ray astrophysics. The complementarity between Fermi-LAT and IACTs has been crucial to glimpse into extreme cosmic accelerators, and to shed light on large-scale properties of the Universe, including the topic of this review: intergalactic magnetic fields (IGMFs).

\bigskip

In the last decades, active galactic nuclei (AGNs) were observed across the whole electromagnetic spectrum, from radio to gamma rays (for reviews see, e.g., Refs.~\cite{padovani2017a, blandford2019a}). Ever since the Whipple Telescope observed the first BL~Lac-type AGN at very-high energies, Mrk~421, in 1992~\cite{punch1992a}, blazars -- a sub-class of AGNs -- have been extensively studied. Because their relativistic jets point approximately towards Earth, their emission can probe the Universe over vast distances. At gamma-ray energies, in particular, they can be used to probe the extragalactic background light (EBL)~\cite{hess2006a, magic2008a, hess2013a, magic2016b, hess2017a, veritas2019a, Acciari:2019zgl} and IGMFs~\cite{magic2010a, hess2014a, veritas2017a}, as well as fundamental physics~\cite{Albert:2007qk,hess2011a,hess2019a}. It is thanks to combined observations of blazars by IACTs and Fermi-LAT that the first studies aiming to constrain IGMFs were possible. 

The process whereby high-energy gamma rays emitted by blazars initiate electromagnetic cascades in the intergalactic space has been known for over half a century~\cite{jelley1966b,gould1966a,1968CaJPS..46..623H, 1974JPhA....7..120S,Berezinsky:1975zz}. It follows immediately from the idea of cascades that magnetic fields can interfere with their development. Despite the numerous works on the topic (e.g.,~\cite{bonometto1971a, wdowczyk1972a, protheroe1986a,Ivanenko1991a,Ivanenko1991b,aharonian1994a}), it was not until later that the potential of electromagnetic cascades as a method to probe IGMFs was fully realised by Plaga~\cite{plaga1995a}, though there have been considerations of the underlying concept even before that \cite{Honda1989}. 

The seminal work of Neronov \& Vovk~\cite{neronov2010a} sparked an avalanche of subsequent investigations along the same lines in the following years (e.g.,~\cite{tavecchio2010a, tavecchio2011a, huan2011a, taylor2011a, vovk2012a, Dolag:2010ni, fermi2018a}), most of which  derived lower bounds on the strength of IGMFs ranging from $B \gtrsim 10^{-18} \; \text{G}$ to $B \gtrsim 10^{-16} \; \text{G}$, depending on the specific details of the analysis, which is in line with more recent works~\cite{finke2015a, fermi2018a, tiede2020a}. So far, only constraints on IGMFs have been derived, as opposed to actual {\it measurements}. Apart from very general considerations, the coherence length ($L_{B}$) of IGMFs has not been constrained by any analysis until very recently, when somewhat weak bounds were obtained based on observations of the neutrino-emitting blazar, TXS~0506+056: $L_{B} \sim$~10~kpc--300~Mpc~\cite{alvesbatista2020a}.

\bigskip
Large-scale cosmic magnetic fields have been investigated using several techniques. For instance, X-ray and radio emission by galaxy clusters have been used to probe the magnetic field in these objects~\cite{asseo1987a, vallee2011a}. Clusters are connected through magnetised ridges observed in radio with instruments like the Low-Frequency Array (LOFAR)~\cite{govoni2019a, osullivan2020a}. At even larger scales, in cosmic voids, measurements are more difficult because of the low density of these regions. This is where high-energy gamma rays from electromagnetic cascades excel: they provide tomographic information of the magnetic fields in these regions. In this case, the short-lived electron-positron\footnote{In the following, unless noted otherwise, we refer to both electrons and positrons as ``electrons''.} pairs produced in electromagnetic cascades are sensitive to the \emph{local} magnetic field. Given the distance scales involved in this type of study, it is probable that, on average, the pairs are formed in cosmic voids, which fill most of the volume of the Universe. Because magnetic fields in voids are virtually unaffected by structure formation, they provide a direct window into the early Universe and the magnetogenesis process (see, e.g., ~\cite{Grasso:2000wj, widrow2002a, ryu2012a, durrer2013a, Vachaspati:2020blt} for reviews). The absence of such fields would indicate that seed magnetic fields originated in astrophysical objects, and were subsequently amplified through dynamo processes until they reached present-day levels of $\sim 1 \; \mu\text{G}$ in galaxies~\cite{ryu2012a, vallee2011a}.

Another way to constrain cosmic magnetic fields (or to explain certain observations) provides only upper bounds. If IGMFs have been generated in the early Universe -- called primordial magnetic fields (PMFs) -- they have an impact on several cosmological aspects. First of all, they represent an additional constituent of the total energy of the Universe and, as such, have an impact on its evolution which results in manifold imprints onto the CMB (see \cite{jedamzik2019a} and the references therein). In fact, they may even be able to reduce the tension between the values of the Hubble constant obtained, on the one hand, from type Ia Supernovae observations and, on the other hand, from Planck measurements of the CMB \cite{Jedamzik:2020krr}. Furthermore, there are claims \cite{DeSimone:2011ek} that, depending on their strength, PMFs created at the electroweak phase transition (EWPT) may prevent the electroweak baryogenesis. Contrarily,  it has been shown recently that Inflation-generated {\it helical} magnetic fields could create the necessary baryon asymmetry in the first place \cite{Kushwaha:2021csq}. Also, strong magnetic fields have an impact on the neutron-proton conversion rate, therefore affecting the rates of the weak reactions which are responsible for the chemical equilibrium of neutrons and protons before Big Bang nucleosynthesis (BBN), thus modifying it~\cite{Grasso:2000wj}.

\bigskip
This article reviews some key results on cosmic magnetism obtained through gamma-ray measurements in the last three decades. First, we present a brief overview of intergalactic magnetic fields, their origin, evolution, and properties, in Sec.~\ref{sec:bfields}. Then, in Sec.~\ref{sec:propagation}, we introduce some gamma-ray sources that have been used for IGMF studies and provide more details on how high-energy gamma rays propagate in the intergalactic space and how they can be used to probe IGMFs, followed by some experimental constraints, in Sec.~\ref{sec:results}. Finally, in Sec.~\ref{sec:outlook}, we reflect upon the status and the main challenges of this particular field, and discuss the prospects for finally measuring IGMFs with gamma-ray telescopes.

\section{Intergalactic Magnetic Fields}
\label{sec:bfields}

Magnetic fields are present on all scales, ranging from small objects like planets to clusters of galaxies and beyond. Galaxies have fields of $B \sim 1 \; \mu\text{G}$~\cite{kulsrud2008a, vallee2011a, ryu2012a, Vazza:2017mbz}, which drive the magnetisation of the circumgalactic medium  via winds~\cite{Bertone:2006mr}. Active galaxies can eject jets of magnetised material into galaxy clusters~\cite{dubois2008a} and even into cosmic voids~\cite{furlanetto2001a}. 
Clusters of galaxies are connected to each other through filaments, whose fields are $B \sim 0.1$--$10. \; \text{nG}$~\cite{vallee2011a, ryu2012a, Vazza:2017mbz}. They compose the cosmic web, whose magnetic properties are poorly known. This is, to a large extent, due to the scarcity of observational data, owing to the intrinsic difficulties in measuring magnetic fields at scales larger than clusters of galaxies. For this reason, numerical simulations play an important role providing the full picture of how magnetic fields are distributed in the cosmic web. For further details, the reader is referred to, e.g., Ref.~\cite{Vazza:2017mbz}.

A natural question that arises is how magnetic fields in galaxies reached the $\mu\text{G}$ level we observe today. One possible explanation is that astrophysical dynamos can amplify seed magnetic fields by many orders of magnitude. In this context, these seed fields are required to have strengths of at least $10^{-22}$ to $10^{-15} \; \text{G}$~\cite{parker1971a, zeldovich1983a}, the actual value depending on the particular model. However, if the seeds are strong enough ($B \gtrsim 10^{-11} \; \text{G}$), one does not need to invoke dynamos. In this case, adiabatic compression \cite{Grasso:2000wj} (potentially together with some stretching and shearing of flows~\cite{Dolag:2002bw}) is sufficient.

Given the distance scales involved in typical IGMF studies using particle probes, only the large-scale distribution of magnetic fields is relevant. In this case, clusters of galaxies fill $\lesssim 10^{-3}$ of the Universe's volume (the so-called volume filling factor), such that their magnetic field is virtually negligible if we are studying how particles propagate over large distances and the effects of magnetic fields upon them.
Filaments are believed to have filling factors $\sim 10^{-3}$--$10^{-1}$, whereas cosmic voids are the most important contribution, with a filling factor $\gtrsim 10^{-1}$. Therefore, magnetic fields in the voids are, to first order, the dominant component that determines how particles propagate over cosmological baselines.

The origin of the seed magnetic fields is one of the main open questions in astrophysics today. In Sec.~\ref{ssec:magnetogenesis}, we briefly mention some of the main mechanisms for magnetogenesis, focusing on providing some estimates of the relevant observables -- field strength, coherence length, and helicity -- that can be probed with high-energy gamma-ray observations. Before that, in Sec.~\ref{ssec:framework}, we provide some important definitions and the mathematical framework required for understanding cosmic magnetism. Conclusively, in Sec.~\ref{sec:FurterConstraints} we present techniques other than gamma-ray astrophysics used to constrain IGMFs and the results originating from them.

\subsection{Statistical Observables}
\label{ssec:framework}

Stochastic magnetic fields can be characterised by a number of observables which correspond to different statistical averages.
The first one is the average magnetic-field strength ($B$). When considering this quantity, it is a common misconception to talk about the \textit{mean} of $B$ because at cosmological scales it is expected that $\langle B_{i} \rangle = 0$ for each individual component $i$, in particular for a Gaussian distribution, which is the typical first-order assumption. The relevant quantity, in this context, is the \textit{root mean square} of the magnetic field, defined from
\begin{equation}
B^{2} \equiv B_{\rm rms}^{2} = \dfrac{1}{V} \int\limits_{V} \mathbf{B}^{2}(\mathbf{r}) \, {\rm d}^{3}r\,,
\end{equation}
where $V$ is the considered volume. 
Another related quantity, the magnetic helicity $H_{B}$, is given by
\begin{equation}
H_{B} = \int\limits_{V} \mathbf{A}(\mathbf{r}) \cdot \mathbf{B}(\mathbf{r}) \, {\rm d}^{3}r\,,
\end{equation}
where $\mathbf{A}(\mathbf{r})$ is the vector potential, i.e.~ $\mathbf{B} = \nabla \times \mathbf{A}$. Originally, $H_{B}$ had been defined for a vanishing normal magnetic field component everywhere at the boundary of $V$, even though it is possible to drop this condition in a more general case~\cite{1984JFM...147..133B}. As the name suggests, $H_{B}$ is directly related to the topology of the magnetic field, more precisely to whether the magnetic field on average is left- or right-handed. It should be noted that magnetic helicity is a well-defined quantity as it is gauge-invariant if the aforementioned requirement of a vanishing normal magnetic field at the border is fulfilled. Furthermore, it is conserved in ideal MHD and plays an important role for the time evolution of the (energy content of the) magnetic field in general (cf.~Sec.~\ref{sec:CosmologicalScenario}).

To define the other quantities, we need to introduce the Fourier transform, which for a given (magnetic) field $\mathbf{B}(\mathbf{r})$ is given by
\begin{equation}
\widetilde{\mathbf{B}}(\mathbf{k}) = \dfrac{1}{(2 \pi)^{\frac{3}{2}}} \int\limits_{V} \mathbf{B}(\mathbf{r}) e^{-i \mathbf{k} \cdot \mathbf{r}} \, {\rm d}^{3}r\,,
\end{equation}
and represents the mode for the wave vector $\mathbf{k}$. We can then determine the statistical connection between any two modes, represented by wave vectors $\mathbf{k}$ and $\mathbf{k}'$, by calculating the corresponding ensemble average (denoted as $\left< ... \right>$), given by
\begin{equation}
\left<\widetilde{B}_{a}(\mathbf{k}) \widetilde{B}_{b}(\mathbf{k}')\right> = (2 \pi)^{3} \delta^{(3)}(\mathbf{k} - \mathbf{k}') \mathcal{P}_{ab}(\mathbf{k}')\,.
\end{equation}
Assuming that the magnetic field is homogeneous and isotropic, the general form of $\mathcal{P}_{ab}$ is
\begin{equation} \label{Bspectrum}
\mathcal{P}_{ab}(\mathbf{k}) \propto \left( \delta_{ab} - \dfrac{k_{a} k_{b}}{k^{2}} \right) M_{k} + \dfrac{i}{c_{H}} \epsilon_{abc} k_{c} \mathcal{H}_{k}\,,
\end{equation}
where $\delta_{ab}$ and $\epsilon_{abc}$ are the Kronecker delta and the Levi-Civita symbol, respectively,  $M_{k}$ is the spectral magnetic energy, $\mathcal{H}_{k}$ is the spectral magnetic helicity density, and $c_{H}$ is a numerical constant which depends on the convention used. It is important to mention here that there is a fundamental relation between $M_{k}$ and $\mathcal{H}_{k}$: one can show (see, for example, \cite{Brandenburg:2004jv}) that for a given $k$ the value of $\mathcal{H}_{k}$ is limited by the corresponding value of $M_{k}$,
\begin{equation} \label{Hmax}
|\mathcal{H}_{k}| \leq \frac{|c_{H}|}{k} M_{k},
\end{equation}
such that the right hand side of Eq.~\ref{Hmax} is also called {\it maximal (spectral) helicity}, and the actual spectral helicity density may be expressed as a fraction $f_{H}$ of it, with $-1 \leq f_{H} \leq 1$.

In general, one assumes
\begin{equation}
M_{k} \propto k^{\alpha_{B} - 1}\,,
\label{eq:Hspectrum}
\end{equation}
which means that the spectrum is given by a power law for which the spectral index $\alpha_{B}$ defines its type.
It is assumed that at small scales (i.e., for large values of $k$), IGMFs have a Kolmogorov ($\alpha_{B} = -2/3$, see \cite{1941DoSSR..30..301K, 1991RSPSA.434....9K}) or an Iroshnikov/Kraichnan ($\alpha_{B} = -1/2$, see \cite{1964SvA.....7..566I,1965PhFl....8.1385K}) spectrum at present time.
Both values for the spectral index were derived from dimensional considerations, with the latter one assumed to be valid under the assumption of a strong mean magnetic field. Still, due to the fact that the numerical values of these two spectral indices are very close to each other, up to the present day it has not been possible to distinguish between them experimentally \cite{2010JGRA..115.2101N,2015FrP.....3...22T}. For large scales (i.e.~small $k$) one expects a Batchelor spectrum ($\alpha_{B} = 5$), as predicted using general causality arguments in \cite{Durrer:2003ja} and confirmed by semi-analytical simulations in \cite{Saveliev:2012ea}. Other works also considered a white noise spectrum ($\alpha_{B} = 3$) \cite{Kahniashvili:2010gp,Jedamzik:2010cy}. On the other hand, IGMFs produced during Inflation (cf.~Sec.~\ref{sec:CosmologicalScenario}) are expected to be scale-invariant, which corresponds to $\alpha_{B} = 0$ (see, for example, \cite{Fujita:2019pmi}). Note that there are different ways to define the spectral index, such that the numerical values in other publications might differ from the one used here while describing the same kind of spectrum.

The last essential characteristic statistical observable considered here is the correlation length ($L_{B}$) which is given by
\begin{equation}
L_{B} = \dfrac{ 2\pi \int k^{-1} M_{k} \, {\rm d}k}{\int M_{k} \, {\rm d}k}\,.
\end{equation}
In a simplified way, $L_B$ can be understood as the average size of the eddies of the magnetic field. Again it should be noted that several different ways to define the correlation length are found in the literature, such that small differences (for example by a factor of a few) are possible. This is discussed, e.g., in~\cite{Harari:2002dy}, where also the power-law case relevant here is addressed in more detail.

\subsection{Origin}
\label{ssec:magnetogenesis}

While the origin of IGMFs is still unknown, there are two classes of scenarios for their magnetogenesis present in the literature \cite{Grasso:2000wj, kulsrud2008a, durrer2013a}. In cosmological scenarios strong seed magnetic fields were created during the early Universe and later decayed to their present state. In astrophysical scenarios, on the other hand, weak seed magnetic fields emerged due to local effects (for example, a battery process) in astrophysical objects, being subsequently amplified by dynamo mechanisms. In the following, we will present possible mechanisms for both of these classes of scenarios. Note that this separation is done here purely for reasons of clarity and comprehensibility. In reality, the situation may be more complex, as the particular mechanism may be the result of (yet) unknown physics, or the actual origin of IGMFs may turn out to be a combination of multiple processes, astrophysical and/or cosmological. 

\subsubsection{Cosmological Scenarios} 
\label{sec:CosmologicalScenario}

The seed magnetic fields of cosmological scenarios, i.e., the PMFs, are thought to be created by some major cosmological effect, such that they permeate the whole Universe. Without any claim to completeness (more details may be found in \cite{Kandus:2010nw,durrer2013a,Vachaspati:2020blt}), we list some of these possibilities below.

\paragraph{Inflation.} Magnetic fields may have been produced during inflation (for a review on Inflation in general, see, e.g.,~\cite{martin2020a}). However, if Maxwellian conformal invariance is preserved, these fields are predicted to be exceedingly weak ($B \lesssim 10^{-50} \; \text{G}$ at the epoch of galaxy formation ~\cite{Kandus:2010nw}), being negligible for all practical purposes~\cite{Turner:1987bw}. Models for inflationary magnetogenesis that are of astrophysical relevance must generate much stronger fields. Because conformally invariant fields are not produced in an expanding conformally-flat spacetime, one has to introduce a coupling of the electromagnetic field with the inflaton, and/or an additional coupling which breaks the conformal or gauge invariance, mainly of the form $R_{\mu\nu\alpha\beta} F^{\mu\nu} F^{\alpha\beta}$ or $R_{\mu\nu} A^{\mu} A^{\nu}$, respectively \cite{durrer2013a} (where $F^{\alpha\beta}$ is the electromagnetic field tensor, $R_{\mu\nu}$ is the Ricci tensor and $R_{\mu\nu\alpha\beta}$ is the Riemann curvature tensor), even though other terms are also possible \cite{Turner:1987bw,Atmjeet:2013yta}. After the seminal publications in the field \cite{Turner:1987bw,Ratra:1991bn} the follow-up works (see, for example, \cite{Martin:2007ue,Subramanian:2009fu,Kunze:2009bs,Motta:2012rn,Jain:2012ga,Domenech:2017caf,Markkanen:2017kmy,Chakraborty:2018dmj,Bamba:2020qdj}) then further explored the idea or investigated more exotic scenarios. Due to a large parameter space the resulting magnetic field strength estimations, even in the simplest models, range from $10^{-65}$ to $10^{-9}$ G \cite{Subramanian:2009fu}.

\paragraph{Post-Inflationary.} It is also possible that magnetic fields emerged between Inflation and the EWPT, for example during or before reheating. The general idea is that the coupling between the electromagnetic and a scalar field breaks the Maxwellian conformal invariance. In particular, the scalar field in question may be an oscillating inflaton, which decays into radiation and reheats the Universe \cite{Kobayashi:2014sga}, resulting in IGMFs with $B \gtrsim 10^{-15}$~G on $\sim {\rm Mpc}$ scales. In another scenario \cite{Long:2013tha}, Majorana neutrino decays result in lepton asymmetries, and ultimately in baryon asymmetries via anomalous processes,  subsequently leading to the violation of lepton/baryon numbers. This then may produce relic hypercharge magnetic fields which are converted to electromagnetic fields during the EWPT, giving $\sim 10^{-18}$~G field strength with $L_{B} \simeq 10$~pc today. More recently, the idea of a Weibel instability emerging and subsequently amplifying a possible inflationary magnetic field during this era has been considered~\cite{Miron-Granese:2021msj}.

\paragraph{Electroweak Phase Transition.} Within the SM, the EWPT is assumed to be rather smooth \cite{Aoki:1996cu}, such that in order to realize a first order transition, mechanisms beyond the Standard Model (BSM) have to be considered \cite{Espinosa:2011ax}. The basic idea of magnetogenesis during the EWPT was first laid out by \cite{Vachaspati:1991nm}. Due to the restrictions of possible values of the vacuum expectation value of the Higgs field, $\Phi$, which breaks the electroweak symmetry, and the fact that it varies with the position in space, we have $\partial_{\mu} \Phi \ne 0$, such that the electromagnetic field strength does not necessarily compensate effects arising from the Higgs field. Hence, we expect a non-vanishing magnetic field after the phase transition. Note, however, that the magnetic field depends on gradients of $\Phi$. Other possible scenarios may be found in \cite{Ellis:2019tjf}, with the general conclusion that magnetic fields of up to $\sim 10^{-11}$~G on scales of $\sim 10$~kpc are possible \cite{Vachaspati:2020blt}. 

\paragraph{Quantum Chromodynamics Phase Transition.} In a similar fashion to the EWPT, it should be mentioned here that within the Standard Model of particle physics (SM) the quantum chromodynamics phase transition (QCDPT) is considered to be of the second order or crossover type \cite{Fodor:2004nz, Aoki:2006we, Bali:2011qj, Bhattacharya:2014ara}, such that, for it to be of first order, a SM extension has to be invoked~\cite{Schwarz:2009ii}. Several works~\cite{Quashnock:1988vs, Cheng:1994yr, sigl1997a} discuss magnetogenesis due to the growth of bubbles of the hadronic phase and, subsequently, charge separation, which ultimately leads to the creation of electric currents and consequently of magnetic fields with an estimated field strength of the order of $\sim 10^{-16}$~G on $\sim$~kpc scales \cite{Cheng:1994yr}. 

\bigskip

It is usually assumed that immediately after magnetogenesis most of the magnetic-field energy is concentrated on a characteristic length called the integral scale. The basic idea, as described in \cite{banerjee2004a}, is that throughout the evolution of the Universe up to the present day, the magnetic energy decays starting with small scales, such that the integral scale is increasing until it reaches the coherence length of IGMFs today. Throughout the years, there has been a large number of simulations, both numerical and (semi-)analytic, which modelled this time evolution for different magnetogenesis scenarios \cite{banerjee2004a,Jedamzik:2010cy,Saveliev:2012ea,Kahniashvili:2012vt,Tevzadze:2012kk,kahniashvili2013a,Subramanian:2015lua,Brandenburg:2017neh}. 

\bigskip

As a final remark it should be pointed out that, due to the fact that magnetic helicity is (nearly) conserved, it plays an important role in the time-evolution of magnetic fields, in particular by causing the so-called inverse cascade of energy, i.e.~the transfer of magnetic energy from small to large scale fluctuations \cite{Brandenburg:2004jv,banerjee2004a,Saveliev:2013uva,Brandenburg:2018ptt,Brandenburg:2020vwp}. These inverse cascades, however, do not seem to be exclusive to helical fields, as shown in recent simulations~\cite{Brandenburg:2014mwa}. 

It is well possible that PMFs actually were helical. One of the first works along these lines was \cite{Vachaspati:2001nb}, suggesting the creation of a left-handed PMF due to a change of the Chern-Simons number. Other possible mechanisms include extra dimensions \cite{Atmjeet:2014cxa}, the coupling to the cosmic axion field \cite{Campanelli:2005ye} or an axion-like coupling \cite{Caprini:2014mja}, the Riemann tensor \cite{Kushwaha:2020nfa}, a spectator field \cite{Fujita:2019pmi}, or an inherently helical coupling \cite{Shtanov:2019civ} during Inflation in the first place. Recently, also the possibility of helicity generation via a chiral cosmological medium around the EWPT has been considered \cite{Vachaspati:2021vam}, however the authors found the effect to be suppressed due to the value of the baryon to entropy ratio.

\subsubsection{Astrophysical Scenarios}
 
A number of possible mechanisms also exists for the astrophysical scenario. They all have in common that magnetic fields are created locally due to some astrophysical process. Some of them are concisely described below.

\paragraph{Biermann Battery.} It is manifestly difficult to create magnetic fields from scratch due to the fact that in classical MHD, if $\mathbf{B}(\mathbf{r}) = \mathbf{0}$ at some instant in time, then this is true for all later times. A way to evade this limitation is through the Biermann battery mechanism \cite{1950ZNatA...5...65B,biermann1951a}, for which the basic idea is that the misalignment of temperature and density gradients induces an electric field which ultimately results in the generation of a magnetic field. Prior to Reionisation, this process produces exceedingly weak fields in the intergalactic space ($B \lesssim 10^{-24} \; \text{G}$)~\cite{Naoz:2013wla}. In protogalaxies, these fields can reach $B \sim 10^{-22}$--$10^{-17} \; \text{G}$~\cite{kulsrud1997a, davies2000a, graziani2015a}. For other astrophysical and cosmological settings see also \cite{Gnedin:2000ax,widrow2002a,Widrow:2011hs,ryu2012a,Attia:2021ywb}.

\paragraph{Galactic Outflows.} One obvious candidate to produce IGMFs are the galaxies themselves, as they eject matter and energy into the intergalactic space. Most authors assume that this can be driven by stars, in particular magnetised winds, or cosmic rays \cite{Bertone:2006mr,Arieli:2011yy,Beck:2012cs,Samui:2017dsz,Garcia:2020kxm}. However, other possibilities exist as well, including the magnetisation of voids by giant radio lobes or bubbles from AGNs, even though energetics requirements generally do not allow for such a substantial effect over the age of the Universe~\cite{Kronberg:2001st,furlanetto2001a, barai2008a, barai2011a}.
   
\paragraph{Cosmic-Ray Return Currents.} In addition to the outflow scenario discussed above, cosmic rays escaping from a galaxy create a charge imbalance resulting in electric fields and, subsequently, return currents. Ultimately, these return currents can produce magnetic fields on scales which are sufficiently large to provide the seed for IGMFs
\cite{Miniati:2010ne,Ohira:2020yvb}.

\paragraph{Photoionisation during the Reionisation Era.} During Reionisation high-energy photons are able to escape from objects like population~III stars, protogalaxies, and quasars into the (then) neutral intergalactic medium (IGM). This causes photoionisation which ultimately causes the generation of radial currents (and electric fields), inducing magnetic fields with strengths $B \sim 10^{-25} - 10^{-20} \; \text{G}$ on scales between $\sim 1 \; \text{kpc}$ and $10 \; \text{Mpc}$ \cite{Durrive:2015cja,durrive2017a,Langer:2018bbk}. Remarkably, this mechanism can generate global magnetic fields through astrophysically-initiated mechanisms. This seeding scheme agrees with results of large-scale cosmological MHD simulations by Garaldi~\textit{et al.}~\cite{garaldi2021a}, although they could, in principle, be subdominant with respect to seeds produced through the Biermann battery.

\paragraph{Primordial Vorticity.} In a seminal paper by Harrison~\cite{Harrison:1973zz}, it was suggested that due to relativistic effects electromagnetic fields are coupled to vorticity\footnote{This mechanism can be viewed as cosmological, since it involves density perturbations. However, the necessary conditions for the vorticity generation involve protogalaxies, so we chose to classify it as an astrophysical magnetogenesis model.}, such that rotating protogalaxies could create primordial vorticity that could generate magnetic fields in the radiation-dominated era. However, vorticity is predicted to decay rather fast in the early Universe, such that more advanced theories based on the same idea, but with vorticity appearing at later stages or using higher-order effects, had to be introduced \cite{1972JETP...34..233M,Matarrese:2004kq,Takahashi:2005nd,Banik:2015owa}.

\bigskip

Several of the mechanisms listed above require a dynamo mechanism in order to amplify the magnetic field strength to the observed present-day values. Especially the small-scale dynamo has attracted major interest in this context (see \cite{Brandenburg:2004jv,Vazza:2017mbz,Donnert:2018lbe,Dominguez-Fernandez:2019ght} for some recent results), even though simulating it numerically poses a challenge due to the size of the scales which have to be resolved.

\subsection{General Constraints} 
\label{sec:FurterConstraints}

\begin{figure}[!ht]
    \centering
    \includegraphics[width=0.8\textwidth]{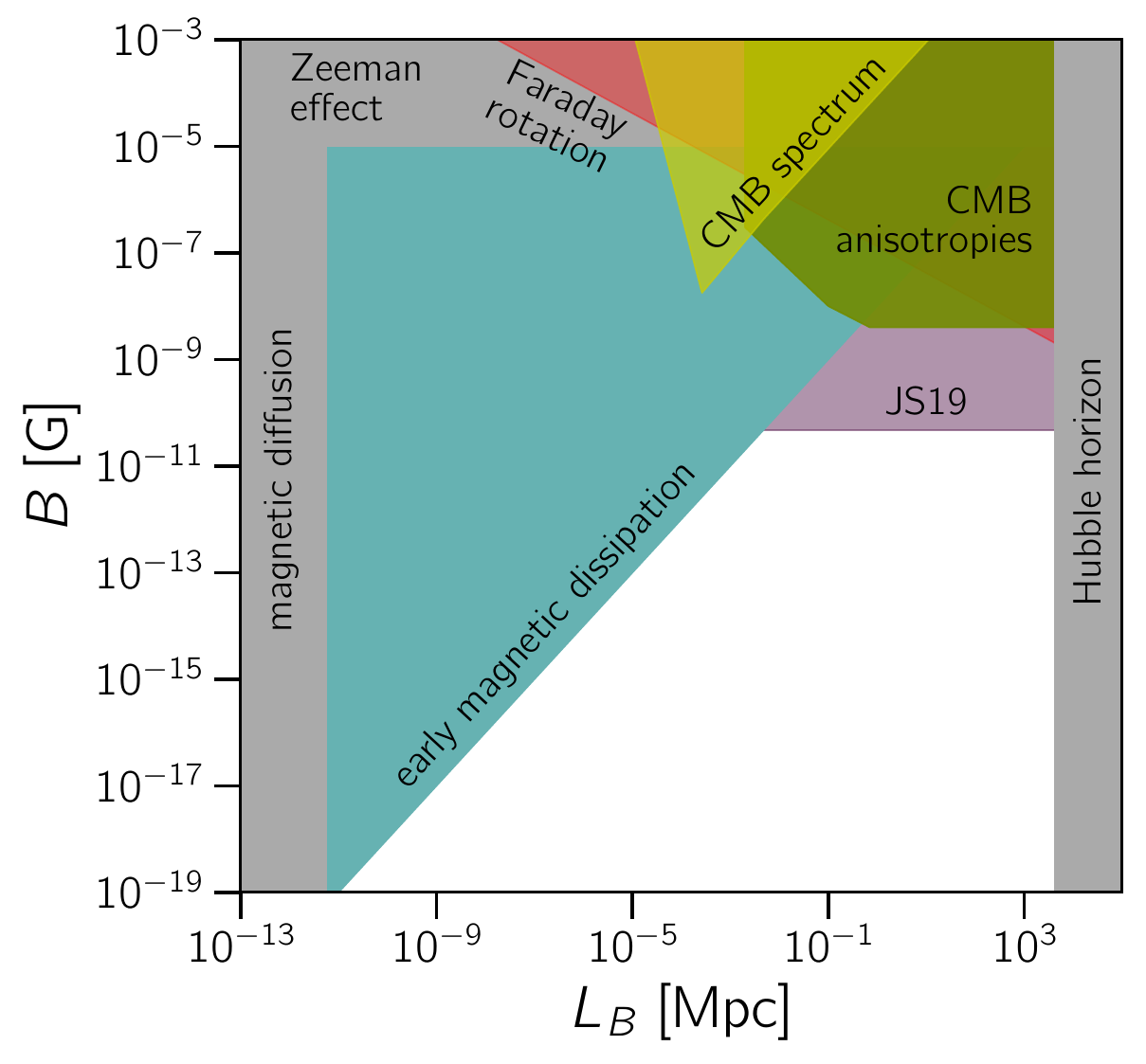}
    \caption{Schematic overview of the main constraints on IGMFs, as discussed in Sec.~\ref{sec:FurterConstraints}. The lower and upper bounds on $L_{B}$ come from the decay of magnetic fields due to magnetic diffusion and the Hubble radius, respectively~\cite{neronov2009a}. The upper bounds are due to Zeeman splitting and Faraday rotation observations of extragalactic objects \cite{neronov2009a}. The `early magnetic dissipation' bound indicates the region of the parameter space excluded by freely decaying MHD in the early Universe \cite{banerjee2004a,durrer2013a}. Other limits from cosmology come from CMB observations (spectrum \cite{Jedamzik:1999bm,durrer2013a} and anisotropies \cite{Barrow:1997mj}); the currently strongest limit~\cite{jedamzik2019a}, labelled `JS19', is shown for the case of a scale-invariant spectrum ($\alpha_{B} = 0$) which leads to the most conservative bounds.}
    \label{fig:GenConstraints} 
\end{figure}

In this review we focus on constraints on IGMFs from gamma-ray observations. However, since IGMFs can interact through various electromagnetic phenomena throughout the Universe, there are other ways to derive bounds on them. In this section we present the general ideas to do so, based on Ref.~\cite{neronov2009a} and including some more recent developments.

\bigskip
First, there is a generic lower and upper limit on the coherence length ($L_{B}$) \cite{neronov2009a}. The latter is given by the size of the observed Universe, i.e., the Hubble radius. On the other hand, the IGMF decays due to magnetic diffusion, such that the lower limit on $L_{B}$ is given by the length scale equivalent to the corresponding decay time, i.e., $L_{B} \gtrsim 2 \times 10^{11} \, {\rm m}$~\cite{Grasso:2000wj}.

As for the magnetic-field strength ($B$), measurements of the Zeeman splitting of H I lines can be used to set upper bounds on this quantity. This can be done either for our own galaxy  or for the radiation from distant quasars \cite{Heiles:2003fa,Wolfe:2008nk,Wolfe:2011ic}, both consistently giving a result of the order of $\sim \mu{\rm G}$. In the latter case, any stronger IGMFs along the line of sight to the object would have a measurable impact on the observations, thus giving a robust upper limit for the IGMF strength. 

Another constraint on IGMFs is derived from Faraday rotation measurements of polarised radio emission from quasars and other extragalactic sources. Faraday rotation describes the (wavelength-dependent) rotation of the polarisation plane of polarised electromagnetic radiation when it traverses a magnetised medium. Therefore, the value of the relevant observable, the so-called rotation measure (RM), may be subdivided into contributions from the host galaxy, the IGM, and the Milky Way. With a rigorous statistical analysis of RM data, one can then identify the impact of the IGMF, and hence derive limits on the IGMF strength which in general depend on $L_{B}$. There are many studies on the topic \cite{Kronberg:2007dy, Bonafede:2010xg, Neronov:2013lta, Akahori:2014qja, Xu:2014hya, Pshirkov:2015tua, OSullivan:2020pll, Amaral:2021mly}, all of which give upper limits ranging from nG to a few $\mu$G. This is also confirmed by other methods, like the interpretation of radio observations as the result of shock acceleration in galaxy clusters~\cite{Bagchi:2002vf,DiGennaro:2020zkz}.
In this context, fast radio bursts (FRBs) can play an important role~\cite{hackstein2019a}, delivering both rotation and dispersion measures. As a consequence, magnetic fields along the line of sight can be better inferred because the use of these two observables reduce the reliance on models on models of the electron density distribution~\cite{akahori2016a}. RMs can be related to the magnetogenesis model, as shown in Ref.~\cite{vazza2018a}.

\bigskip
In addition, an important set of limits can be derived from cosmology considering the cosmological scenarios for magnetogenesis (cf.~Sec.~\ref{sec:CosmologicalScenario}). An indirect, theoretical approach is to consider a given mechanism of magnetic-field generation and derive the corresponding limits on the initial magnetic-field strength and correlation length, and then calculate their time evolution via freely-decaying MHD up to the present day. A detailed description is given in~\cite{banerjee2004a}, while in Fig.~\ref{fig:GenConstraints} we present the region which contains most of these constraints, following~\cite{durrer2013a}. In general, one can state that these limits bound the field strength from above and the coherence length from below, the latter due to the fact that in cosmological scenario IGMFs are generated at small scales (see above).

From an observational point of view, most of the limits on IGMFs from cosmology are derived using the cosmic microwave background (CMB), as there is a large range of effects through which magnetic fields can impact the background radiation. The most basic idea, developed already in \cite{Zeldovich1983}, is to assume a homogeneous field throughout the Universe and then to derive the temperature anisotropies expected from that. Comparing this dataset to data from COBE \cite{Barrow:1997mj} or, more recently, from Planck \cite{Planck2013:XVI}, gives an upper limit of around 4~nG (marked in Fig.~\ref{fig:GenConstraints} as ``CMB anisotropies''). Since then the upper limit has been dramatically improved by using CMB observations in combination with such effects as spectral distortions (in Fig.~\ref{fig:GenConstraints} we present a limit stemming from this phenomenon based on \cite{Jedamzik:1999bm}, denoted ``CMB spectrum''), temperature anisotropies, polarisation, non-Gaussianity and Reionisation (for a concise review see \cite{jedamzik2019a}). The best upper limit so far, $B \sim 10$--$50$~pG \cite{jedamzik2019a}, was derived by considering the change of the Recombination process itself via density fluctuations due to the presence of PMFs. In addition, CMB observations are also interesting because they may also be used to derive constraints on magnetic helicity \cite{Caprini:2003vc,Kahniashvili:2005xe,Hortua:2015jkz}.

\bigskip
Finally, ultra-high-energy cosmic rays (UHECRs), i.e.~nuclei with energies above $10^{18}$~eV, may be used to constrain IGMFs~\cite{rachen1992a, rachen1993a, kronberg1994a, lee1995a, lemoine1997a,neronov2009a}. The general principle used here is that, since UHECRs are charged, they are deflected. Hence, once their sources are identified, the corresponding deflection angle can be measured, providing a direct measure of the magnetic-field strength orthogonal to the line of sight. Ref.~\cite{yueksel2012a} used the observed excess of UHECRs with energies $\sim 10^{20} \; \text{eV}$ in the direction of Centaurus~A to constrain the local extragalactic magnetic field, obtaining $B \lesssim 10^{-8}$~G. This local constrain evidently serves as an upper limit for IGMFs.
Ref.~\cite{Bray:2018ipq} used the anisotropy reported by the Pierre Auger Observatory to associate UHECR detections with extragalactic objects and to derive upper limits of $B \sim 10^{-9}$~G for $L_{B} < 100$~Mpc and $\sim 10^{-10}$~G for $L_{B} > 100$~Mpc. More recently, Ref.~\cite{VanVliet:2021sbc} found that for $B > 6 \times 10^{-10}$~G the Auger anisotropy measurements are in good agreement with the local density of star-forming galaxies. On the other hand, if the local density is treated as a model parameter, the authors found a conservative upper limit of $B L_{B}^{1/2} < 24 \, {\rm nG \, Mpc^{1/2}}$. 
In principle UHECRs may also be used to constrain the helicity of IGMFs, as argued in \cite{Kahniashvili:2005yp} and demonstrated numerically in~\cite{alvesbatista2019b}.

Note, however, that with the direct UHECR observations available today, it is rather difficult to derive IGMF constraints, as their sources would have to be known (see also Secs.~\ref{ssec:sources} and \ref{ssec:UHECRgamma} for an indirect gamma-ray--based approach on how to use UHECRs for deriving IGMF constraints). Moreover, the statistic of events at the highest energies ($E \gtrsim 4 \times 10^{19} \; \text{eV}$) is fairly limited, while the composition (and thus the charge) of UHECRs is only known statistically, and not on an event-by-event basis~\cite{alvesbatista2019d}, posing severe challenges for any attempt to constrain IGMFs with UHECRs. Finally, the distribution of magnetic fields in the cosmic web is more complex than that in cosmic voids, and much more uncertain~\cite{Vazza:2017mbz}. Numerical studies of UHECR propagation in magnetic fields lead to very discrepant results, such that the prospects for UHECR astronomy (and thus IGMF constraints using UHECRs) are far from clear (cf.~\cite{dolag2005a, alvesbatista2017c, hackstein2018a, Garcia:2021cgu}).

\section{Electromagnetic Cascades} 
\label{sec:propagation}

In this section we lay out the theoretical foundations for understanding how electromagnetic cascades develop in intergalactic space. We start off, in Sec.~\ref{ssec:sources}, by describing some classes of astrophysical objects that can emit particles that initiate the electromagnetic cascades. We describe two scenarios, depending on the type of particle that initiate the cascade which ultimately leads to the observed gamma-ray signal. In the first, the cascades are triggered by high-energy gamma rays (or electrons), whereas in the second, they are initiated by ultra-high-energy cosmic rays. After describing how electromagnetic cascades originate, we proceed to Sec.~\ref{ssec:propagationTheory}, where we give a detailed account of how they develop, how they interact with the photon fields that pervade the Universe, and how IGMFs can affect them. In Sec.~\ref{ssec:analytical} we present approximate analytical descriptions for the cascade process, which can also be treated in more detail with numerical codes, as described in Sec.~\ref{ssec:codes} and illustrated in Sec.~\ref{ssec:examples}. In Sec.~\ref{ssec:plasma} we chime into the debate surrounding the role played by plasma instabilities in the development of cascades. Other potentially relevant propagation effects are concisely mentioned in Sec.~\ref{ssec:bsm}. 

\subsection{Origin} 
\label{ssec:sources}

The most common source of high-energy gamma rays used in IGMF studies are blazars. These objects are a sub-class of active galactic nuclei (AGNs) whose relativistic jets point approximately towards Earth~\cite{urry1995a}. Their spectral energy distribution is characterised by a low-energy hump corresponding to synchrotron emission by relativistic electrons~\cite{padovani2017a, blandford2019a}. There is also a second notable hump which, in the case of high and extreme synchrotron-peaked objects, is of interest for IGMF studies, peaks at  $\sim\text{TeV}$ energies~\cite{blandford2019a}. These objects are excellent cosmological probes as the very-high-energy emission assures the production of a substantial electronic component in the cascade, which can be used to probe IGMFs and the EBL~\cite{bonnoli2015a}.

An object widely considered in gamma-ray astronomy to constrain IGMFs is the extreme blazar 1ES~0229+200. It was used, for instance, in Refs.~\cite{hess2007a, neronov2010a, tavecchio2010a, tavecchio2011a, finke2015a, fermi2018a}. In fact, there is a population of extreme blazars like 1ES~0229+200 with hard spectra that have been commonly used for IGMF studies, given the weakness of the $\sim$~GeV contribution with respect to the TeV band~\cite{bonnoli2015a}. Besides 1ES~0229+200, there are other objects that are also employed for this purpose, such as: 
1ES~0347-121~\cite{hess2007b},
1ES~0414+009~\cite{hess2012a},
1ES~1101-232~\cite{hess2007c},  
1ES~1218+304~\cite{veritas2009a},
1ES~1312–423~\cite{hess2013b},
1RXS~J101015.9–311909~\cite{hess2012b},
H~1426+428~\cite{horan2002a},
H~2356–309~\cite{hess2006b},
Mrk~421~\cite{punch1992a},
Mrk~501~\cite{quinn1996a},
PG~1553+113~\cite{magic2007a},
PKS~0548-322~\cite{hess2010a},
PKS~2155-304~\cite{chadwick1999a}
RGB~J0152+017~\cite{hess2008a}, RGB~J0710+591~\cite{veritas2010a},
and VER~J0521+211~\cite{veritas2013a}.
Note that, while typical IGMF studies are done for blazars that can be observed both at $\sim$~GeV and TeV energies, this is not a strict requirement and magnetic-field properties can be inferred solely from the cascade signal.

In general, AGNs are active over time scales of $\mathfrak{T} \sim 10^6$--$10^8 \; \text{years}$~\cite{parma2002a}, which makes it hard to use temporal information for constraining IGMFs since they are, for all practical purposes, quasi-steady sources. However, some objects, such PKS~2155-304~\cite{chadwick1999a}, Mrk~421~\cite{punch1992a, magic2020a} and Mrk~501~\cite{quinn1996a, hegra1997a,Albert:2007zd}, display short-time variability~\cite{hawc2017b}.
This information can, in principle, be used together with light curves in other wavelengths in the context of multimessenger campaigns to improve the constraints on IGMFs via time delays (see Sec.~\ref{sec:results}). Interestingly, for blazars that are slightly misaligned with respect to the line of sight, the GeV gamma rays stemming from the TeV emission could still be observed today over angular scales of $\sim 1^\circ$ even if the objects are no longer high-energy emitters~\cite{neronov2010b}.

\bigskip

Another class of objects that can potentially be used to probe IGMFs are gamma-ray bursts (GRBs). They emit highly collimated relativistic jets of high-energy radiation within a short time. GRBs are the most luminous events known, reaching isotropic-equivalent luminosities of $\sim 10^{54} \; \text{erg}\,\text{s}^{-1}$ (see, e.g., ~\cite{gehrels2009a, piron2016a, schady2017a} for reviews). Only recently were GRBs observed at very-high energies, with the detection of a bright flash from GRB~190114C~\cite{magic2019a}, which was used for IGMF studies~\cite{dzhatdoev2020a, wang2020a}. 

GRBs are interesting cosmological probes because they can be used exactly in the same manner as blazars, while in general providing more accurate temporal information. In this case, the high- and very-high-energy components depend strongly on the properties of intervening IGMFs~\cite{Guetta:2002au, Wang:2003qt,Veres:2013dea}. 
Moreover, if their HE light curve were known, in principle it would be possible to reconstruct a possible TeV emission even in the absence of VHE measurements, based only on the cascade signal at $\sim$~GeV energies, up to high redshifts~\cite{Ichiki:2007nd, takahashi2008a, takahashi2011a}. Note that this argument only holds if the TeV light curve is known from theoretical models, which is not the case~\cite{gehrels2009a, schady2017a}, or if there are well-defined relations between the GeV and TeV light curves.

\bigskip

The shape of the intrinsic spectrum of the sources of interest for this work, whether a blazar or a GRB, is not precisely known. In general, it is assumed to be a power law of the form
\begin{equation}
    \dfrac{\text{d}N}{\text{d}E} \propto E^{-\alpha} f_\text{cut}(E) \,,
    \label{eq:sourceSpec}
\end{equation}
where $f_\text{cut}(E)$ denotes a function that suppresses the spectrum above a given energy $E_\text{max}$, which depends on the mechanism responsible for particle acceleration (and consequently for gamma-ray emission). This function is typically an exponential, log-parabola, or similar~\cite{kardashev1962a, katarzynski2006a, giebels2007a, stawarz2008a, matthews2020a}.
Interestingly, the value of $E_\text{max}$ that could be inferred with observations depends on the opacity of the Universe to gamma-ray propagation, i.e., the distribution of photon fields such as the EBL, as well as on the properties of the intervening IGMFs~\cite{saveliev2021a}.

\bigskip

Cosmic-ray--induced electromagnetic cascades in the intergalactic medium may lead to observational signatures that resemble those initiated by gamma rays. These cascades are evidently affected by intervening IGMFs, as discussed in, e.g., Refs.~\cite{Uryson:2001xb, Ferrigno:2004am, Gabici:2006dv, Kotera:2010xd, Ahlers:2010fw, Aharonian:2010va}.
Therefore, gamma rays from cosmic rays can, in principle, also be used to probe IGMFs. In the case of GRBs, this was suggested by the authors of Ref.~\cite{Waxman:1996zc}.  
Similarly, blazars are prominent contenders to emit UHECRs that can induce electromagnetic cascades in the IGM~\cite{essey2010a, essey2011a}. For highly collimated jets, it could be even possible to distinguish this hadronic scenario from the standard  picture wherein gamma rays from the sources induce the cascades~\cite{takami2013a}.

The cosmic rays relevant for this type of analysis are UHECRs, since they can produce electromagnetic cascades during intergalactic propagation, via photonuclear or hadronuclear interactions. In fact, this type of scenario has been suggested to explain observations of some blazars, as they lead to better agreement with the measurements~\cite{essey2010a, essey2010b, Essey:2010nd, Aharonian:2010va, Ahlers:2011sd, dzhatdoev2017a, Khalikov:2019fbd}.

One process that creates electrons and positrons that trigger cascades is Bethe-Heitler pair production: $\nucleus{X}{A}{Z} + \gamma_\text{bg} \rightarrow  \, \nucleus{X}{A}{Z} + e^- + e^+$,
wherein $\nucleus{X}{A}{Z}$ denotes an arbitrary cosmic-ray nucleus $X$ of atomic mass $A$ with $Z$ protons interacting with a background photon ($\gamma_\text{bg}$). 

Nuclear interactions also produce electrons and photons, starting with the photodisintegration of cosmic-ray nuclei (e.g., $\nucleus{X}{A}{Z} + \gamma_\text{bg} \rightarrow \, \nucleus{X}{A-1}{Z-1} + p$, $\nucleus{X}{A}{Z} + \gamma_\text{bg} \rightarrow \nucleus{X}{A-1}{Z} + n$), possibly producing unstable nuclei ($\nucleus{X}{A}{Z}^*$) which decay as $\nucleus{X}{A}{Z}^* \rightarrow \, \nucleus{X}{A}{Z} + \gamma$. 

The most important hadronic channel for the generation of cascade-inducing particles (electrons and photons) is photopion production. For a cosmic-ray proton, $p + \gamma_\text{bg} \rightarrow \Delta^+ \rightarrow p + \pi^0$ and $p + \gamma_\text{bg} \rightarrow \Delta^+ \rightarrow n + \pi^+$. The decay of the neutral pion produces photons ($\pi^0 \rightarrow \gamma + \gamma$) and the decay of charged pions\footnote{There are other decay channels. For the purposes of this review, we present only the most relevant one. One example is the electronic mode ($\pi^+ \rightarrow e^+ + \nu_e$) that occurs much more rarely ($\lesssim 10^{-3}$) than the main one.} lead to the generation of leptons ($\pi^+ \rightarrow \mu^+ + \nu_\mu$), including muons, whose decays produce electrons ($\mu^+ \rightarrow \nu_e + \bar{\nu}_\mu + e^+$). Note that the cascades stemming from the by-products of pion decays also occur for an arbitrary nucleus $\nucleus{X}{A}{Z}$. In this case, the production rate depends on the number of each nucleonic species (see, e.g., Ref.~\cite{morejon2019a} for further details).

While it is, in principle, possible to constrain IGMFs with UHECR-produced gamma rays, this is not straightforward. Firstly, the sources of UHECRs are not known. Secondly, they are deflected by intervening Galactic and extragalactic magnetic fields, potentially spoiling any correlation between the source direction and the gamma rays. For more details on the cosmic-ray--gamma-ray connection, the reader is referred to some reviews on the topics: \cite{alvesbatista2019d, anchordoqui2019a}. 

\subsection{Theory of Propagation} 
\label{ssec:propagationTheory}

The particle physics aspects relevant for the propagation of high-energy gamma rays are well known. At energies $E \gtrsim 400 \; \text{GeV}$ high-energy gamma rays interact with background photon fields predominantly at infrared frequencies, generating electron-positron pairs: $\gamma + \gamma_\text{bg} \rightarrow e^+ + e^-$. The mean free path for this process is typically of the order of tens to hundreds of Mpc. These pairs up-scatter photons from (mostly) the CMB to high energies via inverse Compton scattering ($e^\pm + \gamma_\text{bg} \rightarrow e^\pm + \gamma$). These new photons, in turn, can either travel straight to Earth or, if their energy is above the threshold for pair production, restart this process, leading to an electromagnetic cascade in the intergalactic medium.

\begin{figure}[hbt]
    \centering
    \includegraphics[width=\textwidth]{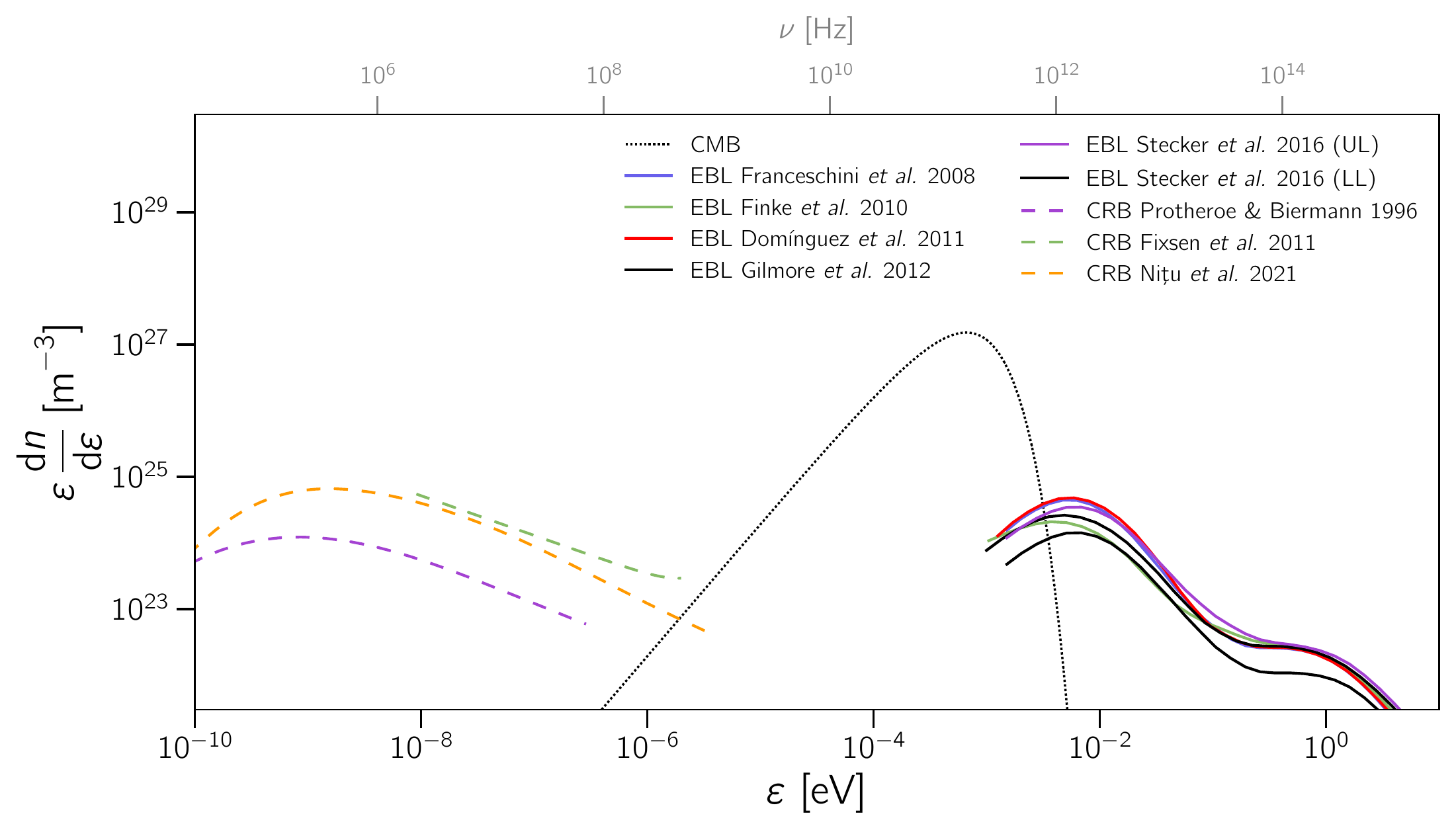}
    \caption{Compilation of the density of background photons ($n(\varepsilon)$) with different energies ($\varepsilon$) at $z=0$. The curves correspond to different backgrounds, radio (dashed lines), microwave (dotted), and infrared/optical (solid). Different colours represent different EBL and CRB models: Franceschini \textit{et al.}~\cite{franceschini2008a}, Finke \textit{et al.}~\cite{finke2010a}, Domínguez \textit{et al.}~\cite{dominguez2011a}, Gilmore \textit{et al.}~\cite{gilmore2012a}, the upper (UL) and lower (LL) limits by Stecker \textit{et al.}~\cite{stecker2016a}, Protheroe \& Biermann~\cite{protheroe1996a}, ~Ni\c{t}u \textit{et al.}~\cite{nitu2021a}, and measurements by ARCADE-2 (Fixsen \textit{et al.}~\cite{fixsen2011a}). The frequencies ($\nu$) corresponding to the photon energies are shown at the top.}
    \label{fig:photonFields}
\end{figure}

The picture outlined in the previous paragraph is theoretically simple, but there are uncertainties that complicate the modelling of electromagnetic cascades in the IGM. The most important one is the distribution of the EBL, which is not precisely known. At extremely high gamma-ray energies ($E \gtrsim 10^{17} \; \text{eV}$), the contribution of the cosmic radio background (CRB) starts to become relevant. A comparison of EBL and CRB models, as well as the density of CMB photons, is illustrated in Figure~\ref{fig:photonFields}.

In general, the inverse of the mean free path $\lambda$ for a particle of energy $E$ and mass $m$ interacting with isotropically-distributed photons of differential number density\footnote{The differential photon number density is defined in a way such that $\int\limits_{0}^{\infty} \varepsilon \dfrac{{\rm d} n(\varepsilon,z)}{{\rm d}\varepsilon} \,{\rm d}\varepsilon$ gives the local energy density of the photon field.} $\frac{{\rm d} n(\varepsilon,z)}{{\rm d}\varepsilon}$ is
\begin{equation}
	\lambda^{-1}(E,z) = \dfrac{1}{8 E^{2}} \int\limits^\infty_0 \int\limits^{s_\text{max}}_{s_\text{min}} \dfrac{1}{\varepsilon^{2}} \dfrac{{\rm d} n(\varepsilon,z)}{{\rm d}\varepsilon} \mathcal{F}(s) \, {\rm d}s {\rm d}\varepsilon  \, ,
	\label{eq:mfpIntegral}
\end{equation}
where $z$ is the redshift (see below), $\varepsilon$ refers to the energy of the background photon, and $\mathcal{F}$ depends on the process of interest, with kinematic limits $s_\text{min}$ and $s_\text{max}$.
For pair production, $\mathcal{F} = s \sigma_\text{PP}(s)$, with $s_\text{min} = 4 m_e^2 c^4$ and $s_\text{max} = 4 E \varepsilon$. For inverse Compton scattering, $\mathcal{F} = \sigma_\text{IC}(s - m_e^2 c^4) / \beta$, wherein $\beta$ denotes the speed of the electrons, in units of the speed of light. The kinematic limits, in this case, are $s_\text{min} = m_e^2 c^4$ and $s_\text{max} = m_e^2 c^4 + 2 E \varepsilon (1 + \beta)$. Here, $\sigma_\text{PP}$ and $\sigma_\text{IC}$ denote, respectively, the cross sections for pair production and for inverse Compton scattering. Note that the minimum and maximum energies are, in principle, unbounded, i.e.,  $\varepsilon_\text{min} \rightarrow 0$ and  $\varepsilon_\text{max} \rightarrow \infty$, but in practice they quickly vanish outside a given energy range. In the case of the EBL, for example, for purposes of calculations,  $\varepsilon_\text{min} \simeq 10^{-4} \; \text{eV}$ and  $\varepsilon_\text{max} \simeq 10 \; \text{eV}$ (see Fig.~\ref{fig:photonFields}).

Following Ref.~\cite{neronov2009a}, we can approximate  Equation~\ref{eq:mfpIntegral} as
\begin{equation} \label{eq:mfpApproxPP}
\lambda_{\rm PP} \simeq 40 \frac{\kappa}{(1 + z_\text{PP})^{2}} \left( \dfrac{E_{\gamma}}{20 \, {\rm TeV}} \right)^{-1} \, {\rm Mpc}
\end{equation}
for pair production, where $\kappa$ is a model-specific parameter of the order of $\kappa \sim 1$, and as 
\begin{equation} \label{eq:mfpApproxIC}
\lambda_{\rm IC} \simeq 32 \frac{1}{(1 + z_\text{PP})^{4}} \left( \dfrac{E_{e}}{10 \, {\rm TeV}} \right)^{-1} \, {\rm kpc} \,,
\end{equation}
for inverse Compton scattering. 

\begin{figure}[htb]
    \centering
    \includegraphics[width=0.495\textwidth]{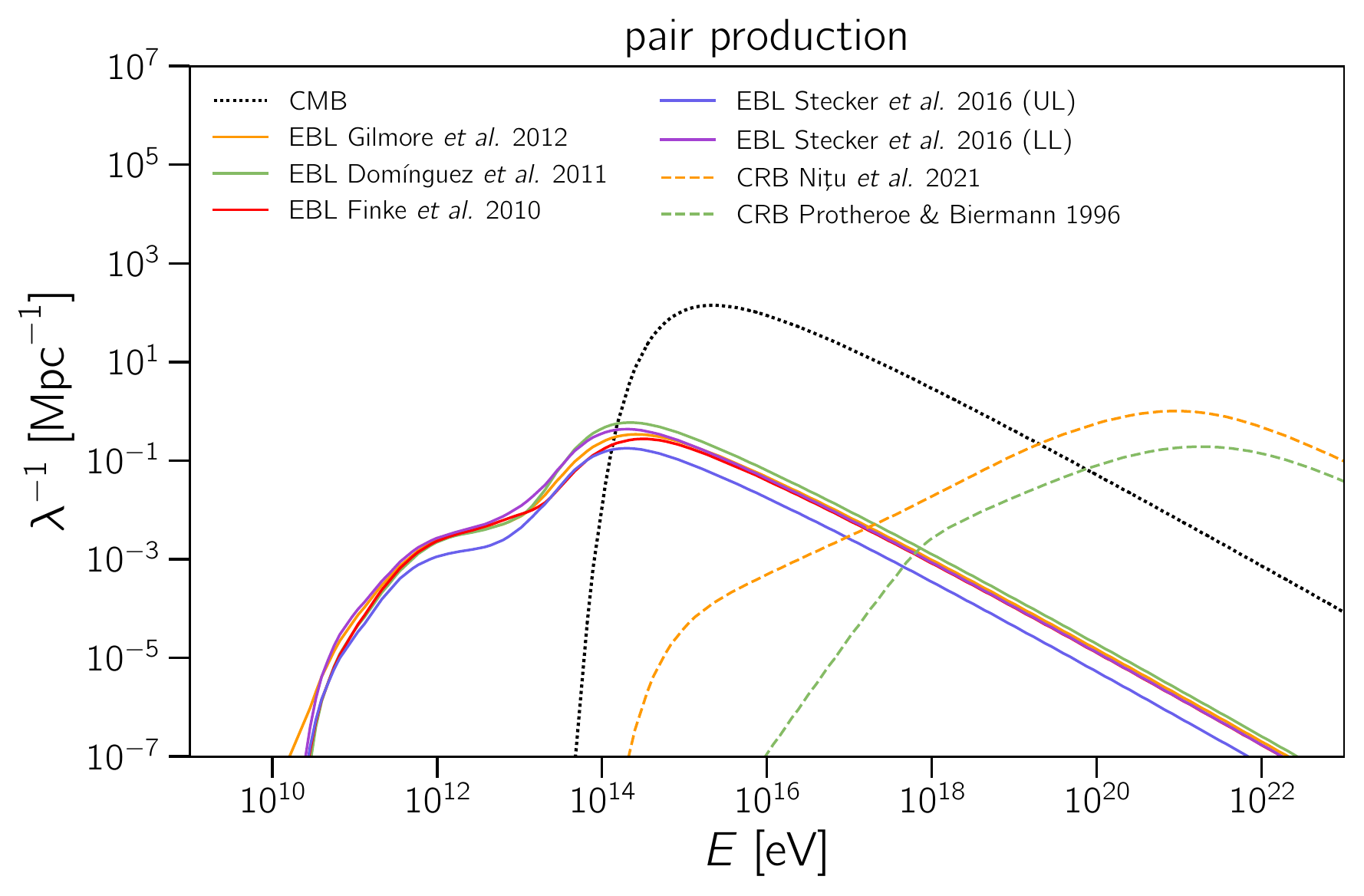}
    \includegraphics[width=0.495\textwidth]{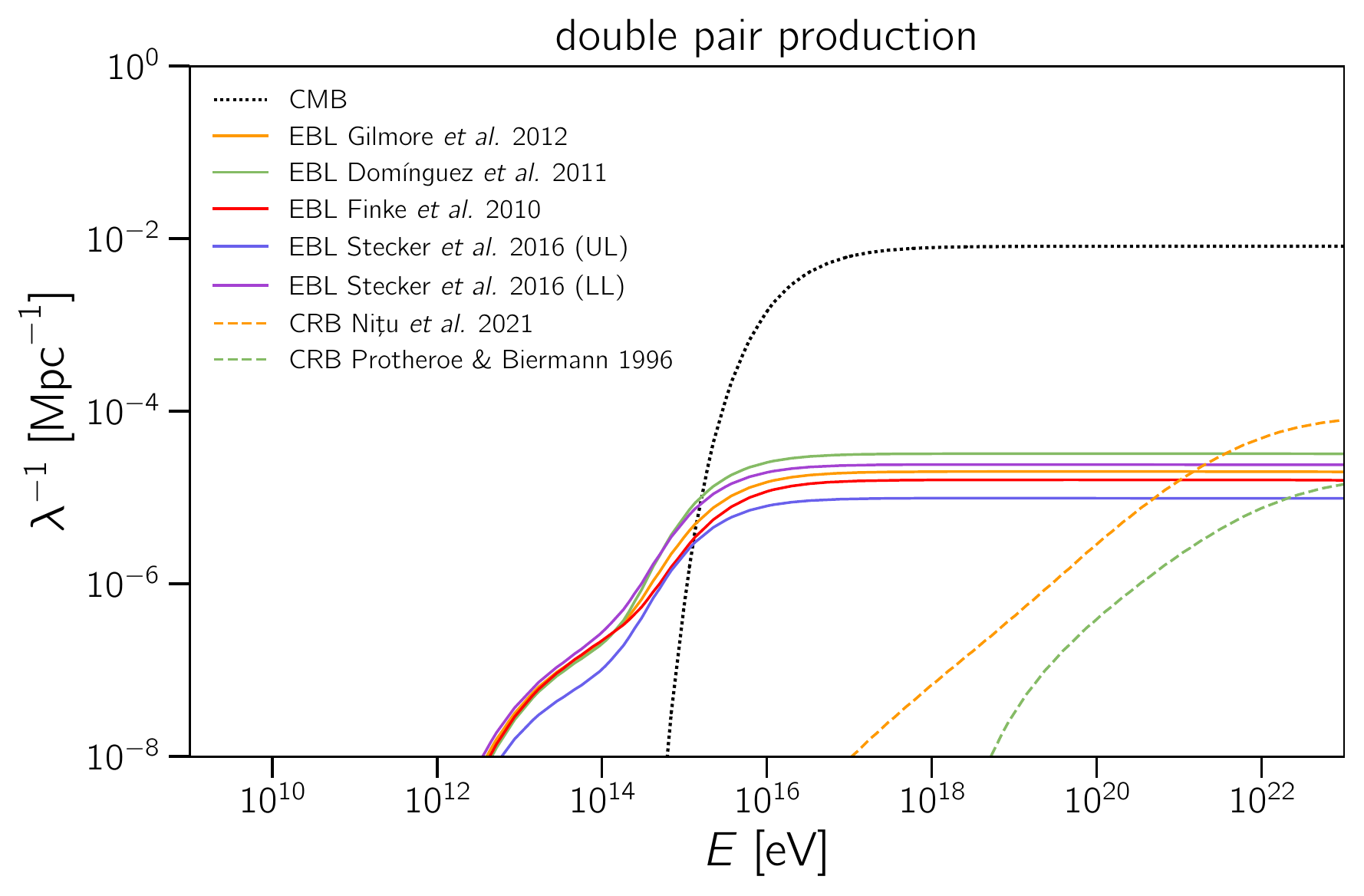}
    \includegraphics[width=0.495\textwidth]{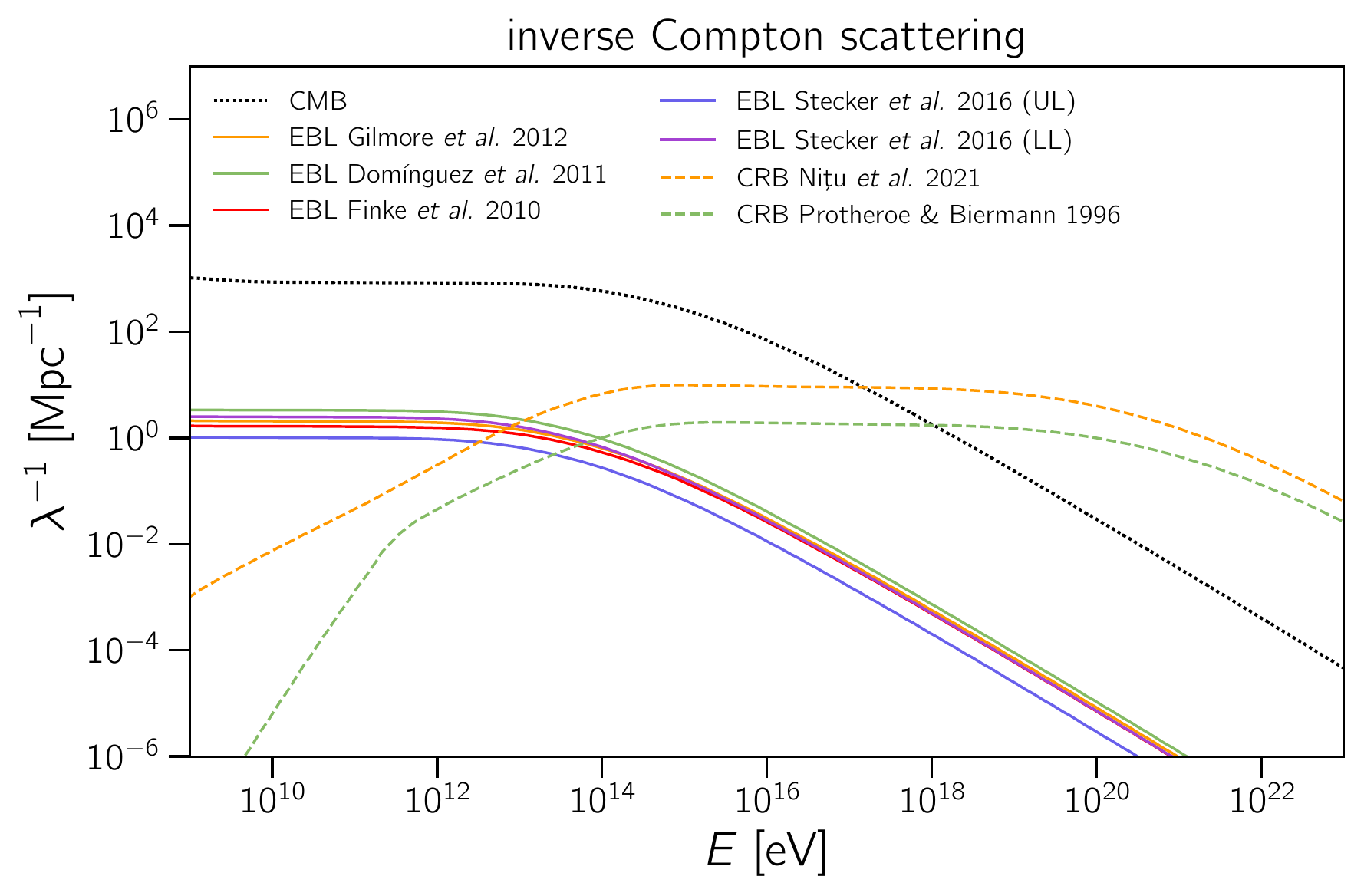}
    \includegraphics[width=0.495\textwidth]{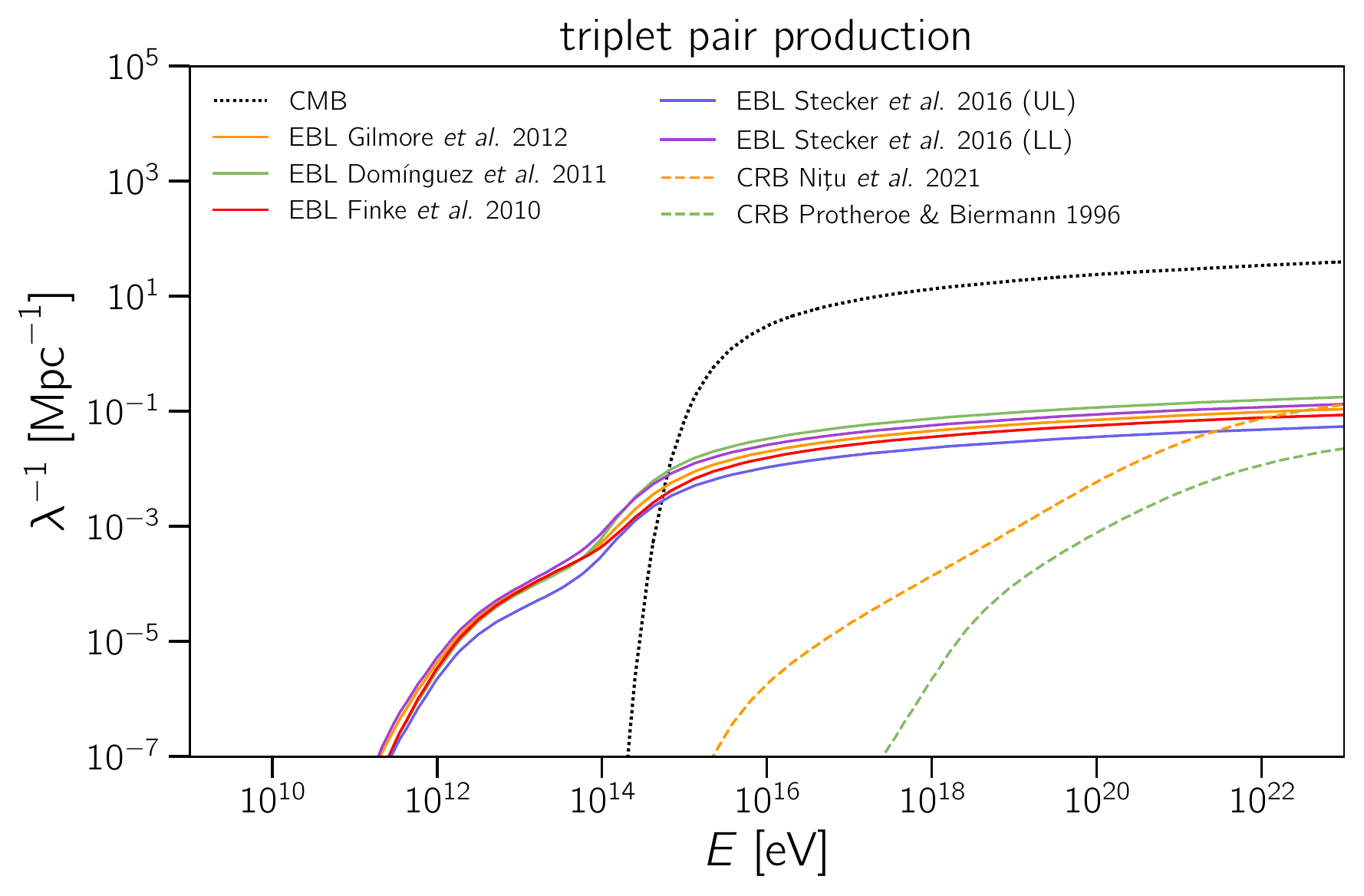}
    \caption{Each panel shows the energy-dependent inverse mean free path at $z=0$ for the processes relevant for the cosmological propagation of gamma rays with energies $E \gtrsim 1 \; \text{GeV}$. Different photon backgrounds are considered. Solid lines correspond to interactions with the EBL, dashed lines are for the CRB, and the dotted line refers to the CMB. The EBL~\cite{finke2010a,dominguez2011a,gilmore2012a,stecker2016a} and CRB~\cite{protheroe1996a,nitu2021a} models used are represented by different colours.}
    \label{fig:interactionRates}
\end{figure}

Typically, gamma rays with $\sim \; \text{TeV}$ energies produce pairs after travelling distances larger than $\sim 100 \; \text{Mpc}$. For inverse Compton scattering, the typical distance TeV electrons travel before they undergo interactions is $\sim 30 \; \text{kpc}$. In Figure~\ref{fig:interactionRates} we show the inverse of the mean free path for these processes, as obtained from Eq.~\ref{eq:mfpIntegral}.

Higher-order processes to pair production and inverse Compton scattering are important for the propagation of gamma rays of $E \gtrsim 10^{15} \; \text{eV}$. In particular, for scenarios wherein UHECRs induce electromagnetic cascades (see Sec.~\ref{ssec:sources}), they are an essential ingredient to understand gamma-ray production.
The higher-order equivalent of the Breit-Wheeler pair production is the double pair production~\cite{cheng1970a, brown1973a} ($\gamma + \gamma_\text{bg} \rightarrow e^+ + e^- + e^+ + e^-$). This process has been extensively studied in various astrophysical contexts, including the propagation of high-energy photons~\cite{demidov2009a, ruffini2016a}. 
Inverse Compton scattering can also occur as a second-order process called triplet pair production ($e^\pm + \gamma_\text{bg} \rightarrow e^\pm + e^+ + e^-$).
Its role in the propagation of high-energy photons has long been recognised~\cite{bonometto1972a, bonometto1974a, dermer1991a, mastichiadis1994a}. This process starts to become important at $E \gtrsim 10^{17} \; \text{eV}$, for cosmological distances. The (inverse) mean free paths for double and triplet pair production are also shown in Fig.~\ref{fig:interactionRates}.

\bigskip
The cosmological propagation of any particle is subject to adiabatic energy losses due to the expansion of the Universe. The change in redshift (${\rm d}z$) for a small propagated distance (${\rm d}x$) can be written as ${\rm d}z = \frac{H(z)}{c} {\rm d}x,$
where $H(z)$ is the Hubble parameter, which in a flat Lambda cold dark matter ($\Lambda$CDM)  universe is given by
\begin{equation}
	H(z) = H_0 \sqrt{\Omega_\Lambda + \Omega_m (1 + z)^3} \, ,
	\label{eq:hubble}
\end{equation}
with $H_0$ denoting the Hubble ``constant'', i.e., the value of the Hubble parameter at present time. Here, $\Omega_m$ and $\Omega_\Lambda$ parameters represent the fraction of the total energy density of the Universe corresponding to matter and dark energy, respectively. According to recent measurements, $H_0 \approx 67.37 \; \text{km} \, \text{s}^{-1} \, \text{Mpc}^{-1}$, $\Omega_m \approx 0.3147$, and $\Omega_\Lambda \approx 0.6853$~\cite{planck2020a}.

Generally, in the presence of a magnetic field the charged component of the electromagnetic cascade (electrons) loses energy through synchrotron emission. 
However, synchrotron losses are small for intergalactic gamma-ray propagation, since $B \lesssim 10^{-9} \; \text{G}$ (see Figure~\ref{fig:GenConstraints}). 

\bigskip

The last theoretical ingredient missing for understanding how electromagnetic cascades propagate is the interaction of its charged component with magnetic fields. The equation of motion for a particle of charge $q$ with velocity $\mathbf{v}$ in a magnetic field $\mathbf{B}$ can be written as
\begin{equation}
\dfrac{{\rm d}\mathbf{p}}{{\rm d}t} = q \mathbf{v} \times \mathbf{B}\,,
    \label{eq:lorentz}
\end{equation}
where $\mathbf{p}$ is the particle momentum. 
As a consequence of this equation, the electrons and positrons will deflect away from each other. This deflection consists of a circular/helical movement (around the magnetic-field lines), characterized by the Larmor radius $r_{\rm L}$, given by
\begin{equation} \label{eq:larmor}
r_{\rm L} = \dfrac{p}{e B}\,,
\end{equation}
where $p$ is the absolute value of the particle momentum.

\begin{figure}[hbt]
    \centering
    \includegraphics[width=\textwidth]{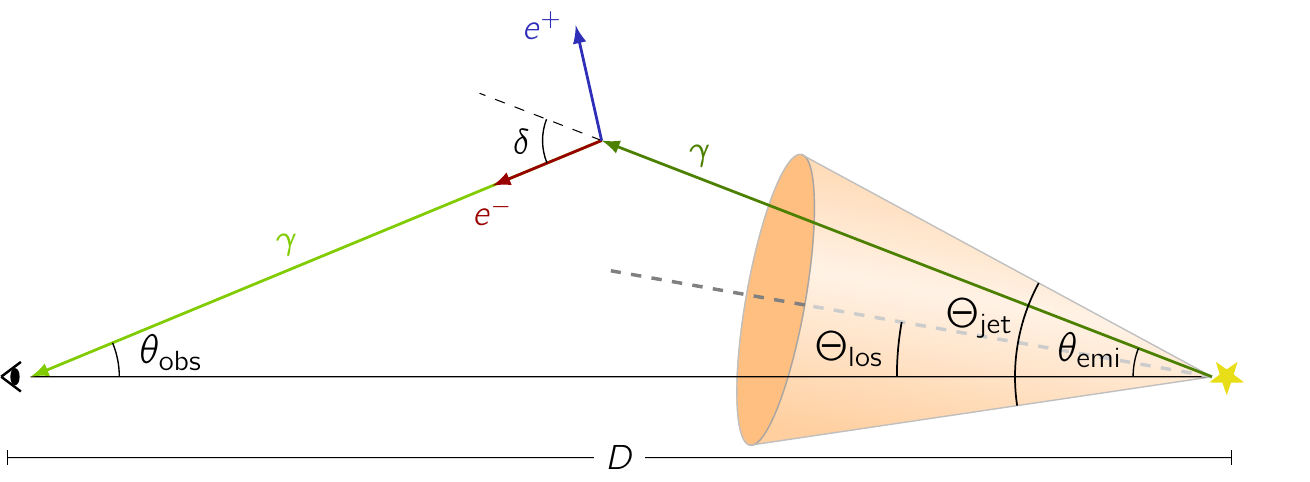}
    \caption{Schematic drawing of the development of an electromagnetic cascade. A source (yellow star) with a jet of opening angle $\Theta_\text{jet}$ tilted -- with respect to the line of sight -- by an angle $\Theta_\text{obs}$ emits high-energy gamma rays (dark green line) forming an angle $\theta_\text{emi}$ with this line. After interaction, it produces an electron-positron pair (blue and red arrows) that, in the presence of IGMFs, is deflected by an angle $\delta$ with respect to the direction of the original gamma ray. These pairs can then up-scatter background photons to high energies (light green line), being detected with an angle $\theta_\text{obs}$.}
    \label{fig:cone}
\end{figure}

\bigskip
We can now draw a general picture of how gamma rays can be used to constrain IGMFs. Consider an object located at a distance $D$ from Earth, corresponding to a redshift $z_\text{src}$, emitting a jet of high-energy gamma rays with an opening angle $\Theta_\text{jet}$, as sketched in Figure~\ref{fig:cone}. Let $\Theta_\text{los}$ denote the angle between the jet axis and the line of sight, i.e., the angle of misalignment. The primary gamma rays are generated at a redshift $z_{\rm src}$ and produce pairs at $z_\text{PP}$, travelling for a distance dictated by the mean free path for pair production ($\lambda_\text{PP}$) for the energy and redshift of interest (see Eqs.~\ref{eq:mfpIntegral} and~\ref{eq:mfpApproxPP}). The pairs produced are deflected by intervening IGMFs, forming an angle $\delta$ with the direction of the parent gamma ray. 
The distance the electrons travel is typically of the order of the mean free path for inverse Compton scattering ($\lambda_\text{IC}$; see Eqs.~\ref{eq:mfpIntegral} and~\ref{eq:mfpApproxIC}). The up-scattered gamma rays can restart the cascade depending on their energy, such that the cascade would have multiple generations of particles. In Figure~\ref{fig:cone} only one generation is shown. Finally, the secondary gamma rays are detected at Earth forming an angle $\theta_\text{obs}$ with respect to the line of sight, i.e., the line connecting the observer and the object. With this scheme in mind, we can estimate the relevant gamma-ray observables, namely the spectrum, arrival directions, and light curves, either analytically (see Sec.~\ref{ssec:analytical}) or numerically (see Sec.~\ref{ssec:codes}).

The secondary photons resulting from the electrons deflected in the presence of IGMFs will be delayed compared to primary gamma rays emitted at the same time. Depending on the distance to the source, the duration of the emission and the properties of the IGMF, some of the secondary gamma rays from the cascade, produced from electrons via inverse Compton scattering, will not be able to arrive at Earth within one Hubble time, leading to an energy-dependent decrease in the flux.
Therefore, the three main gamma-ray observables relevant for IGMF studies are
\begin{itemize}
\item spectral effects; 
\item angular distribution;
\item time delays.
\end{itemize}
Naturally, quite often combinations of these strategies are employed, as will be presented in more detail in Sec.~\ref{sec:results}.

\subsection{Analytical Description of Propagation and Observables}
\label{ssec:analytical}

Neronov \& Semikoz~\cite{neronov2009a} presented a pedagogical model describing how gamma-ray telescopes can be used to probe IGMFs. This model is a suitable approximation for the energy range of interest, between GeV and tens of TeV. Making use of some simplifying assumptions, they derived analytical expressions for the expected signatures of specific combinations of magnetic-field strength ($B$) and coherence length ($L_B$). It is beyond the scope of this review to derive the formulae, but it is certainly worth transcribing the main results and some of the steps required to obtain them.

One can distinguish two regimes of propagation for the charged component of the electromagnetic cascades. They are determined by an interplay between the characteristic scales of inverse Compton scattering ($\lambda_\text{IC}$) and the coherence length of the magnetic field ($L_B$). For $\lambda_\text{IC} \gg L_B$, the propagation is quasi-rectilinear (ballistic), whereas for $\lambda_\text{IC} \ll L_B$, the electrons diffuse before they produce the secondary photons via IC scattering. In the former case, the electrons can be seen as effectively moving in a homogeneous magnetic field, such that in the small-angle approximation $\delta \simeq L_B / r_\text{L}$, wherein $r_\text{L}$ is given by Equation~\ref{eq:larmor}. In the latter case, we have $\delta \simeq \sqrt{\lambda_{\rm IC} L_{B}}/r_{\rm L}$. Together with equations \ref{eq:mfpApproxIC} and \ref{eq:larmor} we then obtain the estimate 
\begin{equation}
\delta \simeq
\begin{cases} \label{delta}
0.03^\circ (1 + z_{\rm PP})^{-\frac{1}{2}} \left( \dfrac{E_{e}'}{10 \, {\rm TeV}} \right)^{-\frac{3}{2}} \left( \dfrac{B}{10^{-15}\,{\rm G}} \right)\left( \dfrac{L_{B}}{1\,{\rm kpc}} \right)^{\frac{1}{2}}  & L_{B} \ll \lambda_{\rm IC}\,,\\
0.003^\circ (1 + z_{\rm PP})^{-2} \left( \dfrac{E_{e}'}{10 \, {\rm TeV}} \right)^{-2} \left( \dfrac{B}{10^{-15}\,{\rm G}} \right) & L_{B} \gg \lambda_{\rm IC}\,,
\end{cases}
\end{equation}
where $E_{e}'$ is the electron energy at redshift $z_\text{PP}$. Note that a more detailed investigation \cite{caprini2015a} shows that the deflection angle also weakly depends on the spectral index $\alpha_{B}$ of the magnetic field (see~Sec.~\ref{ssec:framework}) for $L_{B} \ll \lambda_{\rm IC}$ .

For distant sources, the pairs are produced closer to the source than to Earth ($\lambda_\text{PP} \ll D$). If $\delta \ll 1$, then we can adopt the approximation $z \simeq z_\text{src} \simeq z_\text{PP}$, which allows us to derive an analytic expression for $\theta_\text{obs}$:
\begin{equation} \label{eq:haloSize}
\theta_{\rm obs} \simeq
\begin{cases}
0.07^\circ (1+z)^{-\frac{1}{2}} \left( \dfrac{\tau_{\theta}}{10} \right)^{-1} \left( \dfrac{E_{\gamma}}{0.1 \, \text{TeV}} \right)^{-\frac{3}{4}} \left( \dfrac{B}{10^{-14} \, {\rm G}} \right) \left( \dfrac{L_{B}}{1 \, {\rm kpc}} \right)^{\frac{1}{2}} & L_{B} \ll \lambda_{\rm IC}\,, \\
0.5^\circ (1+z)^{-2} \left( \dfrac{\tau_{\theta}}{10} \right)^{-1} \left( \dfrac{E_{\gamma}}{0.1 \, {\rm TeV}} \right)^{-1} \left( \dfrac{B}{10^{-14} \, {\rm G}} \right) & L_{B} \gg \lambda_{\rm IC}\,,
\end{cases}
\end{equation}
where $\tau_{\theta}$ is the ratio between the angular diameter distance from the observer to the source and the mean free path for pair production, $\lambda_{\rm PP}$. Morphologically, this corresponds to a ``halo'' of secondary photons around the point-like source. Note that, while the morphology of the arrival directions does resemble a halo in the axi-symmetric case, this is not always the case. Depending on the geometry of the jet ($\Theta_\text{los} > 0^\circ$; see Fig.~\ref{fig:cone}) and properties of the intervening magnetic fields (e.g., helical fields), more complex shapes arise. We continue to use the term `halo' nonetheless.

An interesting and somewhat more accurate approach to estimate the size of such haloes was presented in Ref.~\cite{Ahlers:2011jt}, in which the moments of the halo distribution are calculated from diffusion-cascade equations. 
This method is applicable whenever the distribution of gamma rays emitted by the source is isotropic or the jet opening angle ($\Theta_\text{jet}$) is sufficiently large.

\bigskip
Another important quantity when determining IGMFs from electromagnetic cascades is the time delay $\Delta t_B$, defined as the difference between the following two quantities: the cumulative propagation time of the ``reprocessed'' gamma rays resulting from the cascades (see Fig.~\ref{fig:cone}), consisting of the lifetime $t_\text{PP}$ of the primary gamma ray until it results in pair production and of the duration $t_{\rm sec}$ of the cascade from the secondary gamma rays; and the light-travel time ($t_{\rm prim}$) of primary gamma rays. Therefore, one can write the equation
\begin{equation} \label{eq:timeDelay}
\Delta t_{B} = ( t_{\rm PP} + t_{\rm sec} ) - t_{\rm prim}\,.
\end{equation}
For the standard consideration of IGMFs we have $z_\text{PP} \simeq z_\text{src} = z \ll 1$ and $\delta \ll 1$, such that Eq.~\ref{eq:timeDelay} becomes
\begin{equation}
\Delta t_B \simeq
\begin{cases}
7 \times 10^{5} \, {\rm s} \, (1 - \tau_{\theta}^{-1}) (1 + z)^{-5} \kappa \left( \dfrac{E_{\gamma}}{0.1\,{\rm TeV}} \right)^{-\frac{5}{2}} \left( \dfrac{B}{10^{-18}\,{\rm G}} \right)^{2} & L_{B} \ll \lambda_{\rm IC}\,,\\
1 \times 10^{4} \, {\rm s} \, (1 - \tau_{\theta}^{-1}) (1 + z)^{-2} \kappa \left( \dfrac{E_{\gamma}}{0.1\,{\rm TeV}} \right)^{-2} \left( \dfrac{B}{10^{-18}\,{\rm G}} \right)^{2} \left( \dfrac{L_{B}}{1\,{\rm kpc}} \right) & L_{B} \gg \lambda_{\rm IC}\,.
\end{cases}
\end{equation}

\bigskip

The last observable we describe here concerns the probing of magnetic helicity of IGMFs using gamma rays and was first suggested in \cite{Tashiro:2013bxa}. Since then, it has been further extended and investigated in a significant number of publications \cite{tashiro2014a,tashiro2015a,chen2015b,long2015a,vachaspati2017a,alvesbatista2016b,Duplessis:2017rde,alvesbatista2017d,Asplund:2020frm,Kachelriess:2020bjl}. There it was shown that the helical part of the magnetic field spectrum (see Eq.~\ref{Bspectrum}) has a direct impact on the morphology of the halo around the gamma-ray source. In particular, when the magnetic field is helical, the halo becomes ``twisted'', i.e. instead of an (elongated) circular or oval halo, as one would expect from considering the simple analytic formulas derived above, the result is a spiral-like pattern (see Fig.~\ref{fig:helicity}).

This twisted pattern or, more specifically, its handedness, can be measured by the quantity $Q$ introduced in \cite{Tashiro:2013bxa} (and summarised in \cite{Vachaspati:2020blt}) as
\begin{equation}
Q(\hat{\mathbf{n}}_{1},\hat{\mathbf{n}}_{2},\hat{\mathbf{x}}_{\rm los}) = \left( \hat{\mathbf{n}}_{1} \times \hat{\mathbf{n}}_{2} \right) \cdot \hat{ \mathbf{x}}_{\rm los}\,,
\end{equation}
where $\hat{\mathbf{n}}_{1}$ and $\hat{\mathbf{n}}_{1}$ are the unit vectors of the arrival directions of two particles with the respective energies $E_{1}$ and $E_{2}$ (with $E_{1} < E_{2}$), and 
$\hat{ \mathbf{x}}_{\rm los}$ is the unit vector along the line of sight from the observer to the source. Using this one can calculate the so-called $Q$-statistics, given by
\begin{equation}
\overline{Q}(\theta_{\rm obs}^{\rm max}) = \left< Q(\hat{\mathbf{n}}_{1},\hat{\mathbf{n}}_{1},\hat{\mathbf{x}}_{\rm los}) \right>_{\theta_{\rm obs} \leq \theta_{\rm obs}^{\rm max}}\,,
\end{equation}
i.e.~the average over all photons with angles $\theta_{\rm obs}$ up to a value of $\theta_{\rm obs}^{\rm max}$.

If the direction of the line of sight is not known, the arrival direction $\hat{\mathbf{n}}_{3}$ of a third particle with an energy $E_{3}$ (with $E_{3} > E_{2} > E_{1}$) may be considered instead of $\hat{ \mathbf{x}}_{\rm los}$. In fact, by generally considering such triplets of particles from any direction in the sky, one can calculate the generalised $Q$-quantity (and, subsequently, the corresponding statistics) as~\cite{tashiro2015a}
\begin{equation} \label{Qsky}
Q(E_{1},E_{2},E_{3},\theta^{\rm max}) = \dfrac{1}{N_{1} N_{2} N_{3}} \sum_{\hat{\mathbf{n}}_{3}} \sum_{\angle (\hat{\mathbf{n}}_{1}, \hat{\mathbf{n}}_{2}) \leq \theta^{\rm max}} Q(\hat{\mathbf{n}}_{1},\hat{\mathbf{n}}_{2},\hat{\mathbf{n}}_{3})\,,
\end{equation}
where for every particle with the arrival direction $\hat{\mathbf{n}}_{3}$ (and given energy $E_{3}$) the second summation is carried out over all particles with the given energies $E_{1}$ and $E_{2}$ (with $E_{3} > E_{2} > E_{1}$) with arrival directions $\hat{\mathbf{n}}_{1}$ and $\hat{\mathbf{n}}_{2}$, respectively, which lie inside ``patches'' of angular size $\theta^{\rm max}$ around $\hat{\mathbf{n}}_{3}$. Finally, the values $N_{1}$, $N_{2}$, and $N_{3}$ in Eq.~\ref{Qsky} are the corresponding total numbers of particles for each of the three energies.

The final step in connecting the $Q$-statistics (and therefore the handedness of particle arrival directions) with the handedness of the magnetic field (and therefore its helicity) is to consider the case $\theta^{\rm max} \rightarrow \pi/2$. As shown in \cite{tashiro2015a}, this is proportional to the helical part of the spectrum $\mathcal{H}_{k}$, as defined in Eq.~\ref{Bspectrum}.

An alternative to the $Q$-statistics, introduced in \cite{alvesbatista2016b}, is the $S$-statistics which, for a single source, can be used to quantify the spiral shape of the halo.

\subsection{Plasma Instabilities}
\label{ssec:plasma}

The physics of electromagnetic cascades described above is well understood, but it neglects the back-reaction of the intergalactic medium on the cascades. This is a common assumption adopted in most IGMF studies, but if it turns out to be a poor approximation, plasma effects may become dominant. It was suggested~\cite{broderick2012a} that the electrons in the cascade interact with the IGM and lead to the generation of plasma instabilities, losing their energy and consequently heating the IGM. Due to the extreme parameters of the interacting components (for example, a factor of up to $10^{24}$ between the density of the electron beam and the background plasma \cite{0004-637X-758-2-102}), it is practically impossible to exactly calculate the impact of the instabilities on the development of the cascade. Nevertheless, one can rely on approximations and/or extrapolations.

The IGM parameters relevant for plasma instabilities are its temperature, which is typically $T_\text{IGM} \sim 10^4 \; \text{K}$~\cite{broderick2012a}, and the density, which in the cosmic voids is $n_\text{IGM} \sim 0.1 \; \text{m}^{-3}$~\cite{hui1997a}. Another important parameter is density of the gamma-ray beam, which is related to its luminosity.

\begin{figure}[hbt]
    \centering
    \includegraphics[width=0.70\textwidth]{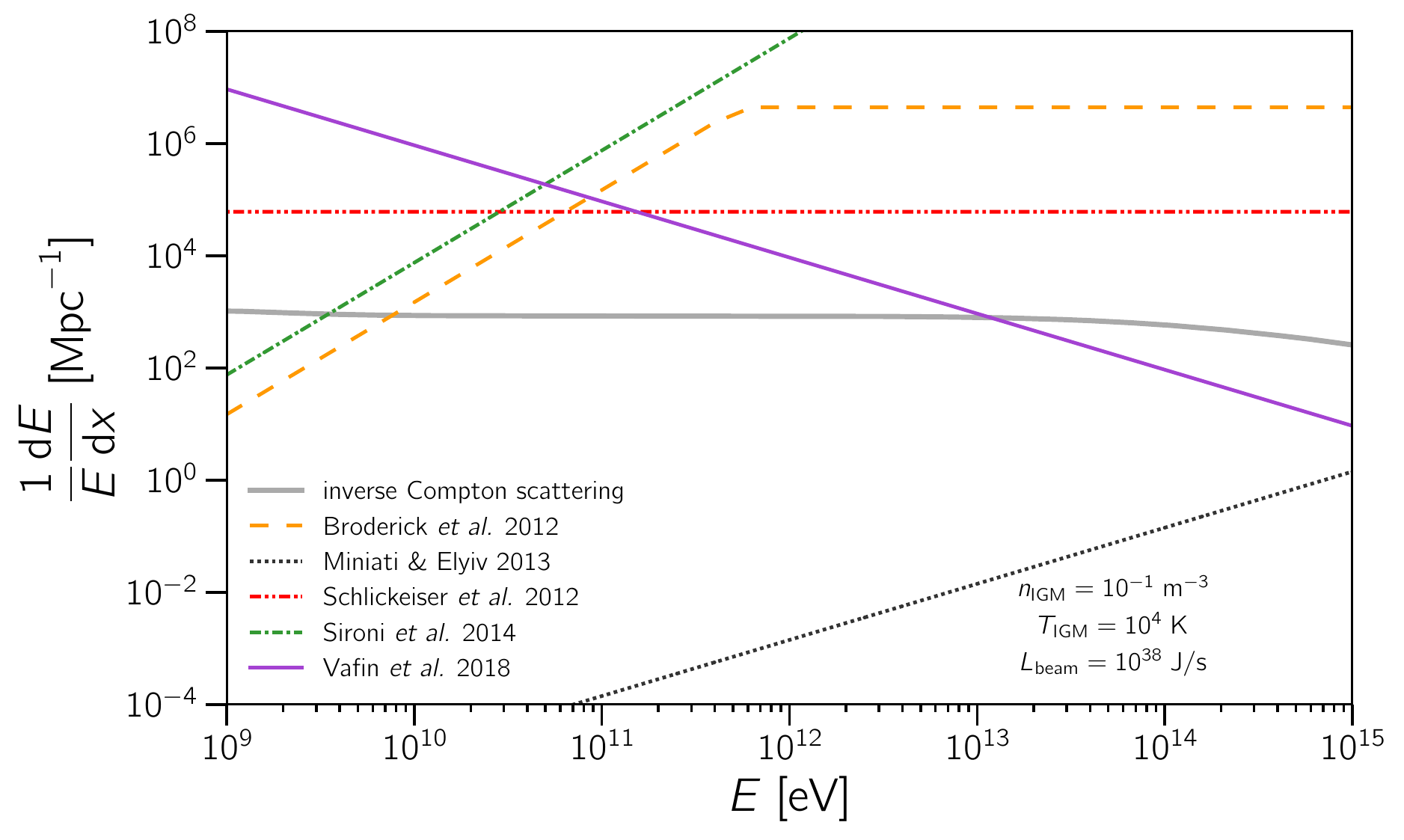}
    \caption{Cooling rates due to plasma instabilities computed at $z=0$, according to different models~\cite{broderick2012a, Miniati:2012ge, Schlickeiser:2013eca,  Sironi:2013qfa, Vafin:2018kox}. This example is for a typical scenario for an IGM density of $n_\text{IGM}=10^{-1} \; \text{m}^{-3}$ and temperature of $10^4 \; \text{K}$, for a blazar beam with luminosity $\mathcal{L}=10^{38} \; \text{J/s}$. We also present the inverse mean free path for inverse Compton scattering in the CMB for comparison.}
    \label{fig:plasma}
\end{figure}

As mentioned above, there is no general agreement on whether plasma instabilities are important for the propagation of electromagnetic cascades. Even if one accepts this assumption, it is not clear which kind of instability could be dominant. In fact, the modulation \cite{0004-637X-758-2-102,Schlickeiser:2013eca,Kempf:2015pvk,Vafin:2018kox}, oblique \cite{broderick2012a,Sironi:2013qfa,Chang:2014cta}, kinetic \cite{Perry:2021rgv}, and longitudinal \cite{Shalaby:2020fnm} instabilities, as well as non-linear Landau Damping \cite{Miniati:2012ge} have been considered in the literature. On the other hand, Ref.~\cite{Rafighi:2017ise} found that even if they are present, the effect of plasma instabilities is too small to cause a significant impact on observations. A comparison of the energy-loss length for different types of instabilities is shown in Fig.~\ref{fig:plasma}.

Several authors subsequently published results of actual simulations of gamma-ray propagation including possible plasma instabilities effects and compared them to actual observations \cite{Saveliev:2013jda,yan2019a, alvesbatista2019g}. The results show that, while the instabilities can, indeed, lead to appreciable deviations from the paradigmatic picture of cascade development, they may not be sufficient to render gamma-ray constraints of IGMFs completely ineffective. In this case, all that would be required is a more detailed modelling of the electromagnetic cascades --- which, understandably, would be more susceptible to uncertainties due to the inclusion of an additional and poorly-understood effect.

There is also another window of opportunity to evade plasma instabilities, even if they majorly disrupt electromagnetic cascades. The growth rates of plasma instabilities are often estimated using simplifying assumptions like a continuous and constant stream of particles. However, if the object in question emits gamma rays in flares, the temporal structure of the resulting charged beam should be considered. In particular, if the duration of the flare is short enough, the instability might not have enough time to fully develop, consequently having no significant impact on the electrons.

\subsection{Other Propagation Phenomena}
\label{ssec:bsm}

So far we have discussed how electromagnetic cascades propagate in the Universe in light of a standard picture entirely contained within the framework of quantum electrodynamics (Sec.~\ref{ssec:propagationTheory}). We also briefly discussed how plasma instabilities could quench electromagnetic cascades propagating in the IGM (Sec.~\ref{ssec:plasma}). In this subsection, we briefly describe how other physical phenomena could affect the development of electromagnetic cascades and, consequently, observations of high-energy gamma-ray sources, with direct implications for IGMF studies. 

\bigskip
One phenomenon that can interfere with the propagation of gamma rays and consequently compromise IGMF constraints is gravitational lensing. Massive objects can significantly deform the space-time surrounding them, altering the path along which particles travel. As a result, in the context of gamma rays, gravitational lenses can significantly deform the morphology of haloes and, in the case of flaring objects, increase the time delays due to this  gravitationally-induced contribution. The first source for which gravitational lensing has been observed at gamma-ray energies (up to 30 GeV) was PKS~1830-211 \cite{barnacka2011a}. Since then, the phenomenon has been detected for this and other  gamma-ray--emitting objects~\cite{cheung2014a, barnacka2015b, fermi2015b, hess2019b}. Ref.~\cite{barnacka2014a} investigated how macrolenses could compromise estimates of the optical depth for pair production, concluding that this effect would not lead to any measurable changes in this observable. This results is corroborated by the more detailed study of~\cite{boettcher2016a}.

\bigskip

Other potentially important phenomena arise in the context of BSM models. The most widely studied BSM processes that could interfere with the gamma-ray--IGMF framework we present here involve Lorentz invariance violation (LIV) and interactions with axion-like particles (ALPs). Because of the potentially important role played by these phenomena in determining the gamma-ray signatures of sources used for IGMF constraints, we briefly touch upon these issues. 

\bigskip

Lorentz invariance violation is a possible consequence of various BSM approaches, especially in the context of quantum gravity. The standard approach, from the field theory side, is to create a minimal SM extension, in particular by introducing additional terms to the SM Lagrangian, resulting in a effective field theory with LIV \cite{Colladay:1998fq}.

In terms dynamics, the main effect when it comes to the propagation of particles is the modification of the dispersion relation, given by
\begin{equation}
E_{\rm LIV}^{2} = E^{2} + \eta \frac{p^{n+2}}{M_{\rm Pl}^{n}}
\end{equation}
where $E_{\rm LIV}$ and $E$ is the particle energy with and without LIV, respectively, $\eta$ is a dimensionless parameter measuring the strength of the LIV, and $M_{\rm Pl}$ is the Planck mass. This, on the one hand, changes the threshold of a given reaction, and, as a consequence, also changes the corresponding propagation length, as it modifies the limits of the integral in Eq.~\ref{eq:mfpIntegral}. In addition, new reactions, which are not possible without LIV, may then be kinematically allowed, such as, to name the ones most relevant in the context of this review, spontaneous photon decay into pairs/photons, the vacuum Cherenkov effect for electrons and charged UHECRs, as well as spontaneous photodisintegration of multi-nucleon nuclei. For an overview on the modifications of the processes in the electromagnetic cascades and in UHECR propagation, see \cite{Mattingly:2005re,Galaverni:2008yj,Martinez-Huerta:2020cut} and \cite{Mattingly:2005re,Bietenholz:2008ni,Saveliev:2011vw}, respectively.

All this may result in a significant modification of particle propagation and, therefore, impact the corresponding observations. In particular, \cite{Vankov:2002gt} showed that LIV might dramatically increase the interaction length of pair production for energies above $100$~TeV and therefore suppress the cascade development. On the other hand, LIV might also imply that the photons' speed is energy-dependent, thus resulting in energy-dependent time delays \cite{Martinez-Huerta:2020cut}.

\bigskip

Axion-like particles, or ALPs, appear in extensions of the SM. They are pseudo Nambu-Goldstone bosons associated with a broken symmetry $U(1)$. They were originally introduced by Peccei and Quinn~\cite{peccei1977a, peccei1977b} as a solution to the strong~CP problem. ALPs couple to standard-model particles via the Lagrangian~\cite{raffelt1988a}
\begin{equation}
    \mathcal{L}_{a\gamma} = -\dfrac{1}{4} g_{a\gamma} \mathbf{E}_\gamma \cdot \mathbf{B}_\text{ext}\, ,
    \label{eq:lagrangianALP}
\end{equation}
where $\mathbf{E}_\gamma$ denotes the electric field of the photon itself, and $\mathbf{B}_\text{ext}$ represents an external magnetic field (in the context of this review, IGMFs). The coupling constant $g_{a\gamma}$ determines how strongly photons, in our case gamma rays, will interact with the ALP field. For a distance $x$, the probability of a photon to convert into an ALP (and vice-versa) is
\begin{equation}
    P_{a \leftrightarrow \gamma} (x) = \sin^2 \left( \frac{1}{2} \arctan \left(\dfrac{2 \Delta_{a\gamma}} {\Delta_a - \Delta_\gamma} \right)\right) \sin^2\left( \dfrac{1}{2} x \sqrt{(\Delta_\parallel - \Delta_a)^2 + 4 \Delta_{a\gamma}} \right) \, ,
    \label{eq:probALP}
\end{equation}
Here the $\Delta$ terms refer to the solution of the equation of motion derived from the Lagrangian (Eq.~\ref{eq:lagrangianALP}). They describe:
the coupling between photons and ALPs ($\Delta_{a\gamma} = g_{a\gamma} B_{T}$) between a photon of energy $E$ and the ALP field for a an external magnetic field $B_T$ transverse to the direction of propagation of the photon; the kinetic term ($\Delta_a = m_a^2 / 2E$) for an ALP of mass $m_a$; the polarisation states of the photon ($\Delta_\parallel$ and $\Delta_\perp$), which, in our case, encompass the contribution of the IGM plasma ($\Delta_\text{pl} = -\omega_\text{pl}/2E$) for a plasma frequency $\omega_\text{pl}$, and the QED vacuum polarisation ($\Delta_\text{QED} \propto B_T^2 $), which depends on the direction ($\Delta_{\text{QED},\perp} = 7 \Delta_\text{QED} / 2$ and $\Delta_{\text{QED},\parallel} = 2 \Delta_\text{QED}$). For more details, the reader is referred to, for example, Ref.~\cite{raffelt1988a}.

Two regimes of propagation can be identified~\cite{deangelis2008a}, depending on whether the gamma-ray energy is larger or smaller than the critical energy ($E_c$), given by:
\begin{equation}
    E_c = \dfrac{m_a^2 - \omega_\text{pl}}{4 \Delta_{a\gamma}} \approx 2.5 \left( \dfrac{\left| m_a^2 - \omega_\text{pl}^2 \right| }{10^{-20} \; \text{eV}^2} \right) \left( \dfrac{10^{-9} \; \text{G}}{B_T} \right) \left( \dfrac{10^{-11} \; \text{GeV}^{-1}}{g_{a\gamma}} \right) \; \text{GeV} \,.
\end{equation}
The limit $E \gg E_c$ corresponds to the so-called strong mixing. In this case, the probability of conversion (see Eq.~\ref{eq:probALP}) does not depend on the energy. If $E \ll E_c$, the energy dependence becomes salient leading to an effective low-energy cut-off. 

The propagation of gamma rays will be affected by ALPs in multiple ways. Firstly, the magnetic fields in the sources will contribute to the total ALP-photon mixing. Secondly, once the gamma rays are injected into the intergalactic space, they may initiate electromagnetic cascades as described in Sec.~\ref{ssec:propagationTheory}. Upon entering the Galaxy, ALP-photon mixing may also occur due to the Galactic magnetic field. The oscillation probability from Eq.~\ref{eq:probALP} will then be a combination of the probabilities in each of these environments, as discussed in, e.g., Refs.~\cite{hochmuth2007a, sanchezconde2009a, deangelis2013a, dobrynina2015a}.

In the case of gamma rays propagating over cosmological distances, the oscillation probability from Equation~\ref{eq:probALP} implies deviations from the expected transparency of the Universe, since gamma rays will be able to travel longer without undergoing pair production. A number of works investigated the possibility that this ``pair production anomaly'' could be related to ALPs~(e.g.~\cite{deangelis2009a, horns2012a, meyer2013a, deangelis2013a, rubtsov2014a}).

The effects of IGMFs on gamma-ray--ALP interconversion has been studied adopting several methods ranging from semi-analytical approaches to more detailed simulations. While the first studies on the topic assumed relatively simple magnetic-field configurations, later studies~\cite{mirizzi2009a, meyer2014a} improved the treatment including turbulent fields (see Sec.~\ref{ssec:framework} for details). Investigations considering the actual distribution of magnetic fields in the magnetised cosmic web have also been performed~\cite{montanino2017a} 

While ALPs are an important ingredient that could play a leading role on the intergalactic propagation of gamma rays in a magnetised Universe, they have not been observed and only constraints exist. Some limits were derived using gamma-ray observations~\cite{meyer2013a, hess2013c, fermi2016b, Buehler:2020qsn}, but much of the parameter space is excluded due to observations of photons in wavelengths other than gamma rays. Interestingly, some works have been obtaining combined limits on IGMFs and ALPs together~\cite{masaki2017a}. For reviews on the status of the field, see, e.g., Refs.~\cite{redondo2011a, horns2016a, meyer2016c}.

\subsection{Propagation Codes}
\label{ssec:codes}

The simulation of the propagation of electromagnetic cascades in the intergalactic medium is often done numerically, employing some approximations to enable (semi-)analytical solutions (e.g.,~\cite{lee1998a, dermer2011a, finke2015a}). In the last decade, Monte Carlo methods have been used to treat this problem~\cite{taylor2011a, kachelriess2012a, alvesbatista2016a, alvesbatista2016b, Fitoussi:2017ble, blytt2020a}. Many codes are now publicly available.

Elmag~\cite{kachelriess2012a, blytt2020a} is a Fortran code that tracks the development of electromagnetic cascades. In the first two versions of the code~\cite{kachelriess2012a}, the effects of magnetic fields on the charged cascade component, namely time delays and deflections, were taken into account using the small-angle approximation. Therefore, this version was limited to low magnetic-field strengths.  The newest version, Elmag~3.01~\cite{blytt2020a}, adds a Lorentz-force solver that enables three-dimensional simulations assuming turbulent magnetic fields generated based to Eqs.~\ref{Bspectrum} and~\ref{eq:Hspectrum}, following Refs.~\cite{giacalone1994a, giacalone1999a}, as well as custom grids.

CRPropa~\cite{armengaud2007a,kampert2013a,alvesbatista2016a} is a well-known code for ultra-high-energy cosmic-ray propagation, written in C++ and with Python bindings (since version 3). The original CRPropa~\cite{armengaud2007a} and CRPropa~2~\cite{kampert2013a} made use of the numerical methods from Ref.~\cite{lee1998a}, namely using transport equations to treat the development of electromagnetic cascades. The newest version include a full treatment of electromagnetic cascades~\cite{alvesbatista2016a, alvesbatista2017d, heiter2018a}. A variety of magnetic-field configurations are available, and the code is flexible enough to handle customisations and arbitrary magnetic-field grids. Moreover, it can generate turbulent magnetic fields on the grid or using grid-less methods~\cite{tautz2013a}, with improvements from~\cite{schlegel2020a}.
Earlier releases of CRPropa~3 supported the propagation of gamma rays with energies $\gtrsim 10^{17} \; \text{eV}$ through the Monte Carlo EleCa code~\cite{settimo2015a}. Due to the computational limitations, EleCa is restricted to the highest energies, but in CRPropa a hybrid approach using the transport-equation treatment of~\cite{lee1998a} was available. Recent developments enable a full Monte Carlo treatment of photons from ultra-high down to GeV energies, which is useful for exploring the UHECR-induced cascade scenarios (see Sec.~\ref{ssec:sources}).

A Fortran code for cascade propagation was developed by Fitoussi~\textit{et al.}~\cite{Fitoussi:2017ble}. It does not rely on any approximations, performing the full three-dimensional propagation of the cascades. In this code, the magnetic field is composed of cells with randomly oriented strengths. A semi-analytical treatment of the cascade development in Mathematica is implemented in $\gamma$-Cascade~\cite{blanco2019a}.

Plasma instabilities are often neglected in simulations, or treated within a dedicated MHD computational framework. In~\cite{alvesbatista2019g}, \texttt{grplinst} was presented. It is a module for the CRPropa code that implements plasma effects on the electrons as an additional energy-loss term of the form
\begin{equation}
	-\dfrac{{\rm d} E_{e}}{{\rm d}x}(E_{e}, x, z) = \dfrac{E_{e}}{c \tau(E_e, x, z)} \, ,
	\label{eq:plasma_dEdx}
\end{equation}
where $x$ is the length of the trajectory described by an electron (or positron) of energy $E_{e}$, $z$ is the redshift, and $\tau$ is the electron energy-loss time due to the plasma instability. Within this simplified treatment, the time scale in which the instability grows ($\mathcal{T}$) is taken to be the electron cooling time ($\tau$). Therefore, Eq.~\ref{eq:plasma_dEdx} is overestimated, since in reality $\mathcal{T} \leq \tau$. More recently, these same parametrisations of \texttt{grplinst}~\cite{alvesbatista2019g} were implemented in Elmag~3.02.


\subsection{Examples}
\label{ssec:examples}

To illustrate the effects of IGMFs, we present the results of Monte Carlo simulations of the development of electromagnetic cascades in the IGM. To this end, we use the CRPropa code~\cite{alvesbatista2016a}, but similar results could have been derived with, e.g., Elmag~\cite{kachelriess2012a, blytt2020a} or the code presented in Ref.~\cite{Fitoussi:2017ble}. 

\begin{figure}[ht]
    \centering
    \includegraphics[width=0.45\textwidth]{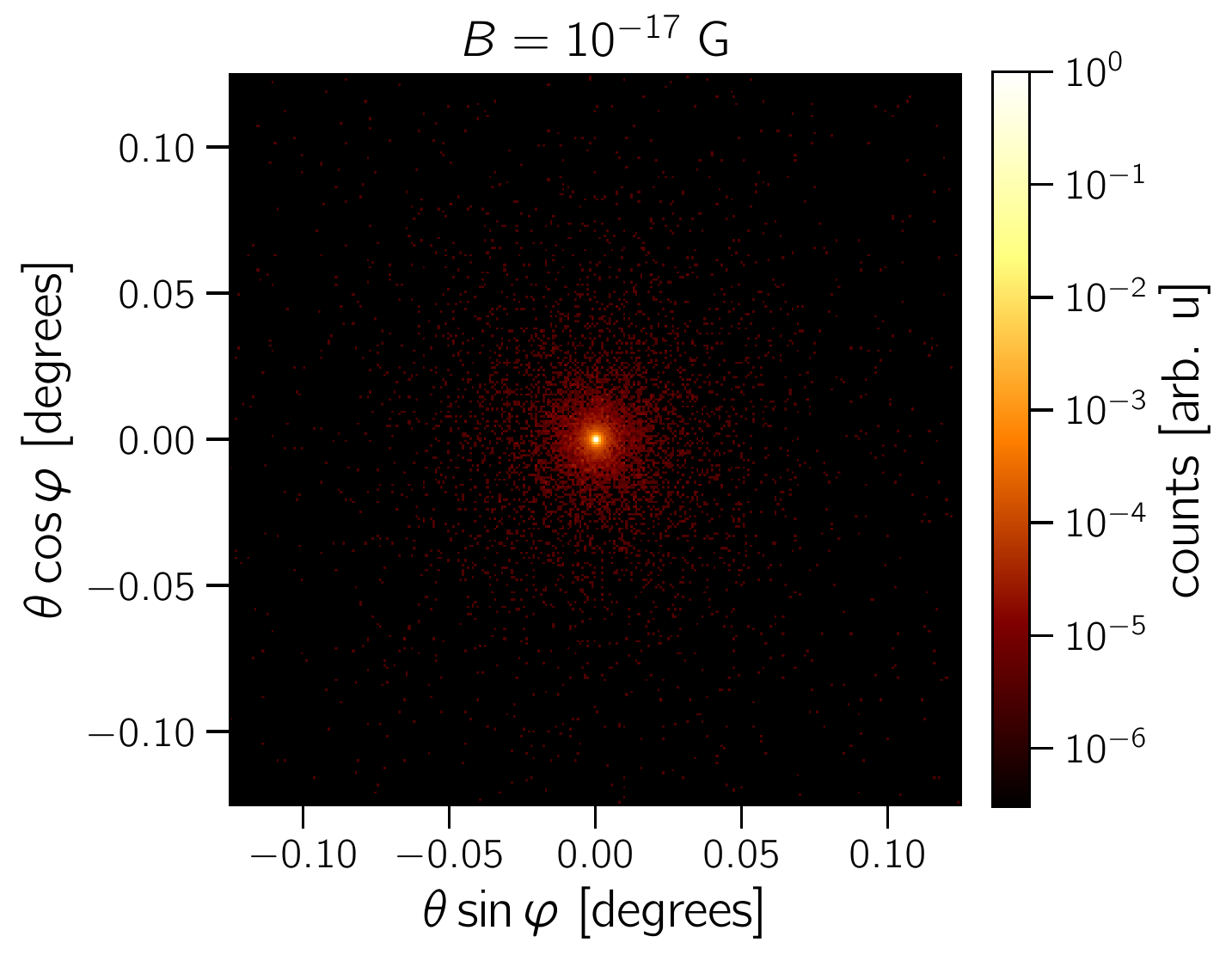}
    \includegraphics[width=0.45\textwidth]{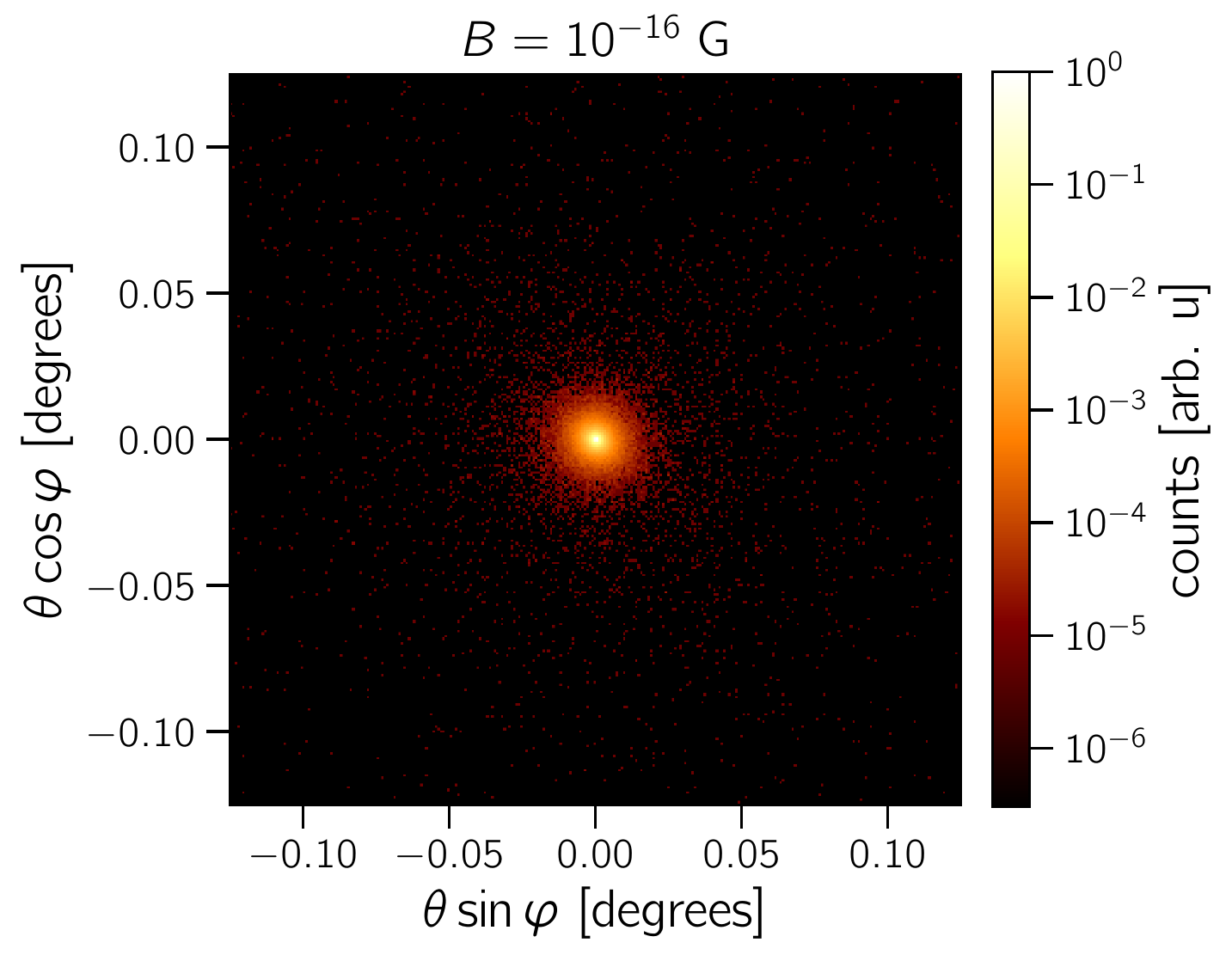}
    \includegraphics[width=0.45\textwidth]{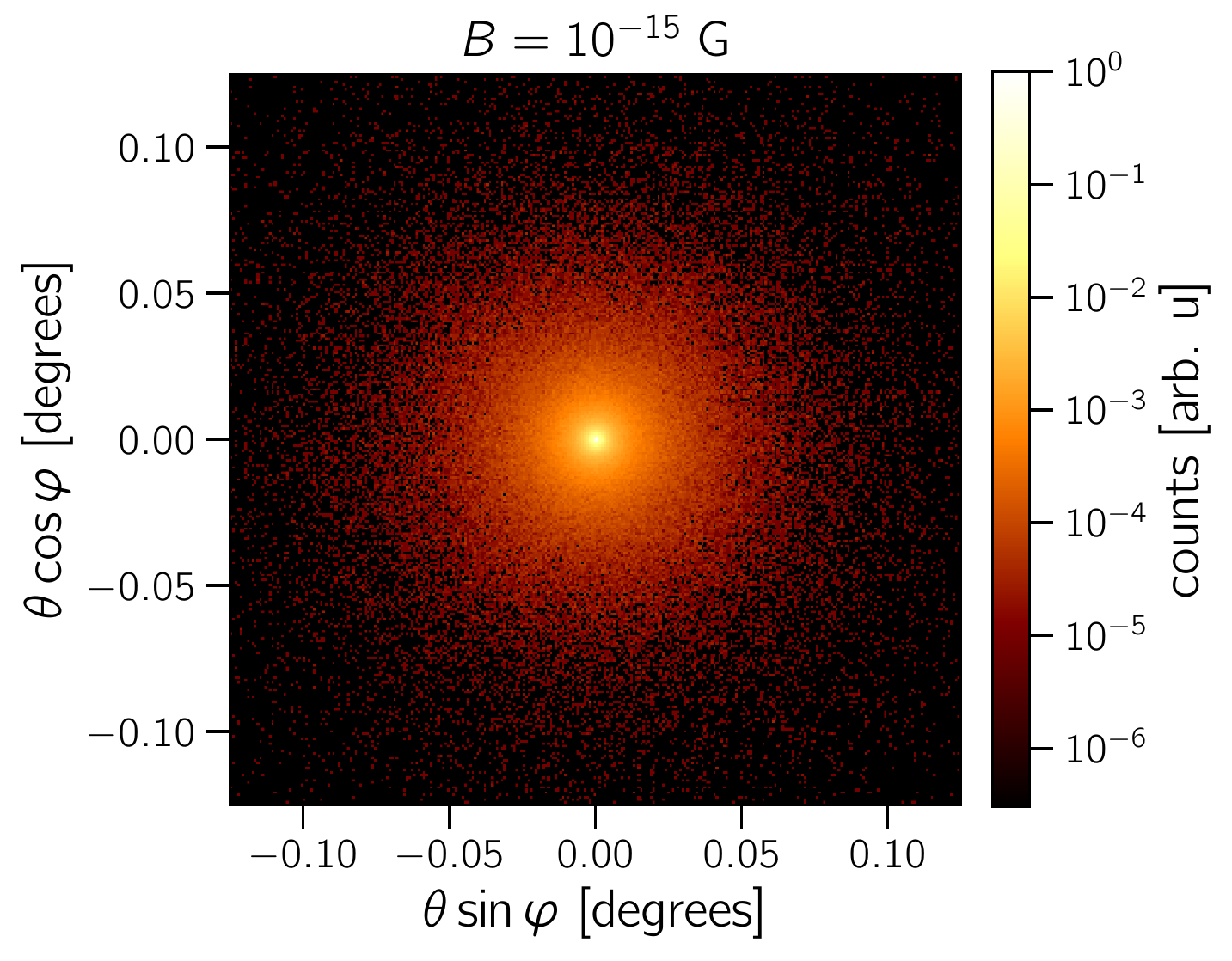}
    \includegraphics[width=0.45\textwidth]{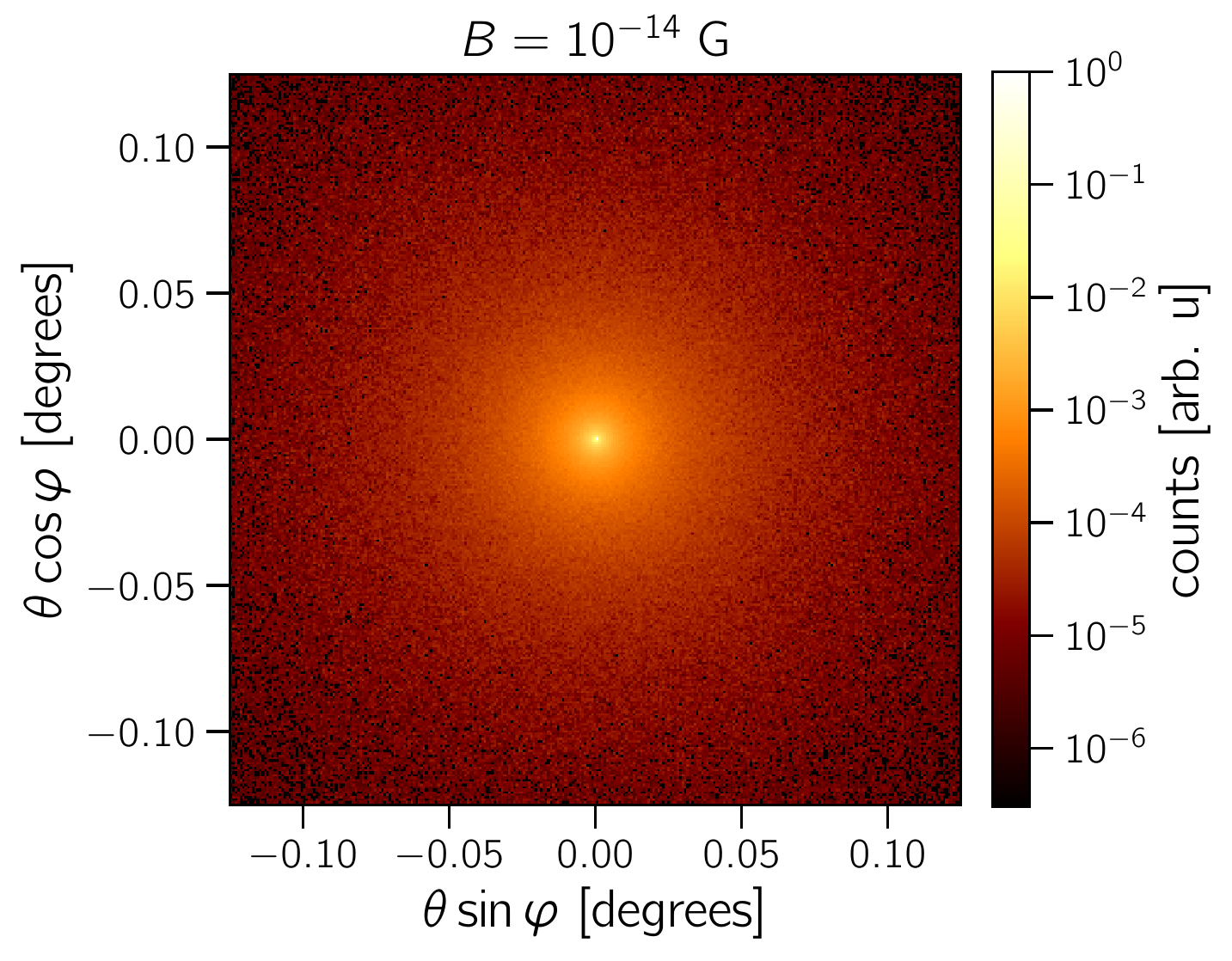}
    \caption{Simulated pair haloes around the blazar 1ES~0229+200, for the  magnetic-field strengths indicated in the figures. The intrinsic source spectrum is a power-law with $\alpha=1.5$ and $E_\text{max}=5 \; \text{TeV}$, following Ref.~\cite{vovk2012a}. The coherence length is assumed to be $L_B=1 \; \text{Mpc}$ in this example. All gamma rays with  $E \gtrsim 1 \; \text{GeV}$ are considered in this plot.}
    \label{fig:simHalo}
\end{figure}

\bigskip
We first select the archetypical blazar 1ES~0229+200~\cite{hess2007a}, used in several IGMF studies (cf.~Sec.~\ref{ssec:sources}). This object is located at a distance corresponding to $z \simeq 0.14$. We fix the coherence length to $L_B = 1 \; \text{Mpc}$ to illustrate the formation of the haloes around the source. This is shown in Fig.~\ref{fig:simHalo}.
Note that these plots are shown in the coordinate system of the simulation, as we would observe from Earth, but they can be immediately converted to another coordinate system, such as Galactic or equatorial.

\begin{figure}[ht]
    \centering
    \includegraphics[width=0.70\textwidth]{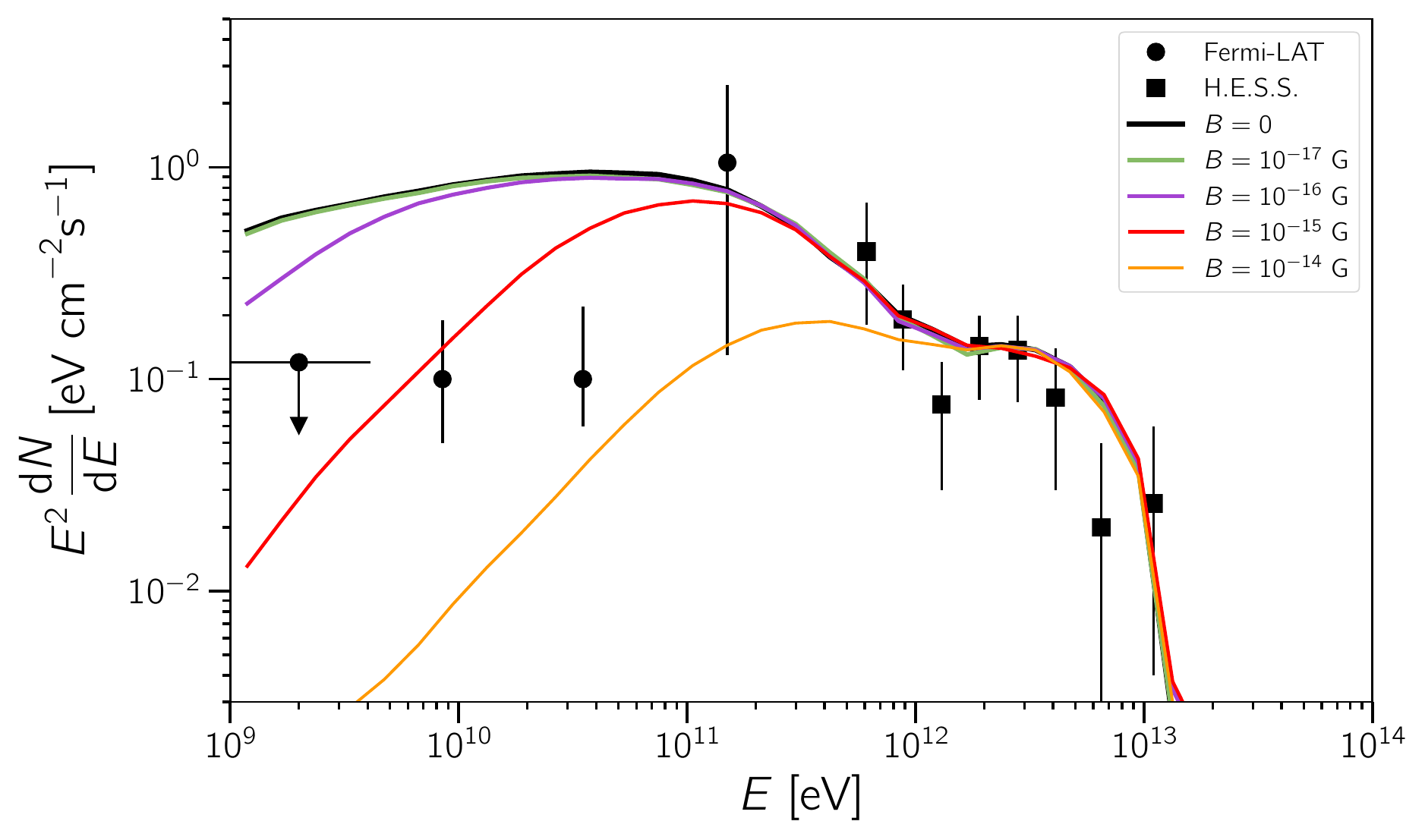}
    \caption{This figure illustrates the expected point-like flux from 1ES~0229+200 obtained with Monte Carlo simulations. The lines corresponds to different magnetic-field strengths, indicated in the legend. The data points represent measurements by Fermi-LAT~\cite{vovk2012a}  and H.E.S.S.~\cite{hess2007a}. The source parameters are the same as in Figure~\ref{fig:simHalo}.}
    \label{fig:simSpec}
\end{figure}

It is evident from Fig.~\ref{fig:simHalo} that a significant fraction of the flux is not contained within a finite-sized containment radius centred at the source. This causes spectral changes with respect to the point-like source flux, as shown in Fig.~\ref{fig:simSpec}.

\begin{figure}[htb]
    \centering
    \includegraphics[width=0.495\textwidth]{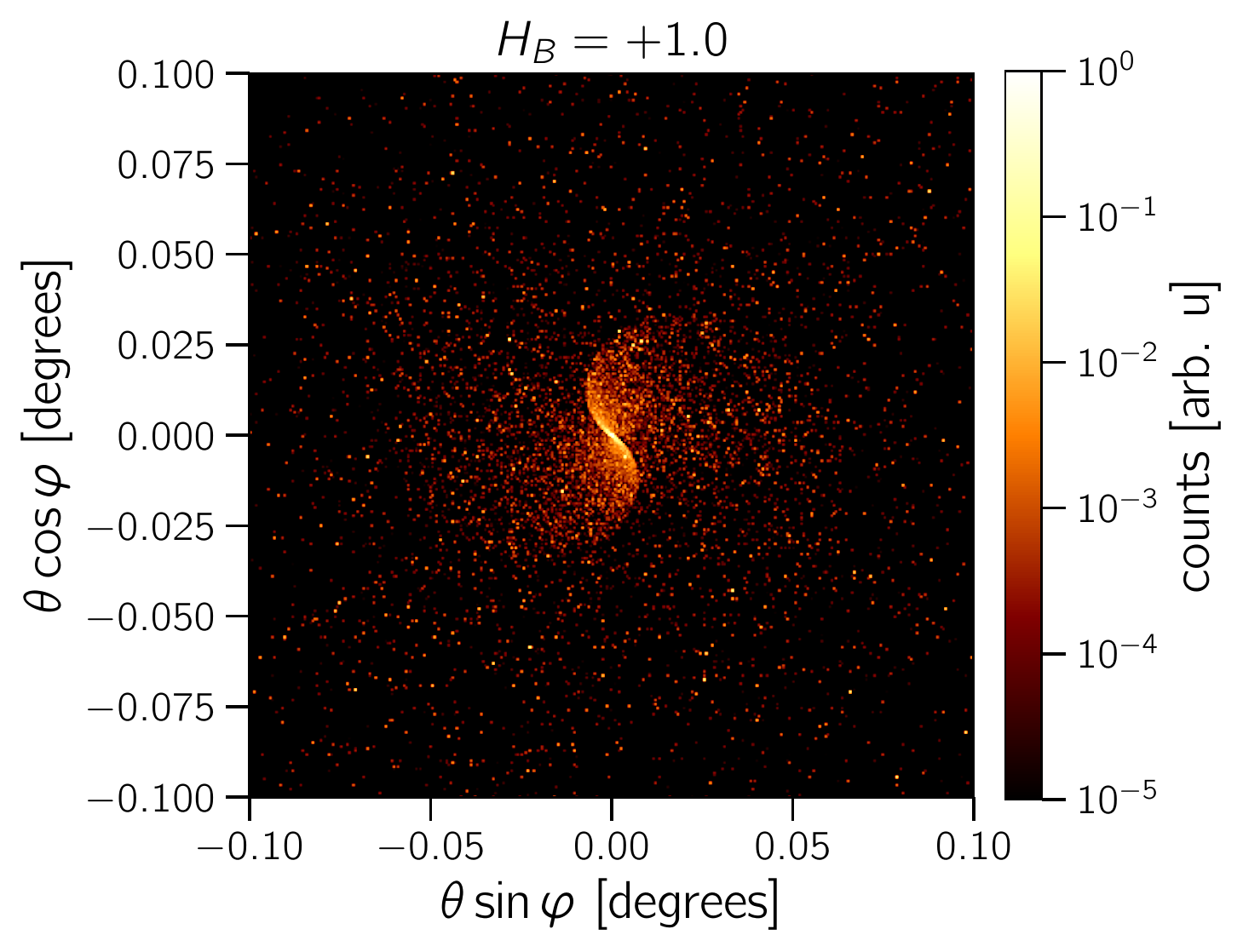}
    \includegraphics[width=0.495\textwidth]{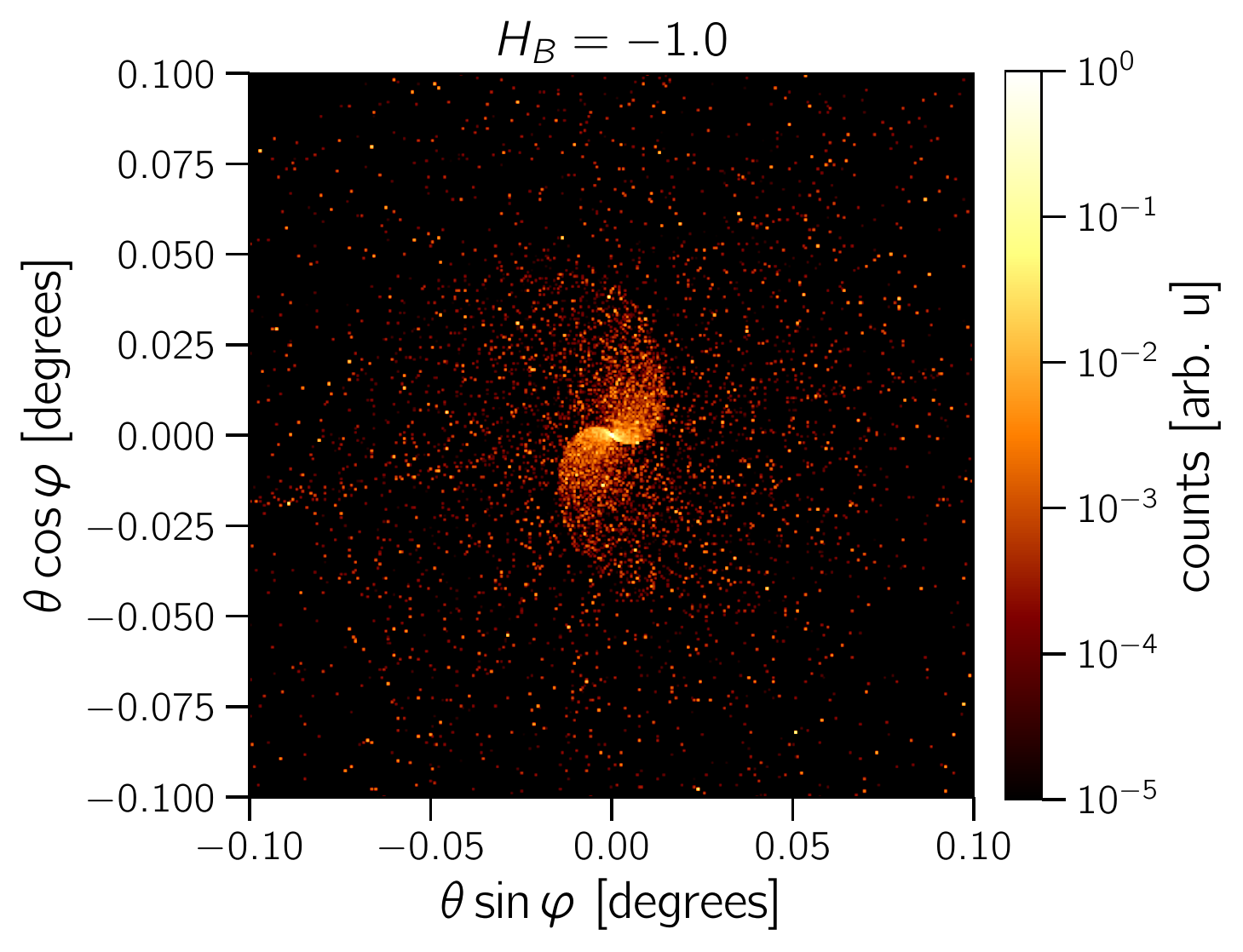}
    \caption{This figure illustrates the expected arrival directions of gamma rays considering arbitrary realisations of a helical turbulent magnetic field of strength $B = 10^{-15} \; \text{G}$ with a Batchelor spectrum ($\alpha_B = 5$) and coherence length $L_B = 200 \; \text{Mpc}$. The left panel corresponds to a realisation with maximally positive helicity ($H_B = +1$), whereas the right one corresponds to another realisation with negative helicity ($H_B = -1$). The source, assumed to be located at $z=0.08$, emits gamma rays with a spectrum $E^{-1.5}$ and an exponential cutoff at $E_\text{max} = 100 \; \text{TeV}$ (see Eq.~\ref{eq:sourceSpec}). Its jet has an opening angle $\Theta_\text{jet} = 5^\circ$ and it points directly at Earth ($\Theta_\text{los}=0^\circ$; see Fig.~\ref{fig:cone}). The colour scale indicates the normalised spectrum-weighted number of detected events in the angular bin.}
    \label{fig:helicity}
\end{figure}

\bigskip
If magnetic fields are maximally helical, then the halo shape shown in Fig.~\ref{fig:simHalo} changes considerably. In fact, the changes can be so drastic that the morphology of the arrival direction pattern is no longer a standard axi-symmetric halo. For a source pointing straight at Earth ($\Theta_\text{los}$), we expect a spiral-like pattern, as shown in Fig.~\ref{fig:helicity} for a hypothetical source at $z=0.08$. In this case, the handedness of the halo reflects the sign of the helicity: left-handed for $H_B > 0$, and right-handed for $H_B < 0$.

The arrival directions of gamma rays can be quantified through the calculation of the $Q$-factors (see Eq.~\ref{Qsky}). The effects of the helicity of IGMFs are more pronounced for large coherence lengths, hence the choice of $L_B = 250 \; \text{Mpc}$ in the example of Fig.~\ref{fig:helicity}. The smaller the ratio between the source distance and the coherence length, the more diluted the signal is, which would reflect in the $Q$-factors (see~\cite{alvesbatista2016b, Kachelriess:2020bjl}).

\section{Results}
\label{sec:results}

Fluxes of distant objects, like the ones used in IGMF studies, are normally computed for a point-like source. The magnetically-induced broadening of the electromagnetic cascade will naturally affect the measured point-like flux, especially at lower (typically $E \lesssim 10 \; \text{GeV}$) energies, since these are predominantly secondary gamma rays if the intrinsic spectrum extends beyond $\sim \; \text{TeV}$ energies. In this case, the larger the angular broadening caused by IGMFs, the more pronounced is the suppression of the gamma-ray flux from a point-like source, since a fraction of the events will leak outside the point spread function (PSF) of the detector.

Gamma-ray sources are observed during a given time. If IGMFs are such that the incurred time delays ($\mathfrak{T}$) exceed this window of observation, then the measured flux will be affected. The suppression will, in general, be stronger as energy decreases because this contribution is likely produced by lower-energy electrons whose Larmor radii increase with energy (see Eq.~\ref{eq:larmor}). The relevance of this effect depends on an interplay between the duration of the emission, which depends on the type of object (see Section~\ref{ssec:sources}), the time window of observation, and the magnetically-induced time delay.

A number of studies attempted to constrain IGMFs based on gamma rays using different methods. One possible way to classify these studies is the number of sources used, such that in Sec.~\ref{ssec:individual} we describe the results of analyses of individual gamma-ray sources, and in Sec.~\ref{ssec:stacked} those of multiple stacked sources. In general, the results are for the magnetic-field strength. However, there have also been attempts to constrain the coherence length and helicity of IGMFs with gamma rays, which we discuss in Sec.~\ref{ssec:coherenceLength} and \ref{ssec:helicity}, respectively, followed by results for IGMFs considering that the cascades are induced by UHECRs in Sec.~\ref{ssec:UHECRgamma}. Finally, in Sec.~\ref{ssec:prospects} we discuss the prospects for IGMF measurements with gamma-ray observatories.

\subsection{Analyses of Individual Sources}
\label{ssec:individual}

\begin{figure}[!ht]
    \centering
    \includegraphics[width=0.80\textwidth]{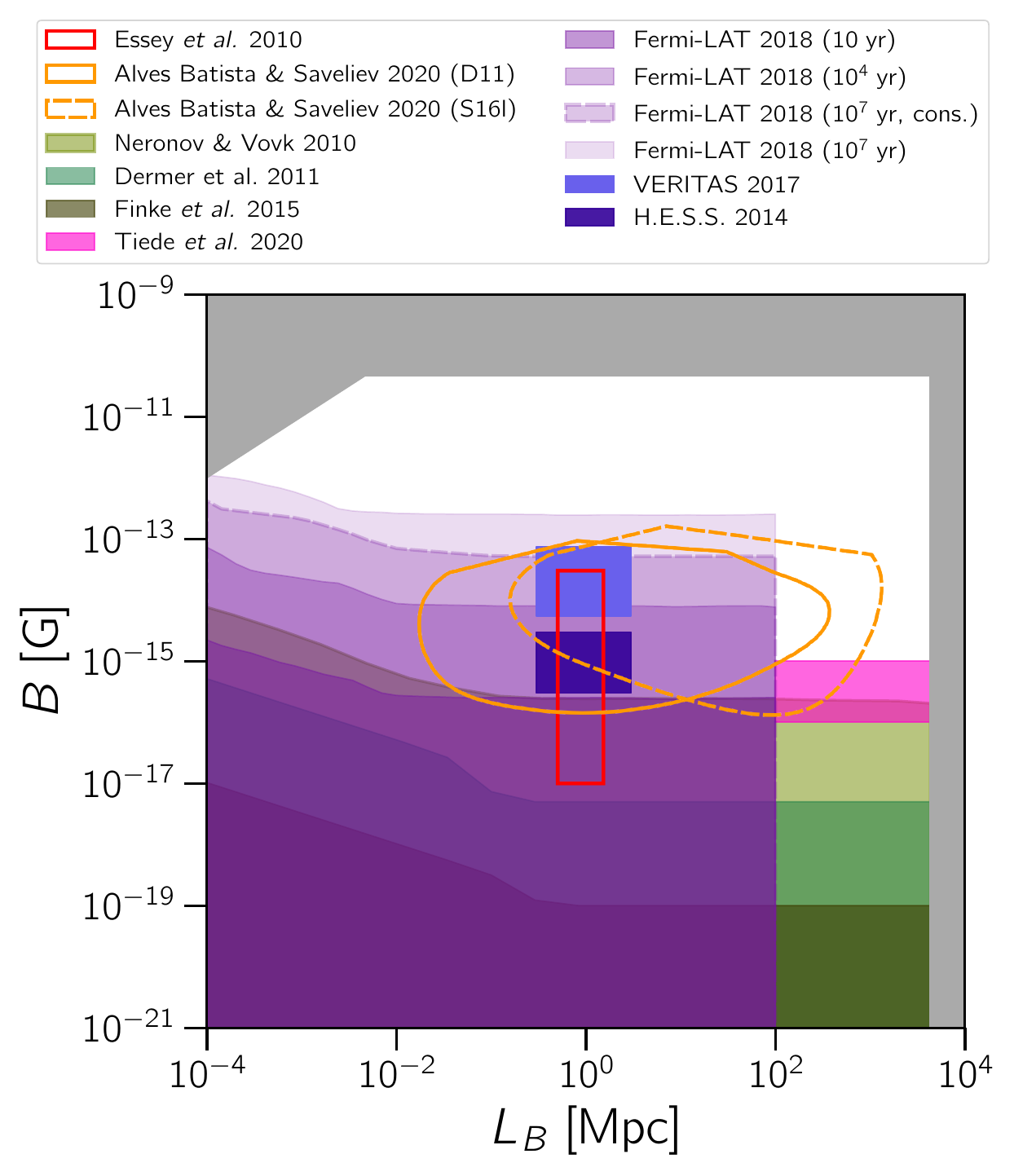}
    \caption{Compilation of some constraints found in the literature. Coloured regions represent \emph{excluded} regions of the parameter space, whereas non-filled bounded by a line indicate \emph{allowed} regions. The regions shown in green are exclusions by Neronov \& Vovk~\cite{neronov2010a}, Dermer \textit{et al.}~\cite{dermer2011a}, and Finke \textit{et al.}~\cite{finke2015a}. The purple regions are bounds derived by Fermi-LAT \& Biteau~\cite{fermi2018a}, for different source acitivity times ($\mathfrak{T}$). The region labelled `conservative' excludes from the analysis the blazars 1ES~0229+200 and 1ES~1208+304 (see text for details). Constraints by VERITAS~\cite{veritas2017a} and H.E.S.S.~\cite{hess2014a} are shown as pinkish rectangles. The red rectangle corresponds to the 95\% C.L. allowed region according to Essey \textit{et al.}~\cite{Essey:2010nd}. The orange lines demarcate the best-fit regions (68\% C.L.) of the parameter space according to Alves~Batista \& Saveliev~\cite{alvesbatista2020a} for the EBL models labelled D11~\cite{dominguez2011a} and the lower-limit S16l model~\cite{stecker2016a}. Note that the regions plotted refer exclusively to the region of the parameter space reported in the corresponding references, without extrapolations to high/low values of the coherence length. The grey region are the combined excluded regions from Fig.~\ref{fig:GenConstraints}, obtained via other methods.}
    \label{fig:BL_summary}
\end{figure}

The first constraints on IGMFs using gamma rays were derived by Neronov \& Vovk~\cite{neronov2010a}, using observations by Fermi-LAT and IACTs of the blazars 1ES~0229+200, 1ES~0347+121, and 1ES~1101-232. The results suggest that $B \gtrsim 10^{-16.5} \; \text{G}$ for $L_B \gtrsim 1 \; \text{Mpc}$ and $B L_B^{1/2} \gtrsim 10^{-16.5}\, \; {\rm G} \, {\rm Mpc}^{1/2}$ for $L_B \ll 1 \; \text{Mpc}$, as shown in Fig.~\ref{fig:BL_summary}. This dependence of the lower limit of IGMFs on the coherence length, $L_B$, follows from the simplified approach commonly used (see Sec.~\ref{ssec:analytical}), and is adopted in most of the works to which we refer below, with a few exceptions.  Most importantly, this work was the first to firmly exclude the case $B = 0$. An earlier investigation~\cite{davezac2007a} of the blazar 1ES~1101-232 concluded that an exceedingly hard intrinsic spectrum for this object would be required to account for observations, unless the EBL was more intense and the IGMF were stronger ($B \gtrsim 10^{-15} \; \text{G}$). Ref.~\cite{Fan:2003qr} argued along the same lines, when interpreting observations of the blazar H1426+428.

Following Neronov \& Vovk's~\cite{neronov2010a} influential work, much attention has been given to this topic, new objects were used in the analyses and other observables were introduced. 
For instance, the MAGIC Collaboration obtained compatible results via the non-observation of pair haloes around Mrk~421 and Mrk~501~\cite{magic2010a}. In~\cite{tavecchio2010a}, the authors analysed gamma-ray observations of 1ES~0229+200, excluding $B \lesssim 10^{-15.5} \; \text{G}$. A more comprehensive study included additional sources (1ES~0347+121, and 1ES~1101-232, RBG~J0152+017, and PKS~0548-322) and showed that, if the emission by these objects is stable over a time scale $\mathfrak{T} \sim 10^{7} \; \text{yr}$, then, in general, $B \gtrsim 10^{-15} \; \text{G}$~\cite{tavecchio2011a}.

\bigskip
A thorough analysis of Fermi-LAT and IACT observations of five blazars was performed in~\cite{finke2015a}, excluding $B \lesssim 10^{-19} \; \text{G}$ (for $L_B \gtrsim 1 \; \text{Mpc}$) at a 5$\sigma$-level, as indicated in Fig.~\ref{fig:BL_summary}. These results are robust with respect to the choice of the EBL model, variability of the source ($\mathfrak{T}$), and jet opening angle ($\Theta_\text{jet}$). Moreover, the authors perform additional checks about the energy range of the Fermi-LAT data used in the analysis, demonstrating that the results are the same regardless of whether a dataset containing gamma rays with energies starting from $100 \; \text{MeV}$ or $1 \; \text{GeV}$ is used. The significance of these results decreases slightly for other EBL models (Refs.~\cite{kneiske2010a, franceschini2008a}). Note that a more detailed treatment of the cascade interactions would increase the flux at lower energies, so that these estimates are actually conservative.

The H.E.S.S Collaboration~\cite{hess2014a} combined its own observations with those of Fermi-LAT. The absence of a detectable halo around PKS~2155-304 excludes $10^{-15.5} \lesssim B / \text{G} \lesssim 10^{-14.5}$ at a 99\% C.L., assuming $L_B \gtrsim 1 \; \text{Mpc}$. Constraints in a similar range ($10^{-15.5} \lesssim B / \text{G} \lesssim 10^{-14.5}$) were obtained by the VERITAS Collaboration~\cite{veritas2017a}, at a 90\% C.L., from 1ES~1218+304. The constraints by both H.E.S.S. and VERITAS are shown in Fig.~\ref{fig:BL_summary}.
In principle, because the coherence length was assumed to be $L_B = 1 \; \text{Mpc}$ in both works, one could be tempted to extrapolate the conclusions to $L_B > 1 \; \text{Mpc}$. However, because the objects used to derive the constraints (1ES~1218+304 and PKS~2155-304) show some intrinsic variability~\cite{hess2010b, veritas2010b, hess2017b}, 
care should be taken extrapolating the bounds to larger values of the coherence length.

The Fermi-LAT Collaboration~\cite{fermi2018a} compiled a catalogue of sources that was used to constrain IGMFs and performed a detailed analysis of this sample of blazars. They found no evidence of extended emission, neither around individual objects, nor in the stacked analysis. This way, they could constrain the allowed values for the strength of IGMFs: $B \gtrsim 10^{-16.5} \; \text{G}$ for $L_B \gtrsim 10 \; \text{kpc}$. This result is conservative, assuming $\mathfrak{T} \simeq 10 \; \text{yr}$. If this condition is relaxed, the bounds are even stronger: $B \gtrsim 10^{-14} \; \text{G}$ and $B \gtrsim 10^{-12.5} \; \text{G}$, for $\mathfrak{T} \simeq 10^{4} \; \text{yr}$ and $\mathfrak{T} \simeq 10^{7} \; \text{yr}$, respectively, as shown in Fig.~\ref{fig:BL_summary}. While these limits were derived for a jet with half-opening angle $\Theta_\text{jet} \leq 10^\circ$ (see Fig.~\ref{fig:cone}), no misalignment was considered ($\Theta_\text{los} = 0^\circ$). These results were derived for a combination of sources which include, among others, 1ES~0229+200 and 1ES~1218+304. Because there are indications that they could be variable~\cite{veritas2014a, veritas2010b}, if these sources are removed from the analysis, limits which are more conservative are obtained.

\bigskip
The time scale over which a given source emits gamma rays ($\mathfrak{T}$) influences the bounds one can derive. For instance, Ref.~\cite{huan2011a} analysed VERITAS and Fermi-LAT data. The lower bounds they obtained for $L_B \gtrsim 1 \; \text{Mpc}$ were $B \gtrsim 3 \times 10^{-18} \; \text{G}$ and $B \gtrsim 2 \times 10^{-16} \; \text{G}$, for source activity periods $\mathfrak{T} \simeq 3 \; \text{yr}$ and $\mathfrak{T} \rightarrow \infty$, respectively. The former is evidently more conservative, as it encompasses exclusively the period for which there are observations of the object (RBG~J0710+591). Similar considerations about $\mathfrak{T}$ were discussed in~\cite{dermer2011a}, which obtained $B \gtrsim 10^{-18} \; \text{G}$ for $L_B \gtrsim 1 \; \text{Mpc}$, for 1ES~0229+200 (see Fig.~\ref{fig:BL_summary}). These results are order-of-magnitude compatible with the lower limits by Ref.~\cite{yang2015a}, which are $B \gtrsim 10^{-16}$--$10^{-18} \; \text{G}$.

The flaring object Mrk~501 drew much interest for IGMF constraints, due to its variability (see, e.g.,~\cite{Stanev:1997zz, dai2002a, yang2008, magic2010a, neronov2012a, takahashi2012a, veritas2017a, fermi2018a, Korochkin:2020pvg}). The prospect for detecting pair echoes from this same object was studied in~\cite{Murase:2008pe}. With an analysis of observations of its 2009 flare by MAGIC, VERITAS, and Fermi-LAT, the authors of~\cite{neronov2012a} argued that its spectrum and time profile could be explained by IGMFs with $B \simeq 10^{-17}$--$10^{-16} \; \text{G}$ (for $L_B \gtrsim 1 \; \text{Mpc}$). Ref.~\cite{takahashi2012a} studied the pair echoes from this same blazar, concluding that $B \gtrsim 10^{-20} \; \text{G}$, assuming $L_{B} \simeq 1 \; \text{kpc}$, at a 90\% confidence level. Making use of similar methods, Ref.~\cite{takahashi2013a} analysed data from ARGO-YBJ and Fermi-LAT for Mrk~421, excluding $B \lesssim 10^{-20.5} \; \text{G}$ for $L_B \simeq 1 \; \text{kpc}$, at a 4$\sigma$ level. The results of the latter analysis are particularly interesting because they do not make assumptions about the intrinsic spectrum of the source during periods when it is not observed. 

A somewhat elaborate treatment of the cascade development was adopted in~\cite{taylor2011a}, in which Monte Carlo simulations were used to derive bounds on IGMFs for a sample of three blazars (1ES~0229+200, RBG~J0710+591, and 1ES~1218+304). The limits obtained depend on the strategy adopted for the analysis: $B \gtrsim 10^{-15} \; \text{G}$ considering the absence of haloes, as observed by Fermi-LAT, and $B \gtrsim 10^{-17} \; \text{G}$ considering time delays. 

\bigskip
An important factor that significantly affects IGMF estimates is the model of the EBL. Ref.~\cite{vovk2012a} studied this dependence for the archetypical extreme blazar 1ES~0229+200. They found $B \gtrsim 10^{-17} \; \text{G}$, which can increase by nearly two orders of magnitude depending on the EBL model. In fact, the EBL is one of the main intrinsic uncertainties that hinders the exclusion of the scenario $B=0$, as argued in~\cite{arlen2014a}. In this analysis, among the seven blazars considered, only one led to $B > 0$ irrespective of the choice of EBL model. However, the uncertainties in the intrinsic source spectrum and EBL model might be unrealistic, as noted in refs.~\cite{durrer2013a, Vachaspati:2020blt}.

\bigskip
The analysis by Dolag \textit{et al.}~\cite{Dolag:2010ni} is interesting because it employed magnetic fields obtained from cosmological magnetohydrodynamical simulations from~\cite{dolag2005a}. These simulations are constrained, i.e., they roughly reproduce the distribution of large-scale structures up to hundreds of Mpc. At larger distances, this cosmological volume was replicated up to the distance of 1ES~0229+200. The authors showed that  more than $\sim$~60\% of the Universe along the line of sight  of this object have magnetic fields with strength $B \gtrsim 10^{-16} \; \text{G}$. Interestingly, this analysis also showed that haloes can be used to probe the maximal energy of the gamma rays emitted by a source, $E_\text{max}$ (see Equation~\ref{eq:sourceSpec}). In fact, there is a considerable correlation between the value of $E_\text{max}$ that could be inferred from fits in the presence and in the absence of IGMFs~\cite{saveliev2021a}.

\bigskip
Ref.~\cite{veres2017a} employed a Monte Carlo code to model the development of electromagnetic cascades initiated by GRBs. However, because GRBs had not been detected at TeV energies until recently, when MAGIC observed GRB~190114C~\cite{magic2019a}, the authors extrapolated the GeV flux of GRB~130427A measured by Fermi-LAT up to TeV energies. If the extrapolation is correct, the lower limit obtained is $B \sim 10^{-17.5} \; \text{G}$ (for $L_B \gtrsim 1 \; \text{Mpc}$). 

With the first observation of VHE emission by a GRB, it was possible to effectively constrain IGMFs with gamma-ray observations. By combining Fermi-LAT and MAGIC data, Ref.~\cite{wang2020a} obtained a lower limit 
of $B \gtrsim 10^{-19.5} \; \text{G}$ for $L_B \gtrsim 100 \; \text{kpc}$. 
Ref.~\cite{dzhatdoev2020a} performed a similar analysis for GRB~190114C using Monte Carlo simulations. They concluded that Fermi-LAT is not sensitive enough to detect the cascade signal from this GRB on time scales of one month. The discrepancy between these two works is due to a combination of factors. Firstly, the former employed a simpler semi-analytical method, whereas the latter performed detailed Monte Carlo simulations using the three-dimensional version of the Elmag code. Moreover, the authors of \cite{dzhatdoev2020a} reconstructed the intrinsic spectrum following~\cite{dzhatdoev2017a}, while in Ref.~\cite{wang2020a} a fixed spectral index $\alpha=2$ (for a power-law distribution $\propto E^{-\alpha}$) was assumed. Yet another difference is the treatment of the time information of the photons. While~\cite{dzhatdoev2020a} accounted only for photons detected more than 62~s after the burst,~\cite{wang2020a} adopted~6~s. This issue is far from simple, as it requires knowledge of the inner workings of gamma-ray bursts. For further details, the interested reader is referred to the works on GRB~190114C by the MAGIC Collaboration~\cite{magic2019a, magic2019b}.

\subsection{Stacked and Diffuse Analyses}
\label{ssec:stacked}

It is rather difficult to observe magnetically-induced haloes from individual sources, as they are normally not bright enough to be detected \cite{aharonian1994a,elyiv2009a,Broderick:2016akd,Broderick:2018nqf}. 
Hence, techniques that are more sensitive are needed. Analyses of stacked samples of blazars could be useful for this purpose, since the signal-to-background ratio increases, easing the identification of any excess to the detector's PSF. 

The authors of~\cite{ando2010a} performed a stacked analysis of 170 AGNs using 11~months of Fermi-LAT data. They claim to have found an excess to the PSF of the detector $\simeq 0.5-0.8^\circ$, at a 3.5$\sigma$ level. Haloes of these sizes are caused by $B \simeq 10^{-15} \; \text{G}$ (see Eq.~\ref{eq:haloSize}). Nevertheless, it was later shown that these results could be attributed to instrumental effects associated to different treatments of photons measured in different parts of the detector~\cite{neronov2011a} (see also Ref.~\cite{fermi2013b}). This was not included in the PSF used in Ref.~\cite{ando2010a}, thus leading to an incorrect estimate of the strength of IGMFs.

The stacked analysis from Ref.~\cite{Chen:2014rsa} considered 24 selected high-synchrotron-peaked BL~Lacs, at $z < 0.5$. Using Fermi-LAT data, the authors found indications of extended emission, consistent with $B \sim 10^{-17}$--$10^{-15} \; \text{G}$. However, an updated analysis using 12 objects of the same population resulted in no compelling evidence for an extended emission, with only a modest $2\sigma$ significance for $B \sim 10^{-15}\,{\rm G}$~\cite{Chen:2018mjd}.

Another stacked analysis of a sample of 394 AGNs, 158 of which present flaring activity, was performed in Ref.~\cite{Prokhorov:2015yja}. Interestingly, the method employed considers temporal information of the sources by comparing the fluxes during low quiescent states and during flaring periods. No evidence for pair haloes was found. The recent analysis by Fermi-LAT~\cite{fermi2018a} corroborates this result, finding no indications for extended emission in the stacked source samples of high-synchrotron peaked BL~Lacs.

Using the method introduced in~\cite{Tiede:2017xng}, Ref.~\cite{Broderick:2018nqf} identified misaligned blazars from a catalogue of radio-loud AGNs, and performed a search for pair haloes around the stacked sample of these objects. They showed that a magnetic field with $B L_B^{1/2} \lesssim 10^{-15} \; \text{G} \, \text{Mpc}^{1/2}$ would lead to specific halo anisotropy patterns that are not observed, thus providing an upper limit on the strength of IGMFs. Note, however, that the assumptions about the intrinsic properties of the considered sources are subject to uncertainties. Considering the available lower limits derived in the works discussed above, the parameter space that would remain available for IGMFs would be tiny or, considering the stronger constrains from the recent results by Fermi-LAT~\cite{fermi2018a}, inexistent. The authors of~\cite{Broderick:2018nqf} then conclude that, if there is indeed no room for IGMFs that can explain the observations, then some other process might be at play that quenches the electromagnetic cascades. They claim that this could be due to, for instance, plasma instabilities (see Section~\ref{ssec:plasma}). More recently, the same group claimed to have found convincing evidence of the non-existence of pair haloes. Using the same method, they exclude $B \sim 10^{-16}$--$10^{-15} \; \text{G}$ with $L_B > 100 \; \text{Mpc}$ at a 3.9$\sigma$ level, and $B \sim 10^{-17}$--$10^{-14} \; \text{G}$ at 2$\sigma$~\cite{tiede2020a} (see Fig.~\ref{fig:BL_summary}). 

\bigskip
An interesting idea to constrain IGMFs is to study their possible imprints on the diffuse gamma-ray background (DGRB)~\cite{Venters:2012bx,yan2012a}, even though the validity of the assumptions used in these analyses is unclear. The presence of IGMFs may suppress the lower-energy diffuse gamma rays measured. Interestingly, the authors of Ref.~\cite{Venters:2012bx} claim that, in fact, the observations by Fermi-LAT already disfavours the scenario with null IGMF.
This agrees with~\cite{yan2012a}, who found that the contribution of cascade gamma rays from blazars to the DGRB changes significantly in the presence of IGMFs, such that for $B \gtrsim 10^{-12} \; \text{G}$ the blazar contribution to the spectrum of the DGRB changes.

\bigskip

In the context of diffuse searches, it is important to keep in mind that, in addition to the uncertainties in the EBL models (see Fig.~\ref{fig:photonFields}), there are fluctuations correlated to the processes that produce the EBL photons. This leads to inhomogeneities in the EBL distribution that can affect the propagation of electromagnetic cascades. However, as shown in ~\cite{abdalla2017a}, this effect is small ($\lesssim 1\%$), so it should have little impact on IGMF measurements using diffuse gamma-ray observations.

\subsection{Bounds on the Coherence Length}
\label{ssec:coherenceLength}

A method to measure the coherence length was suggested in Ref.~\cite{neronov2013a}. In this case, the slope of the light curve of secondary gamma rays would provide an upper limit on $L_B$. More specifically, the time dependence of the flux would be $\propto 1 / \sqrt{\Delta t_B}$ for coherence lengths much larger than the  mean free path for inverse Compton scattering ($L_B \gg \lambda_\text{IC}$), and approximately constant if $L_B \ll \lambda_\text{IC}$. Similarly, the angular profile of the haloes can also retain information about the coherence length. For $L_B \ll \lambda_\text{IC}$, the surface brightness profile is roughly uniform, whereas for $L_B \gg \lambda_\text{IC}$ it decays as the angular distance to the centre of the source increases.

With the first multimessenger observations of high-energy neutrinos from the flaring blazar TXS~0506+056 in coincide with electromagnetic radiation~\cite{icecube2018a, icecube2018b}, Ref.~\cite{alvesbatista2020a} used the cascade signal delayed with respect to the neutrino emission to constrain IGMFs. The derived limits depend on the EBL model, such that the hypothesis of null IGMFs could only be rejected for two out of the four models tested (Domínguez \textit{et al.}~\cite{dominguez2011a} and the lower-limit model by Stecker \textit{et al.}~\cite{stecker2016a}).
Interestingly, while the bounds are not robust, this work derived, \emph{within the investigated parameter space} 
, limits on the coherence length of IGMFs for the first time: $30 \; \text{kpc} \lesssim L_{B} \lesssim 300 \; \text{Mpc}$, at a 90\%~C.L., shown in Fig.~\ref{fig:BL_summary}. Naturally, the significance of this result depends on the reliability of the neutrino--gamma-ray correlation and on the assumptions made, namely that the IGMF has a Kolmogorov power spectrum, and that the intrinsic spectrum of TXS~0506+056 both during the flaring and quiescent periods can be described by a power-law with an exponential cut-off.

\subsection{Constraints on the Magnetic Helicity}
\label{ssec:helicity}

There has been a growing interest in probing the helicity of IGMFs, given its importance for understanding magnetogenesis. All-sky analyses of Fermi-LAT data employing the parity-odd correlators described in Section~\ref{sec:propagation} found indications of IGMFs with $B \sim 10^{-14} \; \text{G}$ at $L_{B} \sim 10 \; \text{Mpc}$ and an overall negative (left-handed) helicity \cite{tashiro2014a, chen2015b}. More recently, a re-analysis of a larger data set showed this result to be a fluctuation stemming from a miscalculation of the statistical significance that neglected the look-elsewhere effect~\cite{Kachelriess:2020bjl}. The same publication also claims that it is currently challenging to detect helicity, both in the fluxes of individual sources and in the diffuse gamma-ray background. In addition, the authors of \cite{Asplund:2020frm} claim that they did not find any handedness using the $Q$-statistics for Fermi data, being, however, unable to state definitively whether there is actually no handedness present or whether the $Q$-statistics is not sensitive enough for measuring it.

The signatures of helical IGMFs on the shape of haloes are unique, with significant deviations from axial symmetry, as illustrated in Fig.~\ref{fig:helicity}. Moreover, the sign of the helicity directly correlates with the handedness of the morphology of the arrival directions of gamma rays.
In Ref.~\cite{long2015a}, the authors employed a semi-analytical method to show that spiral-like patterns are the natural shape of the arriving gamma rays for helical fields. Nevertheless, within this simple framework, IGMFs were assumed to be homogeneous, which is not realistic unless the coherence lengths involved are exceedingly high, comparable to the distance of the source. In the more realistic case of turbulent magnetic fields with coherence lengths possibly shorter than the distance from Earth to the gamma-ray source in question, the spiral pattern could vanish, being diluted into something closer to a typical axisymmetric halo. This was, indeed, observed in a more detailed study using three-dimensional Monte Carlo simulations~\cite{alvesbatista2016b}. 
This work, however, does support the measurement of the helicity of IGMFs for $L_{B} \gtrsim 50 \; \text{Mpc}$, for sources at redshifts $z \lesssim 0.10$.

\subsection{Constraints from UHECR-produced Gamma Rays} \label{ssec:UHECRgamma}

In Sec.\ref{ssec:sources} the model proposed in Refs.~\cite{essey2010a, essey2010b} was presented. 
In this scenario, the flux of some extreme blazars could be attributed to cosmic-ray interactions along the line of sight. Essey \textit{et al.}~\cite{Essey:2010nd} constrained IGMFs considering that the gamma rays observed are a combination of those emitted by the blazars and those stemming from CR interactions. In this case, the combined limits from all three blazars analysed favour $10^{-17} \lesssim B / \text{G} \lesssim 10^{-14.5}$, at 95\% C.L. This result is robust with respect to the choice of EBL model. It is also shown in Fig.~\ref{fig:BL_summary}. Other authors also performed similar investigations (e.g.,~\cite{Prosekin:2012ne, murase2012d, oikonomou2014a, Khalikov:2019fbd, Dong:2021ghu}).

Interestingly, within the UHECR-cascade framework, photons with $E \gtrsim 10 \; \text{TeV}$ could be detected even if the sources are very distant ($z \gtrsim 0.1$). Nevertheless, for IGMFs with $B \gtrsim 10^{-14} \; \text{G}$, significant deviations from a point-like flux would be expected due to magnetically-induced deflections, compromising any constraints that one could derive in the context of this hadronic model~\cite{Essey:2010nd}.

For blazars, the investigations of the role of line-of-sight interactions in gamma-ray measurements~\cite{essey2010a, essey2011a} were also shown to lead to time variabilities that are characteristic for specific magnetic-field properties, of the order of years for $B=10^{-15} \; \text{G}$~\cite{Prosekin:2012ne}. Nevertheless, the variabilities cannot be too short since even for weak IGMFs of the order of $10^{-18} \; \text{G}$ cascade photons with $E=10 \; \text{GeV}$ would be magnetically-delayed by $\sim 10 \; \text{yr}$~\cite{murase2012d}. In the purely leptonic scenario, this timescale is shorter by a ten fold~\cite{takami2013a}.

A detailed account of the effect of magnetic fields on both the electromagnetic cascades as well as on their progenitor UHECRs was presented in~\cite{Khalikov:2019fbd}. Using three-dimensional simulations of the magnetised cosmic web from~\cite{dolag2005a} and detailed numerical methods, the authors found that the cascade broadening could be detected with next-generation gamma-ray telescopes and possibly some of the ones in operation today.

Note that the propagation of cosmic rays is not trivial. There are many uncertainties involved (see, e.g., ~\cite{alvesbatista2015d, alvesbatista2019c} for a discussion), which might compromise the production of gamma rays. Moreover, depending on the location of the blazar in the cosmic web, local magnetic fields (e.g. in filaments) might significantly deflect cosmic rays away from the line of sight~\cite{dolag2005a, alvesbatista2017c, hackstein2018a,Garcia:2021cgu} (see the discussion at the end of Section~\ref{sec:bfields}). 


\subsection{Prospects for Measurements of IGMFs}
\label{ssec:prospects}

From the discussion so far a general picture of IGMFs emerges, wherein gamma rays play a fundamental role in excluding part of the parameter space shown in Fig.~\ref{fig:GenConstraints}, as summarised in Fig.~\ref{fig:BL_summary}. 
It is important to bear in mind that there are many factors that could compromise the derived limits shown in the latter figure. This includes uncertainties regarding the intrinsic source spectrum and possible variability, the knowledge of the EBL, the distribution of magnetic fields in the Universe, the contribution of a putative hadronic component to the cascade, etc. Moreover, plasma instabilities may quench electromagnetic cascades, even if this effect is minor.
The central question that arises is, therefore, if the next generation of gamma-ray telescopes, whether ground- or space-based (or a combination of both), will be able to \emph{unambiguously} detect them. In this subsection we briefly revisit the theory that can be directly connected to the experiments. In particular, we highlight here the requirements for next-generation detectors to be sensitive enough to detect IGMFs.

\bigskip
In general, the detection of haloes depends on two factors. First, the size of the extended emission should be such that it is fully contained within the field of view (FoV) of the detector, of size $\theta_\text{fov}$. Second, this extension must exceed the angular resolution of the detector ($\theta_\text{psf}$). In other words, the signal can be observed if the PSF and FoV of the instruments satisfy $\theta_\text{psf} < \theta_\text{obs} < \theta_\text{fov}$. 

Current-generation IACTs (VERITAS, MAGIC, H.E.S.S.) can resolve scales of $\sim 0.08^\circ$, with a typical FoV of $\lesssim 6^\circ$. 
The upcoming Cherenkov Telescope Array~\cite{cta2019a} will reach angular resolutions as high as $\sim 0.02^\circ$ with a field of view of $\simeq 20^\circ$, improving the possible constraints on IGMFs~\cite{sol2013a, meyer2016b, gate2017a, Alonso:2018ejb}. For instance, 50~hours of observations of the blazar 1ES~0229+200 could be used to probe magnetic-field strengths of $B \sim 10^{-13.5} \; \text{G}$ at a $5\sigma$-level, for $L_{B} \gtrsim 1 \; \text{Mpc}$~\cite{cta2021b}. In particular, with angular resolutions of $\simeq 0.13^\circ$ at $E \gtrsim 100 \; \text{GeV}$, a combination of CTA and Fermi-LAT observations could also be used to probe magnetic helicity~\cite{alvesbatista2016b}, although it is not clear if any measurable signal could be extracted~\cite{Kachelriess:2020bjl}.

\begin{figure}[htb]
    \centering
    \includegraphics[width=0.495\textwidth]{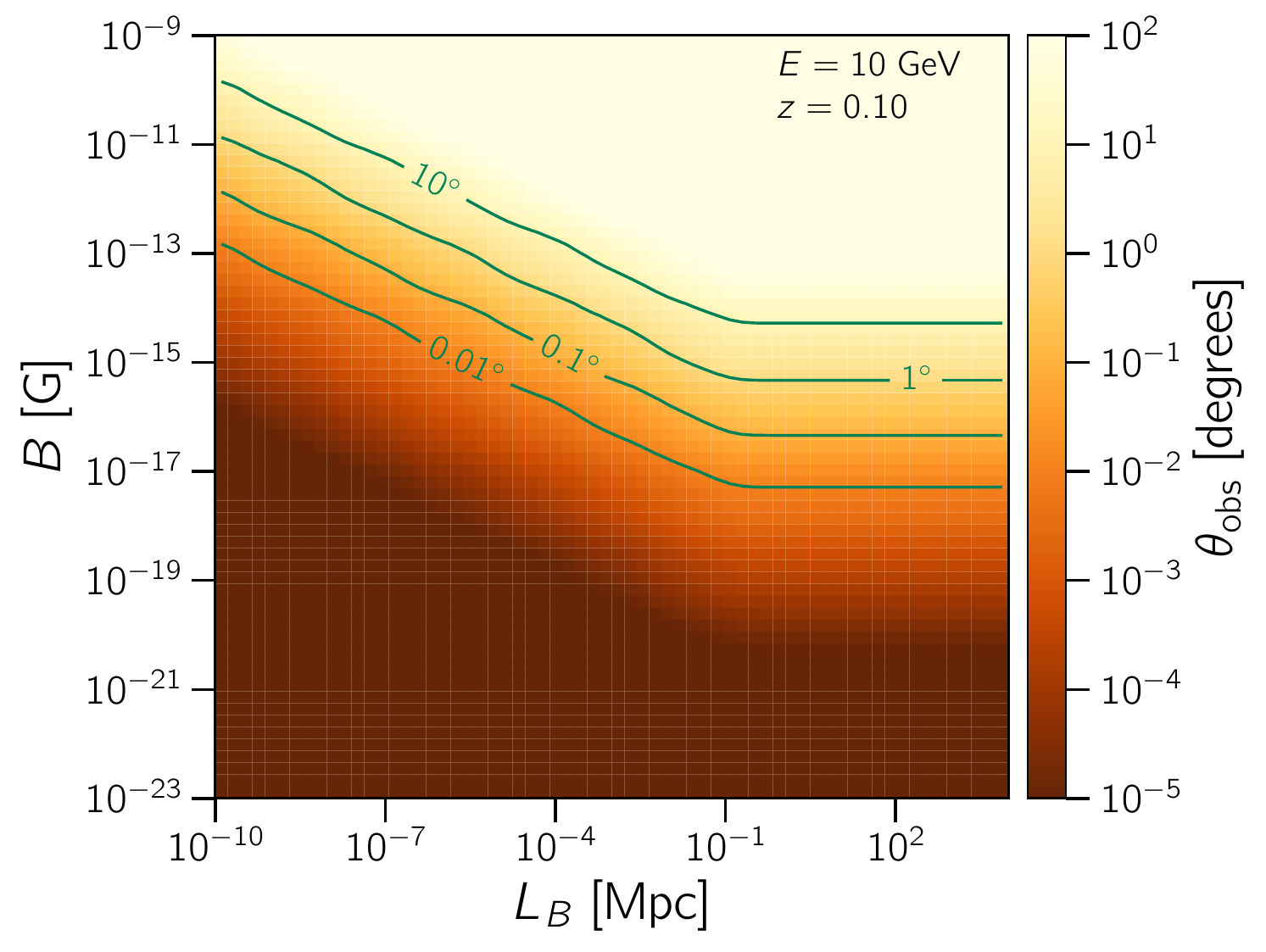}
    \includegraphics[width=0.495\textwidth]{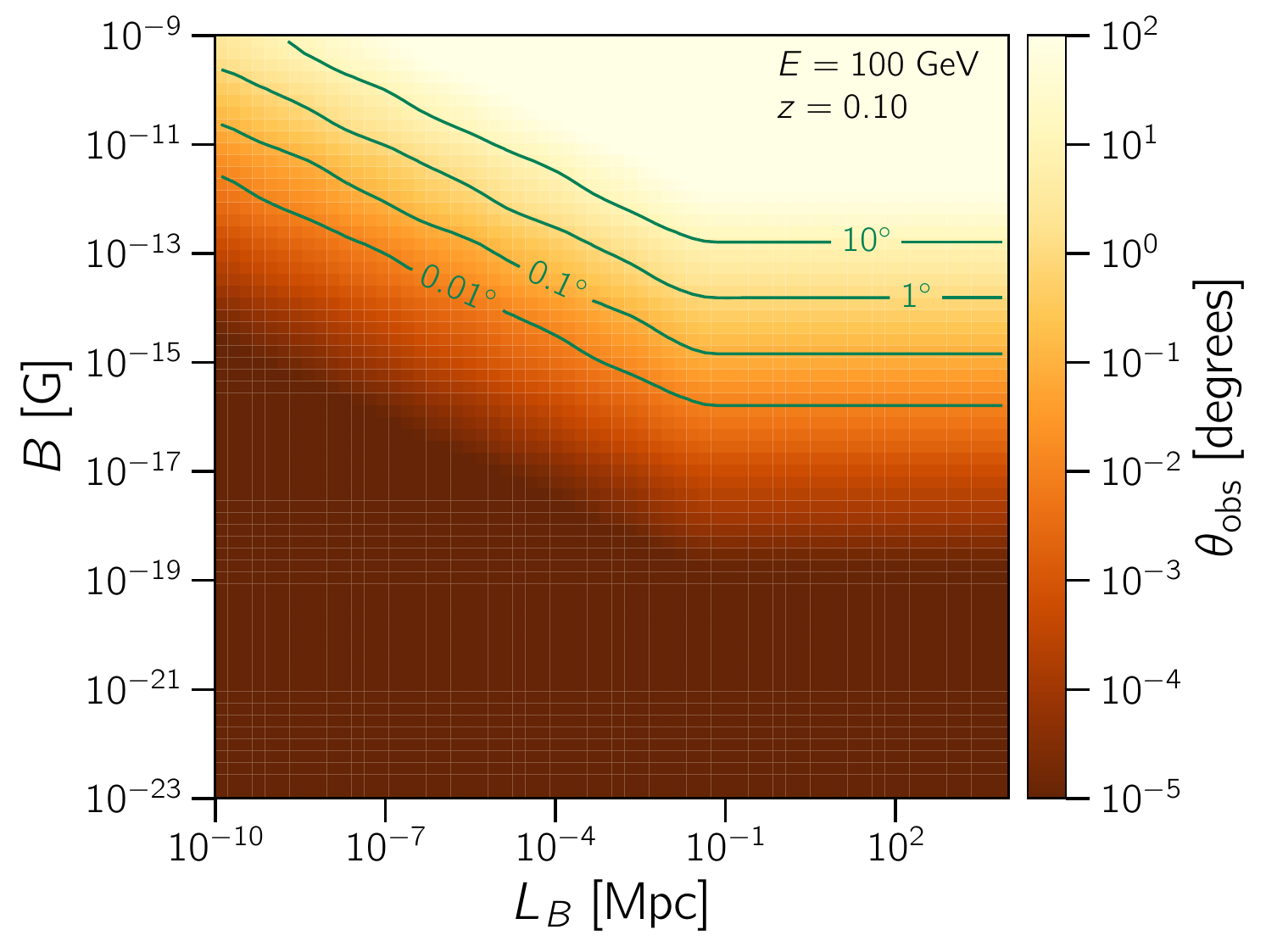}
    \caption{This figure shows the typical size of the extended emission ($\theta_\text{obs}$) for different combinations of $B$ and $L_B$, for gamma rays of energy $10 \; \text{GeV}$ (left panel) and $100 \; \text{GeV}$ (right). This example was calculated using the approximation given by Equation~\ref{eq:timeDelay}~\cite{neronov2009a}. The source is assumed to be located at a distance corresponding to redshift $z=0.10$.}
    \label{fig:constraintsProspectsHalo}
\end{figure}

Fig.~\ref{fig:constraintsProspectsHalo} shows the region of the parameter space that can be probed with the halo strategy. Note that, while the PSF must necessarily be smaller than the size of the halo for any extended emission to be identified, the condition $\theta_\text{fov} > \theta_\text{obs}$ is not a strict requirement. Nevertheless, if the halo is not entirely contained within the field of view of the instrument, it becomes difficult (but not impossible) to reconstruct the image, due to uncertainties stemming from the reconstruction procedure and the motion of the telescope to scan the region surrounding the source. Typically, IACTs have higher angular resolutions near the centre of the FoV,  decreasing radially from that point.


Fig.~\ref{fig:constraintsProspectsHalo} was obtained using simplifying assumptions, in particular Equation~\ref{eq:haloSize}, derived in~\cite{neronov2009a}. If these estimates were improved using detailed Monte Carlo simulations and instrument response functions were accounted for, then the picture could change slightly. Nevertheless, a recent work by the CTA Consortium~\cite{cta2021b} using simulations obtained with the CRPropa code is in qualitative agreement with Fig.~\ref{fig:constraintsProspectsHalo}.

\bigskip

\begin{figure}[htb]
    \centering
    \includegraphics[width=0.8\textwidth]{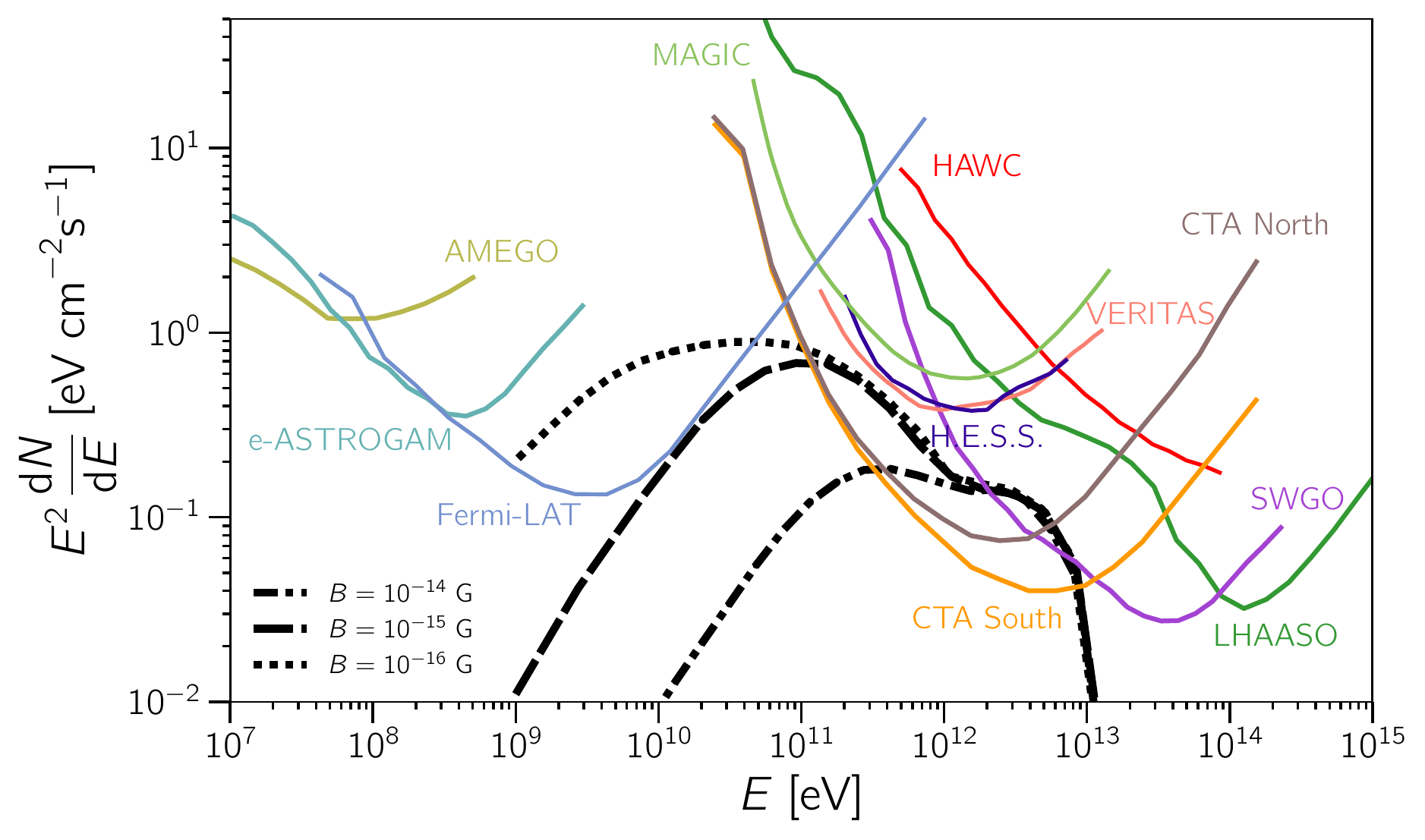}
    \caption{Comparison of the point-source sensitivity for various gamma-ray observatories. The Fermi-LAT band encompasses sources at various positions in the sky, for the \texttt{P8R3\_SOURCE\_V2} instrument response function~\cite{bruel2018a}. The sensitivities for the IACTs, namely VERITAS~\cite{park2015a}, MAGIC~\cite{magic2016a}, H.E.S.S.~\cite{holler2015a}, and CTA~\cite{cta2019a} are given for 50~hours of observations. For SWGO~\cite{swgo2019a} and LHAASO~\cite{lhasso2019a}, the curves shown are for five and one year, respectively. One year is also the observation time used to derive the sensitivity for AMEGO~\cite{mcenery2019a}. For HAWC~\cite{hawc2017c}, the curve corresponds to 507~days (which is equivalent to approximately 3000~hours) of observations. The thick black lines correspond to the simulations from Fig.~\ref{fig:simSpec}. Note that the instrument response functions of each detector are \emph{not} folded into the simulations; the corresponding sensitivities are shown here just for the sake of comparison.}
    \label{fig:sensitivities}
\end{figure}

More generally, IGMF constraints based on gamma-ray observations employing the halo strategy depend on the point-source sensitivity of the instruments, shown in Fig.~\ref{fig:sensitivities}. A simple comparison with the simulations of Fig.~\ref{fig:simSpec} demonstrates that instruments like CTA will be able to probe IGMFs stronger than $\sim 10^{-14} \; \text{G}$, as shown in Ref.~\cite{cta2021b}. The sensitivities shown in Fig.~\ref{fig:sensitivities} are a useful guide for a first assessment of the instrumental capabilities in IGMF studies through  simple comparisons with theoretical expectations (Fig.~\ref{fig:simSpec}). Nevertheless, there are multiple conceivable ways to probe IGMFs with halo searches. The simplest one is the direct search for an extended emission, as we discussed in the preceding paragraphs, but one could also employ methods involving the fit of the halo profile and comparison with the background, for example. This would lead to differences in the sensitivity curves, as discussed in detail in Ref.~\cite{sol2013a} for the case of CTA.

It is worth stressing that facilities operating at slightly higher or lower energies can play an important role in this type of study, despite being seldom considered for IGMF studies. They can be used to constrain putative PeV gamma rays as well as cascade photons in the GeV band. The current and upcoming facilities operating at higher energies, like LHAASO~\cite{lhasso2019a} and the planned Southern Wide-field Gamma-ray Observatory (SWGO)~\cite{swgo2019a}, formerly known as the Southern Gamma-ray Survey Observatory (SGSO), can help in the precise determination of the intrinsic spectrum of the sources and consequently lead to better models. Observatories  such as the planned e-ASTROGAM~\cite{eastrogam2018a} and the All-sky Medium Energy Gamma-ray Observatory (AMEGO)~\cite{mcenery2019a} can detect secondary (cascade) photons in the MeV--GeV band, thus providing additional insights into IGMFs. For the extreme blazars with hard spectra (see discussion in Sec.~\ref{ssec:sources}), in particular, this will ultimately reduce the uncertainties when constraining IGMFs. Interestingly, observations around GeV energies may also probe spectral features expected from some plasma instability models (e.g.~\cite{alvesbatista2019g}; see also Sec.~\ref{ssec:plasma}).

\bigskip

All currently operating instruments can resolve short-duration events from sources at distances closer than $z \sim 1$, probing magnetic fields with strengths $B \lesssim 10^{-17} \; \text{G}$ for $L_{B} \gtrsim 1 \; \text{Mpc}$; note that the exact value of $B$ that can be probed depends on the distance to the source. For stronger magnetic fields, however, it becomes difficult to detect time delays if they are larger than a few years or a decade. In fact, according to Eq.~\ref{eq:timeDelay}, the expected time delay for 10~GeV gamma rays assuming $B \gtrsim 10^{-17} \; \text{G}$ and $L_B \gtrsim 100 \; \text{kpc}$ would already be $\Delta t_B \gtrsim 100 \; \text{yr}$, posing obstacles for measurement within a reasonable time window of a few decades. 

\begin{figure}[htb]
    \centering
    \includegraphics[width=0.495\textwidth]{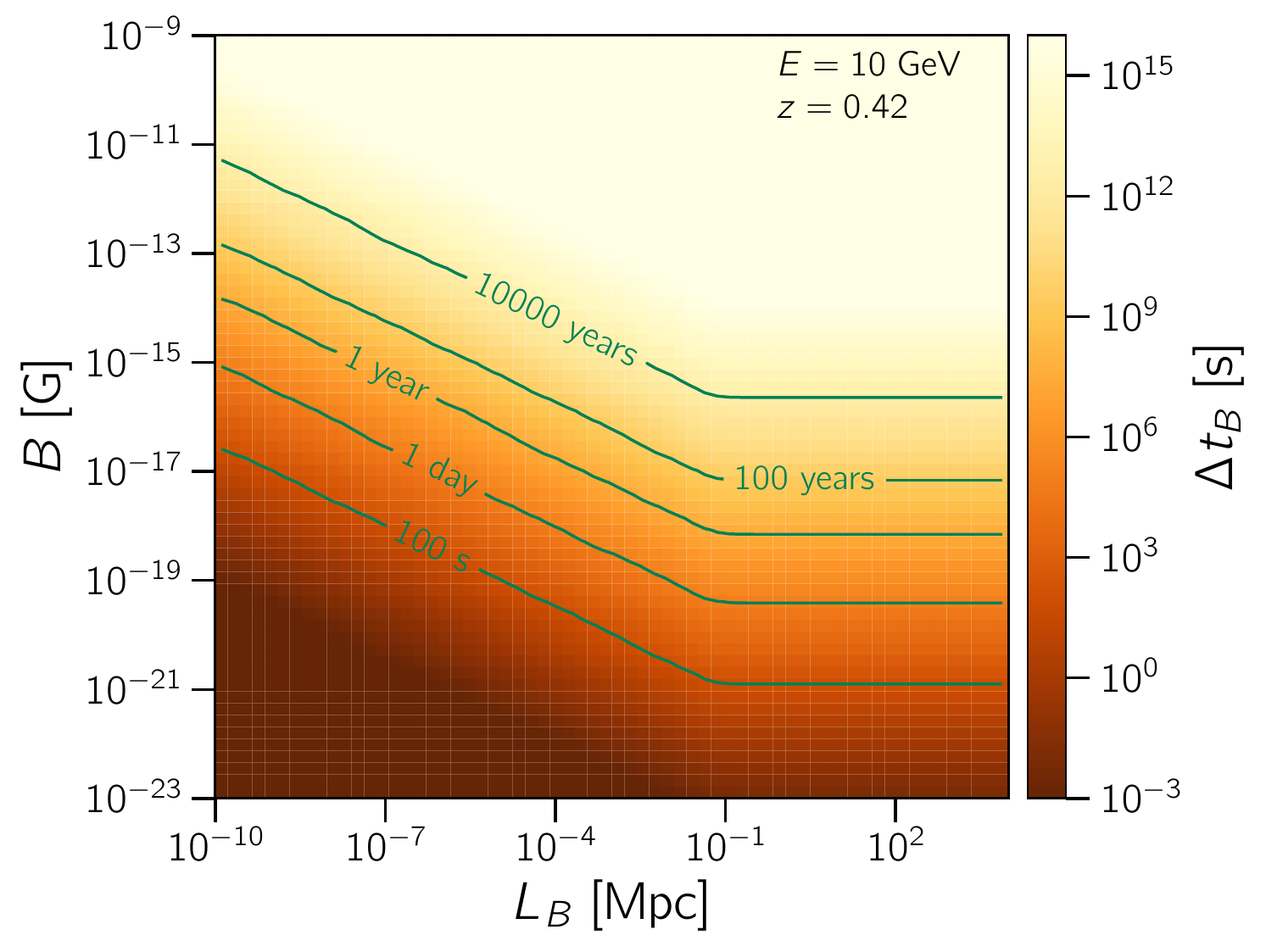}
    \includegraphics[width=0.495\textwidth]{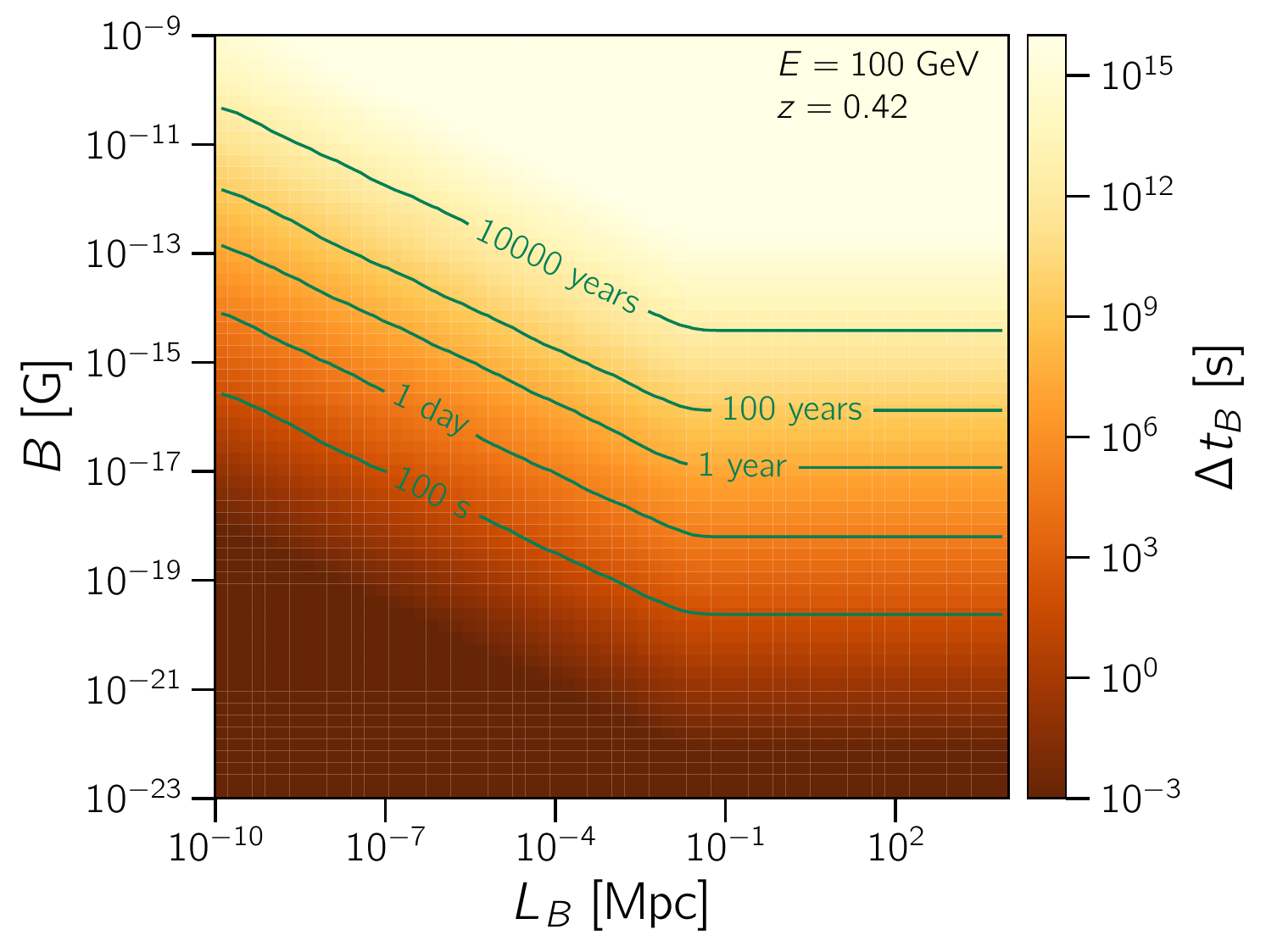}
    \caption{This figure shows the typical size of the extended emission ($\theta_\text{obs}$) for different combinations of $B$ and $L_B$, for gamma rays of energy $10 \; \text{GeV}$ (left panel) and $100 \; \text{GeV}$ (right panel). This example was calculated using the approximation given by Equation~\ref{eq:timeDelay}~\cite{neronov2009a}. The source is assumed to be located at a distance corresponding to redshift $z=0.42$.}
    \label{fig:constraintsProspectsTimeDelay}
\end{figure}

Fig.~\ref{fig:constraintsProspectsTimeDelay} shows the region of the parameter space that can be probed with the time-delay strategy. It is clearly favourable for probing the region of the parameter space corresponding to weaker magnetic fields (compared to Fig.~\ref{fig:constraintsProspectsHalo}). This particular example is for a source at redshift $z=0.42$, the same as GRB~190114C~\cite{magic2019a}.

\bigskip
In a recent work~\cite{Korochkin:2020pvg}, the prospects for measuring strong IGMFs ($B \gtrsim 10^{-12} \; \text{G}$) were analysed using the constrained cosmological simulations of the cosmic web from Ref.~\cite{jasche2013a}, based on gamma-ray observations from both Mrk~421 and Mrk~501. The authors argue that, at least for the latter object, IGMFs with $B \gtrsim 10^{-12}$--$10^{-11} \; \text{G}$ and $L_B \lesssim 10 \; \text{kpc}$ could be measured in the energy range between 1 and 10~TeV via halo searches. Such strong IGMFs could, in principle, be invoked to resolve the Hubble tension~\cite{Jedamzik:2020krr}.
\section{Outlook}\label{sec:outlook}

Following on the footsteps of pioneer ground-based gamma-ray detectors, in particular IACTs like the Whipple Observatory and HEGRA, currently-operating facilities such as H.E.S.S., VERITAS, MAGIC have made outstanding progress in studying the VHE universe. Complemented by space-borne detectors like Fermi-LAT and AGILE at energies below $\sim 100 \; \text{GeV}$, and by ground-based particle detectors such as HAWC, Tibet-AS$\gamma$, and ARGO-YBJ at higher energies ($\sim 100 \; \text{TeV}$), we have in recent years made significant progress towards understanding the Universe at high energies. 
At the dawn of the multimessenger era, the discovery potential of ground-based gamma-ray facilities can be maximised by working with other observatories across the whole electromagnetic spectrum, as well partners measuring cosmic rays, neutrinos, and gravitational waves. In this context, joint studies through multimessenger networks such as the Astrophysical Multimessenger Observatory Network (AMON)~\cite{ayalasolares2020a} are extremely useful to orchestrate campaigns of follow-up observations.
It is through coordinated efforts of multiple of these facilities that we can pave new roads to fully exploit the potential of gamma rays as probes of cosmology and fundamental physics (e.g., Lorentz invariance violation, axion-like particles; see Sec.~\ref{ssec:bsm}).
Within this landscape, we identify a unique opportunity for measuring IGMFs using gamma rays as well as other messengers.

Many challenges lie ahead in the coming decade. Firstly, it is possible that next-generation IACTs like CTA will still not be sensitive enough to enable measurements of magnetically-induced haloes. This limitation is certainly true for other ground-based instruments given the lower angular resolution of water-Cherenkov detectors compared to IACTs. 
Secondly, there are theoretical issues that need to be addressed, including the issue of plasma instabilities (see Sec.~\ref{ssec:plasma}). Moreover, future studies should start relying on more detailed magnetic-field models, capturing also the magnetisation of the cosmic web wherein the gamma-ray sources used as ``lighthouses'' to probe the cosmos are embedded. We are entering an era of precision measurements and, therefore, also require more accurate tools to model the three-dimensional propagation of electromagnetic cascades if we wish to exploit the data as much as possible. Finally, there is room for novel methods to be devised to measure IGMFs, involving, among other messengers, gamma rays. 

New insights into cosmic magnetism will be obtained with the Square Kilometre Array (SKA)~\cite{gaensler2004a, heald2020a}. Through measurements of Faraday of polarised extragalactic sources (e.g., FRBs, GRBs, quasars) SKA will deliver a tomographic map of extragalactic magnetic fields, disentangling part of the IGMF component, and offering clues on the structure and evolution of IGMFs~\cite{vacca2015a}.

\bigskip

Figures~\ref{fig:GenConstraints} and~\ref{fig:BL_summary} neatly summarise the space of parameters for IGMFs allowed by measurements. However, the landscape of cosmic magnetism is more complex than this simple two-dimensional parameter space. Besides the magnetic-field strength ($B$) and the coherence length ($L_B$), IGMFs may be helical, such that a third dimension ($H_B$) should be added to this plot. Moreover, the magnetic power spectrum ($\alpha_B$) can also play a role in the development of electromagnetic cascades, adding a fourth dimension. It is manifestly difficult to scan over all these parameters ($B$, $L_B$, $H_B$, and $\alpha_B$) simultaneously. Still, these caveats should be borne in mind when constraining IGMFs, since there might be degeneracies. In this context, observation of gamma-ray sources can play an important role, given its ability to probe all these parameters. Nevertheless, besides more sensitive gamma-ray observatories, theoretical efforts in this direction are needed.

With the promising prospects for measuring IGMFs using next-generation gamma-ray observatories, we can invert the reasoning presented in Sec.~\ref{sec:FurterConstraints}: from the measurements, assuming IGMFs have a cosmological origin, we could constrain certain aspects of cosmology by inferring the specific parameters that characterise them. In fact, all the IGMF parameters mentioned in the previous paragraph may, in principle, be used for this purpose. With the measurement of $B$, one directly obtains the overall energy content of IGMFs. Like any other form of energy permeating the Universe, this would have an immediate impact in its global evolution, such that it could be necessary to consider this contribution as an addition to the standard $\Lambda$CDM model. Moreover, measurements of both the spectral index ($\alpha_{B}$) and the coherence length ($L_{B}$) could be used to constrain the major processes from which they originate, like Inflation, QCDPT and EWPT. In the case of a phase transition, in particular, these measurements could allow us to infer its order. Finally, the measurement of a non-zero magnetic helicity ($H_{B}$) would strongly hint at a general CP violation in the Universe, with clear implications for various aspects of particle cosmology.

\bigskip
In summary, it is fair to say that gamma rays represent a unique observational window into cosmic magnetism. With the advances in gamma-ray astronomy, we could already  capitalise on this window of opportunity to better understand IGMFs and to start constraining the $B$-$L_B$ parameter space. In the coming decades, the next generation of instruments might improve our understanding of cosmic magnetism more than ever, probing magnetism at cosmological scales and providing us a glimpse into the mechanisms whereby magnetic fields originated.

\bigskip

\vspace{6pt} 




\funding{R.A.B. is currently funded by the Radboud Excellence Initiative. The work of A.S. is supported by the Russian Science
Foundation under grant no.~19-71-10018.}

\acknowledgments{We are grateful to Karsten Jedamzik, Michael Kachelrie\ss, and Tanmay Vachaspati for valuable comments and suggestions which helped us to improve the quality of this review.}

\conflictsofinterest{The authors declare no conflict of interest.} 

\abbreviations{The following abbreviations are used in this manuscript:\\

\noindent 
\begin{longtable}{@{}ll}
AGN & active galactic nucleus \\
ALP & axion-like particle \\
AMEGO & All-sky Medium Energy Gamma-ray Observatory \\
AMON & Astrophysical Multimessenger Observatory Network \\
ARGO-YBJ & Astrophysical Radiation with Ground-based Observatory at YangBaJing \\
ASTRI & Astrofisica con Specchi a Tecnologia Replicante Italiana \\
BBN & Big Bang nucleosynthesis \\
BL~Lac & BL Lacertae \\
BSM & beyoud the Standard Model \\
C.L. & confidence level \\
CMB & cosmic microwave background \\
CRB & cosmic radio background \\
CTA & Cherenkov Telescope Array \\
DGRB & diffuse gamma-ray background \\
DPP & double pair production \\
EBL & extragalactic background light \\
EGRET & Energetic Gamma-Ray Experiment Telescope \\
EWPT & electroweak phase transition\\
Fermi-LAT & Fermi Large Area Telescope \\
FoV & field of view \\
FRB & fast radio burst \\
FSRQ & flat-spectrum radio quasar \\
GRB & gamma-ray burst \\
HAWC & High Altitude Water Cherenkov Experiment\\
H.E.S.S. & High-Energy Stereoscopic System \\
IC & inverse Compton \\
IGM & intergalactic medium \\
IGMF & intergalactic magnetic field \\
$\Lambda$CDM & Lambda cold dark matter \\
LHAASO & Large High Altitude Air Shower Observatory \\
LIV & Lorentz invariance violation \\
LOFAR & Low-Frequency Array \\
MAGIC & Major Atmospheric Gamma Imaging Cherenkov \\
MHD & magnetohydrodynamics \\
PMF & primordial magnetic field \\
PP & pair production \\
PSF & point spread function \\
QCDPT & quantum chromodynamics phase transition\\
SED & spectral energy distribution \\
RM & rotation measure \\
SGSO & Southern Gamma-ray Survey Observatory \\
SKA & Square Kilometre Array \\
SM & Standard Model of particle physics \\
SWGO & Southern Wide-field Gamma-ray Observatory \\
TPP & triplet pair production \\
VERITAS & Very Energetic Radiation Imaging Telescope Array System  \\
VHE & very-high-energy \\
UHECR & ultra-high-energy cosmic ray
\end{longtable}}



\externalbibliography{yes}
\bibliography{bib.bib}


\end{document}